\documentclass[12pt]{article}
\pdfoutput=1

\usepackage{bm,amsmath,amssymb,latexsym,graphicx,euscript,multirow,color,url,verbatim}
\usepackage{epsf}
\usepackage[compress,numbers]{natbib}
\usepackage{enumerate}
\usepackage[colorlinks=true,urlcolor=blue,anchorcolor=blue,citecolor=blue,filecolor=blue,linkcolor=blue,menucolor=blue,pagecolor=blue]{hyperref}

\allowdisplaybreaks

\makeatletter
\renewcommand{\@dotsep}{1000}
\makeatother

\newcommand{\captionfonts}{\small}
\makeatletter
\long\def\@makecaption#1#2{%
  \vskip\abovecaptionskip
  \sbox\@tempboxa{{\captionfonts #1: #2}}%
  \ifdim \wd\@tempboxa >\hsize
    {\captionfonts #1: #2\par}
  \else
    \hbox to\hsize{\hfil\box\@tempboxa\hfil}%
  \fi
  \vskip\belowcaptionskip}
\makeatother

\newcommand{\Sla}[1]%
{\kern0.12em{\raise.15ex\hbox{$/$}\kern-.74em #1}}% Feynman slash

\numberwithin{equation}{section}

\long\def\symbolfootnote[#1]#2{\begingroup%
\def\thefootnote{\fnsymbol{footnote}}\footnote[#1]{#2}\endgroup}

\newcommand{\be}{\begin{equation}}
\newcommand{\ee}{\end{equation}}

\newcommand{\bea}{\begin{eqnarray}}
\newcommand{\eea}{\end{eqnarray}}
\newcommand{\mat}{\begin{pmatrix}}
\newcommand{\rix}{\end{pmatrix}}

\renewcommand{\l}{\left}
\renewcommand{\r}{\right}

\renewcommand{\bar}{\overline}
\renewcommand{\slash}[1]{#1\!\!\!/}

\newcommand{\go}{{\tilde g}}
\newcommand{\Bo}{{\tilde B}}
\newcommand{\Wo}{{\tilde W}}
\newcommand{\Ho}{{\tilde H}}

\newcommand{\cho}{{\tilde \chi}}
\newcommand{\sq}{{\tilde q}}
\newcommand{\st}{{\tilde t}}
\newcommand{\sbo}{{\tilde b}}
\newcommand{\slep}{{\tilde\ell}}
\newcommand{\snu}{{\tilde\nu}}
\newcommand{\stau}{{\tilde\tau}}
\newcommand{\q}{\quad}
\newcommand{\qq}{\qquad}

\newcommand{\beqa}{\begin{eqnarray}}
\newcommand{\eeqa}{\end{eqnarray}}

\newcommand{\beq}{\begin{equation}}
\newcommand{\eeq}{\end{equation}}
\newcommand{\hc}{\mbox{h.c.}}

\newcommand{\order}[1]{{\cal O}\left(#1\right)}

\newcommand{\met}{{\slash E_T}}
\newcommand{\bs}{\boldsymbol}

\begin{document}

% =============================================================================
% Title page
% =============================================================================
\begin{titlepage}

\begin{flushright}
\small{RUNHETC-2012-18}\\
\end{flushright}

\vspace{0.5cm}
\begin{center}
\Large\bf
LHC Coverage of RPV MSSM with Light Stops
\end{center}

\vspace{0.2cm}
\begin{center}
{\sc Jared A. Evans\symbolfootnote[4]{jaredaevans@gmail.com} and Yevgeny Kats\symbolfootnote[1]{kats@physics.rutgers.edu}
}\\
\vspace{0.6cm}
\small \textit{New High Energy Theory Center\\
Rutgers University, Piscataway, NJ 08854, USA}
\end{center}

\vspace{0.5cm}
\begin{abstract}
\vspace{0.2cm}
\noindent

We examine the sensitivity of recent LHC searches to signatures of supersymmetry with $R$-parity violation (RPV). Motivated by naturalness of the Higgs potential, which would favor light third-generation squarks, and the stringent LHC bounds on spectra in which the gluino or first and second generation squarks are light, we focus on scenarios dominated by the pair production of light stops. We consider the various possible direct and cascade decays of the stop that involve the trilinear RPV operators.  We find that in many cases, the existing searches exclude stops in the natural mass range and beyond.  However, typically there is little or no sensitivity to cases dominated by UDD operators or LQD operators involving taus.  We propose several ideas for searches which could address the existing gaps in experimental coverage of these signals.

\end{abstract}
\vfil

\end{titlepage}

\tableofcontents

% =============================================================================
\section{Introduction}
\label{sec:intro}
% =============================================================================

The very successful first years of running at the Large Hadron Collider (LHC) have created an exciting atmosphere for particle physics.  With 5~fb$^{-1}$ of 7~TeV data analyzed by ATLAS and CMS, the LHC experiments have significantly tightened bounds on models of new physics.  For decades, supersymmetric (SUSY) theories have been some of the leading candidates to resolve the few tensions with the Standard Model (SM).  The minimal supersymmetric standard model (MSSM) has been viewed as one of the most well-motivated candidates for TeV scale physics, so the LHC's overwhelming validation of the SM and the apparent absence of MSSM signals thus far has proven quite startling.  At the same time, the current null results constrain parameter space, providing the community with a vision of where to search for the MSSM, if it is present.

In the MSSM, obtaining the electroweak symmetry breaking scale without fine tuning requires that the SUSY breaking soft mass parameters of the right-handed stop and the left-handed stop-sbottom doublet are not too large.  This condition, naturalness, suggests that the two stops and at least one of the sbottoms should not be heavier than roughly 500~GeV (for a recent review, see~\cite{Papucci:2011wy}). However, results from the LHC indicate that, in most models, gluinos and most of the squarks must be beyond this natural range of masses. Depending on the details of the spectrum, the lower bounds on the masses of these particles vary between about 600~GeV and above 1~TeV for both $R$-parity conserving (see, e.g.,~\cite{CMS-PAS-SUS-11-016,CMS-SUSY-URL,ATLAS-SUSY-URL,Kats:2011qh,Papucci:2011wy}) and $R$-parity violating (RPV) scenarios (see, e.g.,~\cite{Allanach:2012vj,CMSmultileptons,Asano:2012gj}). This motivates us to focus our attention on scenarios in which the third generation squarks are much lighter than all other colored superpartners.

Additionally, the discovery of a particle consistent with a Higgs boson at $m_h \approx 126$~GeV suggests a splitting of more than a few hundred GeV between the two stops, however this does not need to be the case in extensions of the MSSM (see, e.g.,~\cite{Hall:2011aa}).  Extensions of the MSSM are strongly motivated by the fact that a Higgs of this mass makes the MSSM relatively fine-tuned (unnatural) for any choice of the soft parameters. Depending on the nature of the extension, production and decay processes may be affected. However, since no very compelling extension has been suggested thus far and it is conceivable that the processes we consider will not be affected, we believe that studying MSSM processes would be most useful at this stage.  Without knowledge of whether the heaviness of the Higgs is due to new states or whether the required level of fine-tuning in the MSSM is realistic, it is reasonable to consider scenarios with either a large or small splitting between the two stops, yielding minimal spectra in which either a single stop, or both stops and a sbottom, are light.

Recently, there have been many theoretical studies on the collider signatures of light directly-produced third-generation squarks in the $R$-parity conserving MSSM (e.g.,~\cite{Papucci:2011wy,Kats:2011it,Essig:2011qg,Brust:2011tb,Bai:2012gs,Plehn:2012pr,Han:2012fw,Alves:2012ft,Kaplan:2012gd}), feeding into a vast experimental program devoted to hunting for such stops and sbottoms. Much less work has been done (e.g.,~\cite{Brust:2012uf}) on exploring the rich variety of possible signatures of third-generation squarks in RPV scenarios.

One of the reasons that RPV scenarios have been under-studied is the larger parameter space~-- RPV models add several dozen new parameters to those of the MSSM. However, by restricting ourselves to the light stop, a general study of the RPV phenomenology is feasible. Another reason that $R$-parity conserving scenarios have received more attention is that $R$-parity avoids unacceptably large proton decay rates and a plethora of other disasters (for reviews, see~\cite{Dreiner:1997uz,Barbier:2004ez}).  Additionally,  $R$-parity conservation makes the lightest superpartner stable, allowing it (if neutral) to serve as a dark matter candidate.  Still, $R$-parity conservation is merely a phenomenologically motivated assumption.  An alternative to forbidding RPV couplings completely is setting many of them to be very small.  There is no \emph{a priori} reason that a technically natural hierarchy of RPV couplings, which satisfy phenomenological constraints, cannot be realized in nature (in fact, various theoretical frameworks have been shown to give rise to such situations, e.g.,~\cite{Hall:1983id,Csaki:2011ge}). It is therefore important to experimentally address all possible signatures of the RPV MSSM, especially if ongoing searches exclude the naturally light stop in the $R$-parity preserving MSSM.

In this paper, we will present a classification of the RPV scenarios with a light stop, using simplified models in which a single trilinear RPV coupling dominates, and examine to what extent each model is constrained by current LHC searches.  We focus on the case where one of the stops is somewhat lighter than the other third generation squarks, so that it dominates the production, since typically this would be the most conservative scenario.  Production of the second stop and a sbottom will be studied as well, although with less generality.  Many of the signatures we will study are sufficiently simple that one could also imagine them arising outside the context of the RPV MSSM.  In effect, the RPV MSSM can be viewed as a ``signature generator'' with the various possible final states viewed as simplified topologies~\cite{Alves:2011wf} for generic models of new physics.  We hope that our results will serve as guidance for designing new searches to cover the regions of parameter space where existing analyses are not sensitive.  In addition, our results can be useful for guiding theoretical model building by elucidating the experimental status of the various RPV scenarios, very few of which have been explicitly analyzed by the experiments thus far.

% OUTLINE OF PAPER
Section~\ref{sec:RPV} provides a brief overview of $R$-parity violation and the simplifying assumptions that will be made in the context of this study.  Section~\ref{sec:RPVstops} discusses the ways in which the stop may decay, either directly or through other superpartners.  In section~\ref{sec:limits-singlestop}, we use a large set of recent ATLAS and CMS searches to derive bounds on the various scenarios from direct stop pair production.  We then, in section~\ref{sec:limits-twostops}, study how the bounds are improved if the second stop and a sbottom are sufficiently light to be readily produced. Section~\ref{sec:Discussion} both discusses the scenarios in which naturally light stops are not excluded by the existing searches and proposes strategies which may better address these signatures.

% =============================================================================
\section{$R$-parity violation in the MSSM}
\label{sec:RPV}
% =============================================================================

When one lists all possible renormalizable couplings in the superpotential that respect the gauge symmetries of the MSSM, four types of ``disastrous'' terms appear,\footnote{Here the $L_i$ are the left-handed lepton doublets, $E^c_i$ the right-handed leptons, $Q_i$ the left-handed quark doublets, $U^c_i$ and $D^c_i$ the right-handed quarks, and $H_u$ is the Higgs that gives mass to the up-type quarks; $i,j,k=1,2,3$ are generation indices. While the couplings are potentially complex, the phase is not important because only the magnitude, $\left| \lambda \right|^2$, enters in the processes we study in this paper.}
\beq
\label{eq:superpot}
W \supset \frac12\lambda_{ijk} L_iL_j E^c_k + \lambda_{ijk}' L_iQ_j D^c_k + \frac12\lambda_{ijk}'' U^c_iD^c_j D^c_k + \mu_i L_i H_u
\eeq
These terms are disastrous because they all violate baryon and/or lepton number. Together, they can lead to proton decay and other processes which have very strict bounds. Therefore, these couplings are often assumed to vanish, which can be obtained by assuming the existence of the ``$R$-parity'' symmetry.  More generally, some of these couplings may be non-zero; one only needs to assume that the combinations of couplings that have stringent bounds are sufficiently small. Additionally, soft SUSY-breaking RPV couplings are possible:
\beq
\label{eq:softRPV}
\mathcal L \supset \frac12 A_{ijk} \tilde{L}_i\tilde{L}_j \tilde{\ell}^c_k + A_{ijk}' \tilde{L}_i\tilde{Q}_j \tilde{d}^c_k + \frac12 A_{ijk}'' \tilde{u}^c_i\tilde{d}^c_j \tilde{d}^c_k + B_i \tilde{L}_i h_u + {\tilde m}^2_{di} h_d^\dagger \tilde{L}_i  +\hc
\eeq
See~\cite{Barbier:2004ez} for a thorough review of $R$-parity violating operators.

%ignore bilinear couplings
In this work, we will focus on the trilinear couplings $\lambda$, $\lambda'$ and $\lambda''$, which give rise to vertices coupling a sfermion to two Standard Model fermions.  The bilinear $\mu_i$ terms in the superpotential, which mix the lepton and Higgs superfields, can be rotated away, adding contributions to the $\lambda$ and $\lambda'$ terms (primarily $\lambda_{i33}$ and $\lambda'_{i33}$). The soft bilinear couplings, $B_i$ and ${\tilde m}^2_{di}$, may introduce further mixings, but we will assume them to be negligible.\footnote{If this is not the case, two-body decays of neutralinos and charginos to a charged lepton or neutrino, and a $W$, $Z$, or Higgs, become possible. The focus of this work is on the trilinear couplings, and exploration of this possibility is left for future work.}
These soft couplings can be absent at the messenger scale if, for example, SUSY breaking is mediated by gauge interactions.  The couplings remain small at the weak scale if the messenger scale is sufficiently low.

The RPV contributions to the supersymmetric scalar potential, and the soft RPV ``$A$-terms'' of eq.~(\ref{eq:softRPV}), $A$, $A'$ and $A''$, will also be ignored. Since these couplings always involve at least one additional squark while we assume the stop to be the lightest colored superpartner, the resulting decays are unlikely to be important because they would be at least three-body and suppressed by at least one additional RPV coupling.

%single RPV coupling dominance
While one could consider many different kinds of hierarchies between the RPV couplings, or be guided by a specific hypothesis such as minimal flavor violation (MFV)~\cite{Nikolidakis:2007fc,Csaki:2011ge}, we will base our study on simplified models in which a single RPV coupling, with a single choice of the generation indices $i,j,k$,\footnote{Barring the antisymmetry conditions $\lambda_{ijk} = -\lambda_{jik}$ and $\lambda''_{ijk} = -\lambda''_{ikj}$, the couplings are independent.} is assumed to dominate the phenomenology. These models are automatically viable from the point of view of indirect constraints, since bounds on individual couplings are quite weak~\cite{Chemtob:2004xr,Barbier:2004ez,Kao:2009fg}: for TeV scale sfermions, with only a few exceptions, the bounds are $\order{0.1}$ or higher. Each simplified model directly corresponds to a small number of particular collider signatures, making the connection with experimental searches straightforward.   To obtain bounds on a scenario in which decays through more than one RPV coupling are relevant, one may rescale our cross section limits by two powers of the branching fraction.

% =============================================================================
\section{Overview of signatures}
\label{sec:RPVstops}
% =============================================================================

As was discussed in the Introduction, the combination of naturalness arguments with the null results in current LHC new physics searches suggests that one should consider spectra in which the two stops and at least one sbottom are much lighter than most of the other colored superpartners. This motivates us to make the conservative assumption that the dominant SUSY production mechanism is the pair production\footnote{If the UDD coupling $\lambda''_{312}$ is sufficiently large, single stop production, $\bar d\bar s \to \st$, can also be observable. We will not consider it here. See~\cite{Desai:2010sq} for a recent study of this possibility.} of the lighter stop ($\st_1$, which we will denote simply as $\st$).  In section~\ref{sec:limits-twostops}, we will include the production of the heavier stop ($\st_2$) and the sbottom ($\sbo_1$) and show that even if these particles are not very much heavier than $\st_1$, the improvements in the limits need not be drastic.  For that reason, the limits we derive for $\st_1\st_1^\ast$ production are realistic despite being conservative. We will start by considering all possible decay topologies of the stop in the RPV MSSM.

\subsection{Stop decay modes}

The stop may decay directly via an RPV coupling as
\be
\st \to \ell j \qq\mbox{(LQD with $\lambda'_{i3k}$)}
\ee
or
\be
\st \to j j \qq\mbox{(UDD with $\lambda''_{3jk}$)}
\ee
where $j$ is a quark or antiquark (a jet, possibly a $b$-jet) and $\ell$ is a charged lepton ($e$, $\mu$ or $\tau$).

\begin{figure}[t]
\begin{center}
\includegraphics[scale=1.3]{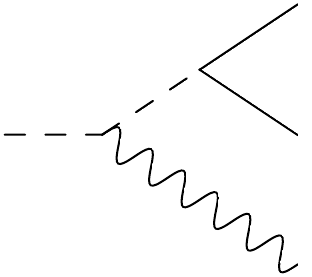}
\begin{picture}(30,0)(0,0)
\large
\put(-130,50){$\st$}
\put(-3,100){$f_1$}
\put(-3,50){$f_2$}
\put(-3,0){$W^+,H^+$}
\put(-73,73){$\sbo_R$}
\put(75,50){$\st$}
\put(230,100){$t,b$}
\put(230,65){$f_2$}
\put(230,35){$f_3$}
\put(230,0){$f_1$}
\put(170,50){$\tilde{f}$}
\put(130,22){$\tilde{X}$}
\normalsize
\end{picture}
\hspace{1.5cm}
\includegraphics[scale=1.3]{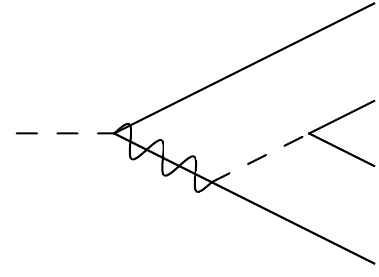}
\begin{picture}(10,0)(0,0)\end{picture}
\begin{minipage}[t]{7.0in}%
\caption{\label{fig:3b4bstop}  Left: Stop decays through an off-shell right-handed sbottom.
Right: Four-body cascade decays of the stop, where the intermediate particles ($\tilde X = \cho^0$, $\cho^+$, or $\go$, and $\tilde f = \snu$, $\slep$, or $\sq$) may or may not be on-shell.  For $\tilde X$ much heavier than the stop, diagrams with a helicity flip on the $\tilde X$ propagator will dominate, as discussed in appendix~\ref{app:helicity}. The symbols $f_1$, $f_2$, $f_3$ denote Standard Model fermions.}
\end{minipage}
\end{center}
\end{figure}

Alternatively, if the Higgsino or one of the gauginos is lighter than the stop and/or the RPV couplings that allow direct decays happen to be sufficiently small, the stop decay may proceed via (on-shell or off-shell) superpartners. The possibilities are:
\be
\st \;\to\;
\cho^0\,t \,,\q
\cho^+\,b \,,\q
\go\,t \,,\q
\sbo_R\,W^+ ,\q
\sbo_R\,H^+
\ee
with the superpartners decaying as
\bea
\cho^0 \;&\to&\;
\ell\ell\nu \;\mbox{(LLE)}  \q\mbox{or}\q
\ell j j,\, \nu j j\;\mbox{(LQD)}  \q\mbox{or}\q
j j j \;\mbox{(UDD)}
\label{eq:N-decay}
\eea
\bea
\cho^+ \;&\to&\;
\ell\ell\ell,\, \ell\nu\nu \;\mbox{(LLE)}  \q\mbox{or}\q
\ell j j,\, \ell t j,\, \nu j j \;\mbox{(LQD)}  \q\mbox{or}\q
j j j \;\mbox{(UDD)}
\label{eq:C-decay}
\eea
\bea
\go \;&\to&\;
\ell j j,\, \nu j j\;\mbox{(LQD)}  \q\mbox{or}\q
j j j \;\mbox{(UDD)}
\eea
\bea
\sbo_R \;&\to&\;
\ell j,\, \nu j \q\mbox{(LQD with $\lambda'_{ij3}$)}  \q\mbox{or}\q
j j \q\mbox{(UDD with $\lambda''_{ij3}$)}
\eea
Here, the sbottom decays directly through an RPV coupling (see figure~\ref{fig:3b4bstop}, left),\footnote{We only consider a right-handed sbottom, since RPV couplings through which a left-handed sbottom can decay allow the stop (which we assume to be at least somewhat mixed) to decay directly.} while the ``inos'' (Higgsinos and gauginos), which are not present in any of the trilinear RPV terms, decay through a sfermion, as illustrated in figure~\ref{fig:3b4bstop}, right.   As an example, with the $\lambda'_{123}$ coupling (that is, first generation lepton doublet, second generation quark doublet and a right handed $b$), one finds the six competing final states of $\st \to t \Wo^{0(*)}\to t e^- c \bar b$, $ t e^+ \bar c b$, $ t \nu_e s \bar b$ and $ t \bar\nu_e \bar s b$ for neutral wino mediated decays and $\st \to b \Wo^{+(*)}\to b \nu_e c \bar b$ and $ b e^+ \bar s b$ for charged wino mediated decays.

If the stop decays most of the time into an on-shell neutralino or chargino while the RPV couplings involving the stop are non-negligible, these inos may preferentially decay back through the original stop (i.e., $\tilde f$ in figure~\ref{fig:3b4bstop} is $\st$).  If only the neutralino is present (the predominantly bino case), this allows for four-top final states
\be
\st \to \cho^0\,t,\,\q
\cho^0 \;\to\;
t\ell j \q\mbox{(LQD with $\lambda'_{i3k}$)}  \q\mbox{or}\q
t j j \q\mbox{(UDD with $\lambda''_{3jk}$)}
\label{eq:N-stop-decay}
\ee
which would sometimes be easier to discover than the direct RPV decays through the same couplings. Analogously, the decay to a chargino in the wino or Higgsino case would give rise to
\be
\st \to \cho^+\,b,\,\q
\cho^+ \;\to\;
\ell b j \q\mbox{(LQD with $\lambda'_{i3k}$)}  \q\mbox{or}\q
b j j \q\mbox{(UDD with $\lambda''_{3jk}$)}
\label{eq:C-stop-decay}
\ee
which yields events with a large number of $b$-jets. Of course, diagrams where the ino decays through other sfermions may also contribute, with final states as in (\ref{eq:N-stop-decay})--(\ref{eq:C-stop-decay}), but, in the LQD case, with the additional final states of
\be
\cho^0 \;\to\; b\nu j,\qq
\cho^+ \;\to\; \nu t j \q\mbox{(LQD with $\lambda'_{i3k}$)}
\ee

Multistage cascade decays that involve several inos are also possible. To keep the parameter space manageable, we will only treat the possibilities that may appear naturally within the wino system, which contains both a chargino and a neutralino, or the Higgsino system, which contains a chargino and two neutralinos.  If there are no light sfermions to which the inos can decay, it is plausible for a heavier member of the ino multiplet to transition into a lighter one (emitting a couple of soft SM particles~-- the decay products of an off-shell $W$ or $Z$), which will only then undergo an RPV decay through an off-shell sfermion.  If the transition is between two neutralinos, the RPV decay products remain the same, so there is no need to consider such cases separately. On the other hand, if the transition is between a chargino and a neutralino,\footnote{We assume the chargino to be heavier than the neutralino. Scenarios with the opposite hierarchy are possible, but rare.}
\be
\cho_1^\pm \to \cho_1^0\, W^{(\ast)\pm}
\label{eq:ino-cascade}
\ee
the resulting final state of the stop decay is different. Whether the chargino decays as in (\ref{eq:ino-cascade}) or via a sfermion and its RPV coupling as in (\ref{eq:C-decay}), is model-dependent, so we will study both possibilities.  Similar transitions between on-shell sleptons are also possible, but we will not consider the possibility that they dominate over the two-body RPV decays of the sleptons.

Even with only a single RPV coupling, several decay paths (through different mediators) can potentially contribute, each leading to a different final state. Which of these dominates depends on the masses and couplings of the intermediate particles and the available phase space.  We stress that the $R$-parity conserving decay of the stop into an on-shell ino, which subsequently decays via RPV, is a very realistic scenario.  Naturalness suggests the Higgsinos should not be much heavier than $\sim 200$~GeV (see, e.g.,~\cite{Papucci:2011wy}),\footnote{In extensions of the MSSM, the Higgsinos may be heavier (e.g., up to $\sim 350$~GeV~\cite{Hall:2011aa}) and still be natural.} and nothing forbids the bino and winos from being light.   In order to span the various possibilities, we will construct a separate simplified model for each RPV coupling and each type of mediator, with several benchmark points for the mediator's mass (both lighter and heavier than the stop), as will be described in more detail in the next section.

\subsection{Lifetime}

Depending on the details of the model, the decay length of the stop (or the superpartners it decays into) can vary from prompt (even in the case of four-body decays through off-shell mediators) to detector-stable. For example, a generic expression for the four-body decay rates of the stop would be
\be
\Gamma \sim \frac{\lambda^2\alpha^2}{3072\pi^3}\l(\frac{m_\st}{m_{\tilde X}}\r)^2 \l(\frac{m_\st}{m_{\tilde f}}\r)^4 m_\st
\ee
where $\lambda$ is the relevant RPV coupling, and $4\pi\alpha_i = g_i^2$ for gauginos and the product of the relevant effective Yukawa couplings for Higgsinos. The resulting decay length can be written as
\be
c \tau \sim \l(6\times 10^{-5}\mbox{ mm}\r) \l(\frac{0.01}{\alpha}\r)^2 \l(\frac{0.1}{\lambda}\r)^2 \l(\frac{m_{\tilde X}}{m_\st}\r)^2 \l(\frac{m_{\tilde f}}{m_\st}\r)^4 \l(\frac{300\mbox{ GeV}}{m_\st}\r)
\ee
If the intermediate particles are not much heavier than the stop and the RPV coupling is relatively large, the stop will decay promptly, while for a small RPV coupling and/or heavy mediators the stop can easily be long-lived.

Detector-stable stops have already been excluded by LHC searches in the natural range of stop masses~\cite{Chatrchyan:2012sp,ATLAS-CONF-2012-075}.  The same searches could likely set very strong limits on scenarios where the stop decays promptly to a detector-stable charged particle, such as a charged slepton or a chargino; it would be useful for these searches to recast their analysis to cover such scenarios. Cases with a long-lived neutral particle are covered by $\slash E_T$-based searches for $R$-parity conserving SUSY with prompt decays. The intermediate case of displaced decays within the detector is less straightforward. Interpreting the results of the existing displaced decay searches~\cite{CMS-PAS-EXO-11-101,ATLAS-CONF-2012-022,ATLAS:2012jp,Aad:2012zx} in the context of the variety of decays that we will consider here is beyond the scope of this work.  Doing so reliably would require a close familiarity with the details of the detector and the subtleties of the particular analyses. Also, it is not obvious that gluinos and/or all the squarks are excluded from being as light as the stops in such case.\footnote{For a recent discussion on searches for displaced decays in the context of a specific RPV model, see~\cite{Graham:2012th}.} We will therefore restrict ourselves to the case of prompt decays. We hope that our discussion of the various stop decay signatures will motivate the experiments to include them as additional benchmark models in searches for displaced decays.

The possible longevity of the stop in RPV scenarios may allow processes through which a stop oscillates to an antistop (or vice versa) after hadronization~\cite{Sarid:1999zx}, similar to the oscillations of quarks within neutral mesons (e.g., $K^0$--$\bar K^0$ oscillations). As a result, the relative charges of the decay products of the two stops in the event can be different from the na\"{\i}ve expectation. The minimum lifetime required for oscillations to occur is model-dependent, and may allow prompt decays. We will study scenarios with oscillations in section~\ref{sec:limits-oscillation}.

% Stoponium
For sufficiently light stops, there is an additional signature, the annihilation signal of ``stoponium,'' a near-threshold stop-antistop bound state.  The stoponium would decay primarily by annihilation if the stop decay rate is somewhat suppressed, and lead to a prompt diphoton resonance at about twice the stop mass (see, e.g.,~\cite{Drees:1993yr,Drees:1993uw,Martin:2008sv,Kahawala:2011pc,Barger:2011jt}).  However, even if stoponium is observed, studying the (displaced or prompt) decays of the stop itself will still be important for verifying its identity, the presence or absence of $R$-parity violation, and potentially other properties of the model.

% =============================================================================
\section{Limits on simplified models with $\st_1$ production}
\label{sec:limits-singlestop}
% =============================================================================

\begin{table}[t]
\begin{center}\small{
\begin{tabular}{|c|c|c|c|c|}\hline
\textbf{LLE}       & \multicolumn{2}{|c|}{mediators} & \multicolumn{2}{|c|}{final state (of each stop)} \\\hline
$ijk$              &\q first \q& second & $\cho^0,\cho^\pm\to\mbox{RPV}$ & \begin{tabular}{c} $\cho^\pm\to\cho^0 W^{\ast\pm}$ \\ $\cho^0\to\mbox{RPV}$ \end{tabular} \\\hline\hline
\multirow{2}{*}{121, \textbf{122}}
                   & $\Bo$ & $\snu/\slep_L/\slep_R$ & $\ell\ell\nu t$ & \\
                   & $\Wo$ & $\snu/\slep_L$         & $\bs{\ell\nu\nu b}$, $\ell\ell\nu t$, ($\bs{\ell\ell\ell b}$) & $\ell\ell\nu b$ \\\hline
\multirow{3}{*}{\begin{tabular}{c} 131, \textbf{231},\\ 132, 232\end{tabular}}
                   & $\Ho$ & $\snu_\tau/\stau_L$    & $\bs{\ell\ell\tau b}$, $\ell\tau\nu t$ & $\ell\tau\nu b$ \\
                   & $\Bo$ & $\snu/\slep_L/\slep_R$ & $\ell\ell\nu t$, $\ell\tau\nu t$ & \\
                   & $\Wo$ & $\snu/\slep_L$         & $\bs{\ell\nu\nu b}$, $\ell\ell\nu t$, $\ell\tau\nu t$, ($\bs{\ell\ell\tau b}$) & $\ell\ell\nu b$, $\ell\tau\nu b$ \\\hline
\multirow{3}{*}{\textbf{123}}
                   & $\Ho$ & $\stau_R$              & $\bs{\ell\nu\nu b}$, $\ell\tau\nu t$ & $\ell\tau\nu b$ \\
                   & $\Bo$ & $\snu/\slep_L/\stau_R$ & $\ell\tau\nu t$ & \\
                   & $\Wo$ & $\snu/\slep_L$         & $\bs{\tau\nu\nu b}$, $\ell\tau\nu t$, ($\bs{\ell\ell\tau b}$) & $\ell\tau\nu b$ \\\hline
\multirow{4}{*}{133, \textbf{233}}
                   & $\Ho$ & $\snu_\tau/\stau_L$    & $\bs{\ell\tau\tau b}$, $\tau\tau\nu t$ & $\tau\tau\nu b$ \\
                   & $\Ho$ & $\stau_R$              & $\bs{\ell\nu\nu b}$, $\bs{\tau\nu\nu b}$, $\ell\tau\nu t$, $\tau\tau\nu t$ & $\ell\tau\nu b$, $\tau\tau\nu b$ \\
                   & $\Bo$ & $\snu/\slep_L/\stau_R$ & $\ell\tau\nu t$, $\tau\tau\nu t$ & \\
                   & $\Wo$ & $\snu/\slep_L$         & $\bs{\tau\nu\nu b}$,  $\ell\tau\nu t$, $\tau\tau\nu t$, ($\bs{\ell\tau\tau b}$) & $\ell\tau\nu b$, $\tau\tau\nu b$ \\\hline
\end{tabular}}\end{center}
\caption{Simplified models with LLE couplings. Couplings that will be analyzed explicitly are indicated in bold in the first column. We denote $\ell = e,\mu$. If certain final states dominate due to phase-space suppression of final states with tops, they are shown in bold. Final states that are suppressed for heavy inos due to helicity considerations (see appendix~\ref{app:helicity}) are shown in parentheses. In all final states with a single charged lepton ($\ell$ or $\tau$), its charge is correlated with that of the stop.}
\label{tab:LLE}
\end{table}

\begin{table}[t]
\begin{center}\small{
\begin{tabular}{|c|c|c|c|c|}\hline
\textbf{LQD}       & \multicolumn{2}{|c|}{mediators} & \multicolumn{2}{|c|}{final state (of each stop)} \\\hline
$ijk$              &\q first \q& second & $\cho^0,\cho^\pm\to\mbox{RPV}$ & \begin{tabular}{c} $\cho^\pm\to\cho^0 W^{\ast\pm}$ \\ $\cho^0\to\mbox{RPV}$ \end{tabular} \\\hline\hline
\multirow{3}{*}{\begin{tabular}{c}111, 112, 121, 122,\\ 211, 212, \textbf{221}, 222 \end{tabular}}
                   & $\go$ & $\sq$                      & $\ell t jj$, $\nu t jj$ & \\
                   & $\Bo$ & $\sq/\snu/\slep_L$         & $\ell t jj$, $\nu t jj$ & \\
                   & $\Wo$ & $\sq/\snu/\slep_L$         & $\bs{\nu bjj}$, $\ell t jj$, $\nu t jj$, ($\bs{\ell bjj}$) & $\ell bjj$ {\scriptsize[SS]}, $\nu bjj$ \\\hline
\multirow{5}{*}{113, \textbf{123}, 213, 223}
                   & $\go$      & $\sq$                      & $\ell tbj$, $\nu tbj$ & \\
                   & $\Ho$      & $\sbo_R$                   & $\ell tbj$, $\nu tbj$ & $\ell bbj$ {\scriptsize[SS]}, $\nu bbj$ \\
                   & $\Bo$      & $\sq/\snu/\slep_L$         & $\ell tbj$, $\nu tbj$ & \\
                   & $\Wo$      & $\sq/\snu/\slep_L$         & $\bs{\nu bbj}$, $\ell tbj$, $\nu tbj$, ($\bs{\ell bbj}$) & $\ell bbj$ {\scriptsize[SS]}, $\nu bbj$ \\
                   & $\sbo_R$   & ---                        & $\ell W j$, $\nu W j$ & \\\hline
\multirow{4}{*}{311, 312, \textbf{321}, 322}
                   & $\go$ & $\sq$                   & $\tau tjj$, $\nu tjj$ & \\
                   & $\Ho$ & $\snu_\tau/\stau_L$     & $\tau t j j$, $\bs{\tau b j j}$ & $\tau bjj$ {\scriptsize[SS]} \\
                   & $\Bo$ & $\sq/\snu_\tau/\stau_L$ & $\tau tjj$, $\nu tjj$ & \\
                   & $\Wo$ & $\sq/\snu_\tau/\stau_L$ & $\bs{\nu b jj}$, $\tau tjj$, $\nu tjj$, ($\bs{\tau bjj}$) & $\tau bjj$ {\scriptsize[SS]}, $\nu bjj$ \\\hline
\multirow{6}{*}{313, \textbf{323}}
         & $\go$      & $\sq$                    & $\tau t b j$, $\nu t b j$ & \\
         & $\Ho$      & $\sbo_R$                 & $\tau t b j$, $\nu t b j$ & $\tau b b j$ {\scriptsize[SS]}, $\nu b b j$ \\
         & $\Ho$      & $\snu_\tau/\stau_L$      & $\tau t b j$, $\bs{\tau b b j}$ & $\tau b b j$ {\scriptsize[SS]} \\
         & $\Bo$      & $\sq/\snu_\tau/\stau_L$  & $\tau t b j$, $\nu t b j$ & \\
         & $\Wo$      & $\sq/\snu_\tau/\stau_L$  & $\bs{\nu bbj}$, $\tau t b j$, $\nu tbj$, ($\bs{\tau bbj}$) & $\tau b b j$ {\scriptsize[SS]}, $\nu b b j$ \\
         & $\sbo_R$   & ---                      & $\tau W j$, $\nu W j$ & \\\hline
\end{tabular}}\end{center}
\caption{Simplified models with LQD couplings (continued in table~\ref{tab:LQD-i3k}). Couplings that will be analyzed explicitly are indicated in bold in the first column. We denote $\ell = e,\mu;\; j=u,d,c,s$. If certain final states dominate due to phase-space suppression of final states with tops, they are shown in bold. Final states that are suppressed for heavy inos due to helicity considerations (see appendix~\ref{app:helicity}) are shown in parentheses. For final states with a single $\ell$, $\tau$, $t$ or $W$, cases with potentially same-sign dilepton events are indicated with [SS].}
\label{tab:LQD}
\end{table}

\begin{table}[t]
\begin{center}\small{
\begin{tabular}{|c|c|c|c|c|}\hline
\textbf{LQD}       & \multicolumn{2}{|c|}{mediators} & \multicolumn{2}{|c|}{final state (of each stop)} \\\hline
$ijk$              &\q first \q& second & $\cho^0,\cho^\pm\to\mbox{RPV}$ & \begin{tabular}{c} $\cho^\pm\to\cho^0 W^{\ast\pm}$ \\ $\cho^0\to\mbox{RPV}$ \end{tabular} \\\hline\hline
\multirow{7}{*}{131, 132, 231, \textbf{232}}
                   &  ---  &  ---  & $\ell j$ & \\
                   & $\Ho$ or $\Wo$ & $\st$ & $\ell t t j$, $\bs{\ell b b j}$ & $\ell tbj$ \\
                   & $\Ho$ or $\Wo$ & $\sbo_L$ & $\nu t b j$ & $\nu bbj$ \\
                   & $\Wo$ & $\sq/\snu/\slep_L$ & $\ell t t j$, $\bs{\ell b b j}$, $\nu t b j$ & $\ell tbj$, $\bs{\nu b b j}$ \\
                   & $\Bo$ & $\st$ & $\ell t t j$ & \\
                   & $\Bo$ & $\sbo_L$ & $\nu t b j$ & \\
                   & $\Bo$ & $\sq/\snu/\slep_L$ & $\ell t t j$, $\bs{\nu t b j}$ & \\\hline
\multirow{7}{*}{133, \textbf{233}}
                   &  ---  &  ---  & $\ell b$ & \\
                   & $\Ho$ or $\Wo$ & $\st$ & $\ell t t b$, $\bs{\ell b b b}$ & $\ell tbb$\\
                   & $\Ho$ or $\Wo$ & $\sbo_L$ & $\nu t b b$ & $\nu bbb$ \\
                   & $\Ho$ & $\sbo_R$ & $\ell t t b$, $\bs{\nu t b b}$ & $\ell t b b$, $\bs{\nu b b b}$ \\
                   & $\Wo$ & $\snu/\slep_L$ & $\ell t t b$, $\bs{\ell b b b}$, $\nu t b b$ & $\ell tbb$, $\bs{\nu b b b}$ \\
                   & $\Bo$ & $\st$ & $\ell t t b$ & \\
                   & $\Bo$ & $\sbo_L$ & $\nu t b b$ & \\
                   & $\Bo$ & $\snu/\slep_L/\sbo_R$ & $\ell t t b$, $\bs{\nu t b b}$ & \\\hline
\multirow{8}{*}{331, \textbf{332}}
                   &  ---  &  ---  & $\tau j$ & \\
                   & $\Ho$ or $\Wo$ & $\st$ & $\tau t t j$, $\bs{\tau b b j}$ & $\tau tbj$ \\
                   & $\Ho$ or $\Wo$ & $\sbo_L$ & $\nu t b j$ & $\nu bbj$ \\
                   & $\Ho$ & $\snu_\tau/\stau_L$ & $\tau t t j$, $\bs{\tau bbj}$ & $\tau t b j$ \\
                   & $\Wo$ & $\sq/\snu_\tau/\stau_L$ & $\tau t t j$, $\bs{\tau b b j}$, $\nu t b j$ & $\tau tbj$, $\bs{\nu b b j}$ \\
                   & $\Bo$ & $\st$ & $\tau t t j$ & \\
                   & $\Bo$ & $\sbo_L$ & $\nu t b j$ & \\
                   & $\Bo$ & $\sq/\snu_\tau/\stau_L$ & $\tau t t j$, $\bs{\nu t b j}$ & \\\hline
\multirow{8}{*}{\textbf{333}}
                   &  ---  &  ---  & $\tau b$ & \\
                   & $\Ho$ or $\Wo$ & $\st$ & $\tau t t b$, $\bs{\tau b b b}$ & $\tau tbb$ \\
                   & $\Ho$ or $\Wo$ & $\sbo_L$ & $\nu t b b$ & $\nu bbb$ \\
                   & $\Ho$ & $\snu_\tau/\stau_L$ & $\tau t t b$, $\bs{\tau bbb}$ & $\tau t b b$ \\
                   & $\Ho$ & $\sbo_R$ & $\tau t t b$, $\bs{\nu t b b}$ & $\tau t b b$, $\bs{\nu b b b}$ \\
                   & $\Wo$ & $\snu_\tau/\stau_L$ & $\tau t t b$, $\bs{\tau b b b}$, $\nu t b b$ & $\tau tbb$, $\bs{\nu b b b}$ \\
                   & $\Bo$ & $\st$ & $\tau t t b$ & \\
                   & $\Bo$ & $\sbo_L$ & $\nu t b b$ & \\
                   & $\Bo$ & $\snu_\tau/\stau_L/\sbo_R$ & $\tau t t b$, $\bs{\nu t b b}$ & \\\hline
\end{tabular}}\end{center}
\caption{Simplified models with LQD couplings through which the stop can decay directly (therefore, only cases with on-shell inos are considered). Same conventions as in table~\ref{tab:LQD} apply.}
\label{tab:LQD-i3k}
\end{table}

\begin{table}[t]
\begin{center}\small{
\begin{tabular}{|c|c|c|c|c|}\hline
\textbf{UDD}       & \multicolumn{2}{|c|}{mediators} & \multicolumn{2}{|c|}{final state (of each stop)} \\\hline
$ijk$              &\q first \q& second & $\cho^0,\cho^\pm\to\mbox{RPV}$ & \begin{tabular}{c} $\cho^\pm\to\cho^0 W^{\ast\pm}$ \\ $\cho^0\to\mbox{RPV}$ \end{tabular} \\\hline\hline
112, \textbf{212}  & $\go$ or $\Bo$ & $\sq$    & $tjjj$ & \\\hline
\multirow{3}{*}{113, 123, \textbf{213}, 223}
                   & $\go$ or $\Bo$ & $\sq$    & $tbjj$ & \\
                   & $\Ho$          & $\sbo_R$ & $tbjj$ & $bbjj$ \\
                   & $\sbo_R$       &  ---     & $Wjj$ & \\\hline\hline
\multirow{3}{*}{\textbf{312}}
                   & ---            & ---      & $jj$ & \\
                   & $\Ho$          & $\st$    & $ttjj$, $\bs{bbjj}$ & $tbjj$ {\scriptsize[SS]} \\
                   & $\Bo$          & $\sq$    & $ttjj$ & \\\hline
\multirow{4}{*}{313, \textbf{323}}
                   & ---            & ---      & $bj$ & \\
                   & $\Ho$          & $\st$    & $ttbj$, $\bs{bbbj}$ & $tbbj$ {\scriptsize[SS]} \\
                   & $\Ho$          & $\sbo_R$ & $ttbj$ & $tbbj$ {\scriptsize[SS]} \\
                   & $\Bo$          & $\sq$    & $ttbj$ & \\\hline
\end{tabular}}\end{center}
\caption{Simplified models with UDD couplings. Couplings that will be analyzed explicitly are indicated in bold in the first column. We denote $j=u,d,c,s$. If certain final states dominate due to phase-space suppression of final states with tops, they are shown in bold. For final states with a single $t$ or $W$, cases with potentially same-sign dilepton events are indicated with [SS]. The bottom part of the table includes couplings through which the stop can decay directly (as with the LQD$i3j$ couplings of table~\ref{tab:LQD-i3k}, only cases with on-shell inos are considered).}
\label{tab:UDD}
\end{table}

\begin{table}[t]
\begin{center}\small{
\begin{tabular}{|c|c|c|c|}\hline
final state  & collaboration & $\cal L$ (fb$^{-1}$) & ref. \\\hline\hline
\multirow{2}{*}{pairs of dijets}
                       & ATLAS & 0.034, 4.6 & \cite{Aad:2011yh,ATLAS-CONF-2012-110} \\
                       & CMS   & 2.2  & \cite{CMS-PAS-EXO-11-016} \\\hline
\multirow{2}{*}{leptoquark pairs}
                       & CMS   & 5.0  & \cite{CMS-LQ} \\
                       & CMS   & 4.8  & \cite{CMS-PAS-EXO-12-002} \\\hline
\multirow{2}{*}{$t\bar t$}
                       & ATLAS & 0.70 & \cite{ATLAS:2012aa} \\
                       & CMS   & 2.0-2.3  & \cite{:2012bt,Chatrchyan:2012vs} \\\hline
$t\bar t$ + jet        & CMS   & 5.0  & \cite{CMS-ttbarjet} \\\hline
$t\bar t$ + $m_T$      & ATLAS & 1.04 & \cite{Aad:2011wc} \\\hline
leptonic $m_{T2}$      & ATLAS & 4.7  & \cite{:2012uu} \\\hline
\multirow{2}{*}{$\ell$ + jets + MET}
                       & CMS   & 4.7  & \cite{CMS-PAS-SUS-12-010} \\
                       & ATLAS & 4.7  & \cite{ATLAS-CONF-2012-041,ATLAS-CONF-2012-140} \\\hline
\multirow{2}{*}{OS $\ell\ell$ + MET}
                       & CMS   & 4.98 & \cite{CMS-OS-DIL} \\
                       & ATLAS & 1.04, 4.7  & \cite{Aad:2011cwa,Aad:2012ms} \\\hline
SS $\ell\ell$ + MET    & ATLAS & 1.04, 2.05 & \cite{Aad:2011cwa,ATLAS:2012ai} \\\hline
SS $\ell\ell$          & ATLAS & 1.6, 4.7 & \cite{Aad:2012cg,ATLAS:2012mn} \\\hline
SS $\ell\ell$ (+ MET)  & CMS   & 4.98 & \cite{CMS-SS-DIL,CMS-SSSF-DIL} \\\hline
SS $\ell\ell$ + $b$ (+ MET) & CMS & 4.98 & \cite{CMS-SS-DIL-b} \\\hline
$b'$ (SS $\ell\ell$ or $3\ell$ + $b$) & CMS & 4.9 & \cite{Chatrchyan:2012yea} \\\hline
$b'$ (SS $\ell\ell$)   & ATLAS & 4.7 & \cite{ATLAS-CONF-2012-130} \\\hline
3 or 4 $\ell$          & ATLAS & 1.02 & \cite{ATLAS-CONF-2011-158,ATLAS-CONF-2011-144} \\\hline
3 $\ell$ + MET         & ATLAS & 2.06, 4.7 & \cite{1204.5638,ATLAS3lMET} \\\hline
4 $\ell$ + MET         & ATLAS & 2.06 & \cite{ATLAS-CONF-2012-001} \\\hline
3 or 4 $\ell$ (+ MET)  & CMS   & 4.98 & \cite{CMSmultileptons} \\\hline
1 or 2 $\tau$ + jets + MET & ATLAS & 2.05, 4.7 & \cite{1204.3852,1203.6580,:2012ht} \\\hline
\multirow{2}{*}{$\tau$ + $\ell$ + jets + MET}
                       & ATLAS & 4.7 & \cite{:2012ht} \\
                       & CMS   & 5.0 & \cite{CMS-OS-DIL} \\\hline
\multirow{2}{*}{$b$ + jets + MET}
                       & ATLAS & 2.05, 4.7 & \cite{ATLAS:2012ah,ATLAS-3b} \\
                       & CMS   & 1.1, 4.98 & \cite{CMS-PAS-SUS-11-006,:2012rg} \\\hline
\multirow{2}{*}{$b$ + $\ell$ + jets + MET}
                       & ATLAS & 2.05 & \cite{ATLAS:2012ah} \\
                       & CMS   & 4.96-4.98 & \cite{CMS-PAS-SUS-11-027,CMS-PAS-SUS-11-028} \\\hline
\multirow{2}{*}{$Z$ + jets + MET}
                       & CMS   & 4.98 & \cite{Chatrchyan:2012qka} \\
                       & ATLAS & 2.05 & \cite{Aad:2012cz} \\\hline
\multirow{2}{*}{jets + MET}
                       & ATLAS & 4.7  & \cite{ATLAS:2-6jets,ATLAS:6-9jets} \\
                       & CMS   & 1.1, 4.98 & \cite{CMS-PAS-SUS-11-004,CMS:2012mf} \\\hline
($b$)-jets with $\alpha_T$ & CMS & 1.14, 4.98 & \cite{Chatrchyan:2011zy,CMS-PAS-SUS-11-022} \\\hline
\end{tabular}}
\end{center}
\caption{7~TeV LHC searches used for inferring limits.}
\label{table:searches}
\end{table}

As discussed in the previous section, even for a fixed RPV coupling, the stop may decay in several different ways, either directly to a two-body final state or through an on- or off-shell Higgsino $\Ho$ (which includes two neutralinos and a chargino), bino $\Bo$ (neutralino), wino $\Wo$ (a neutralino and a chargino), gluino $\go$ or sbottom $\sbo_R$. To cover a broad range of possibilities, we construct a separate simplified model for each of these decay channels. For on-shell $\Ho$ and $\Wo$, we present separate models for the situations in which a chargino ($\cho^\pm$) transitions to a neutralino ($\cho^0$) before decaying. Our simplified models for stop decays and the final states obtained in each case are listed in tables~\ref{tab:LLE} (LLE couplings), \ref{tab:LQD}--\ref{tab:LQD-i3k} (LQD couplings), and~\ref{tab:UDD} (UDD couplings).\footnote{Throughout the text, $L_i Q_j D^c_k$ will be referred to as LQD$ijk$, etc.} These will be described in more detail in sections~\ref{sec:2body}--\ref{sec:4body}. In section~\ref{sec:limits-oscillation}, we will examine the same scenarios with stop-antistop oscillations.

% On the searches
To determine to what extent the existing LHC searches provide coverage of these models, we have simulated all available 7~TeV ATLAS and CMS searches which could potentially be relevant and easily recast for our study. These are summarized in table~\ref{table:searches}.\footnote{New analyses relative to v1 of our preprint are~\cite{:2012bt,:2012uu,ATLAS-CONF-2012-140,ATLAS:2012mn,ATLAS-CONF-2012-130,:2012ht} and the dilepton channels of~\cite{Aad:2012ms}. Potentially relevant searches that are not included are~\cite{ATLAS-CONF-2012-108,CMS-PAS-SUS-12-004}.}  Details of the simulation are described in appendix~\ref{simdetails}, where we also explain how we eliminate unreliable limits from searches probing the tails of our signal distributions. The resulting limits are presented in figures~\ref{fig:2body}--\ref{fig:LQD-i3k}. For couplings related to each other by exchanging electrons and muons, or first- and second-generation quarks (all grouped together in tables~\ref{tab:LLE}--\ref{tab:UDD}), we present results for just one representative case.\footnote{Exchanging electrons and muons occasionally does have an effect on the limits due to differences in the trigger, identification and isolation requirements and fake rates, and different fluctuations in data in analyses that have separate search regions for electrons and muons. However, this effect is typically small.}  In the figure captions, we indicate which of the searches from table~\ref{table:searches} turned out to be most relevant for each scenario.

% ------------------------------------------------------------------------------------------------------------------------
\subsection{Direct two-body decays\label{sec:2body}}
% ------------------------------------------------------------------------------------------------------------------------

\begin{figure}[t]
\begin{center}
\includegraphics[scale=0.65]{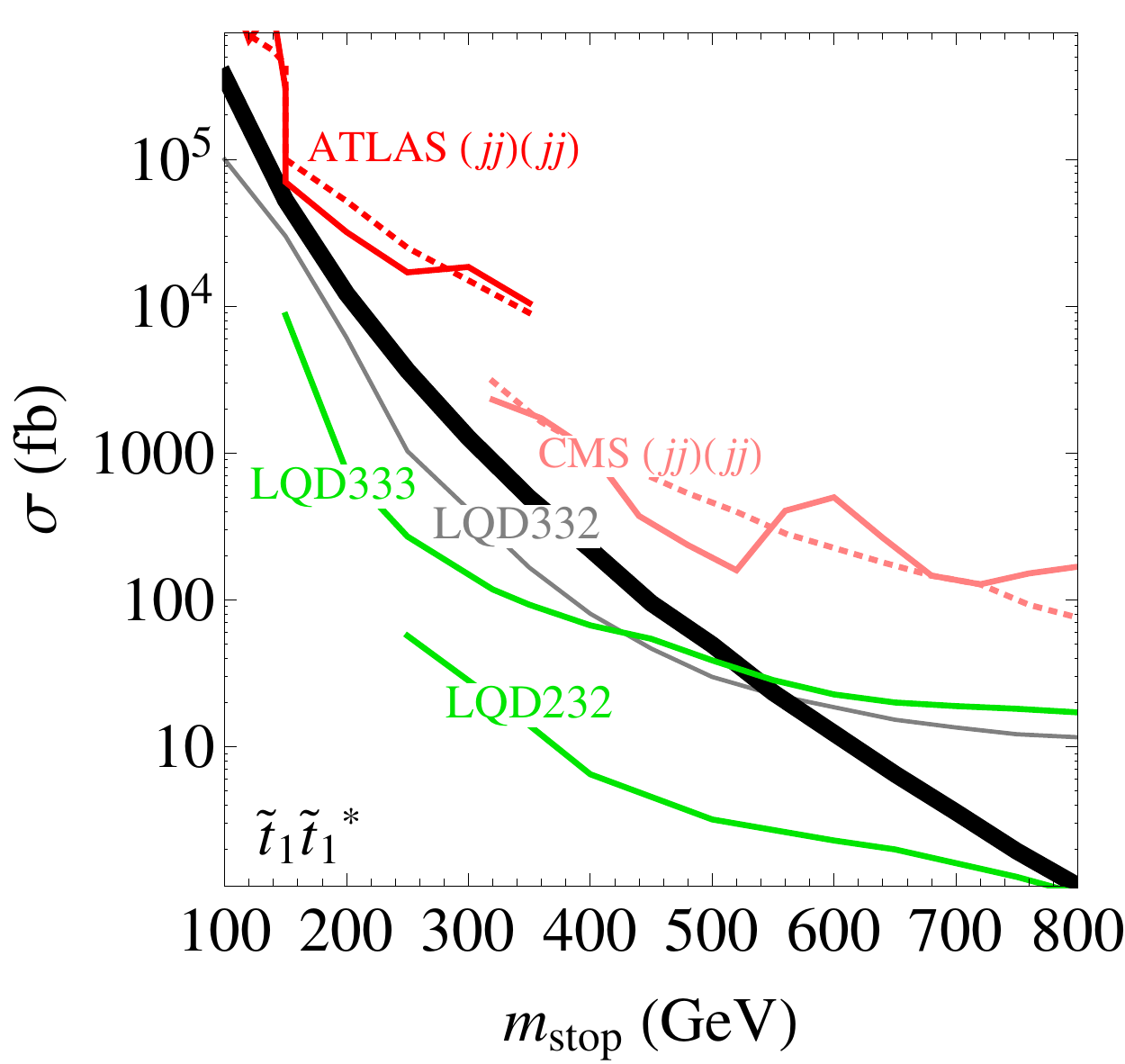}\qq
\includegraphics[scale=0.73]{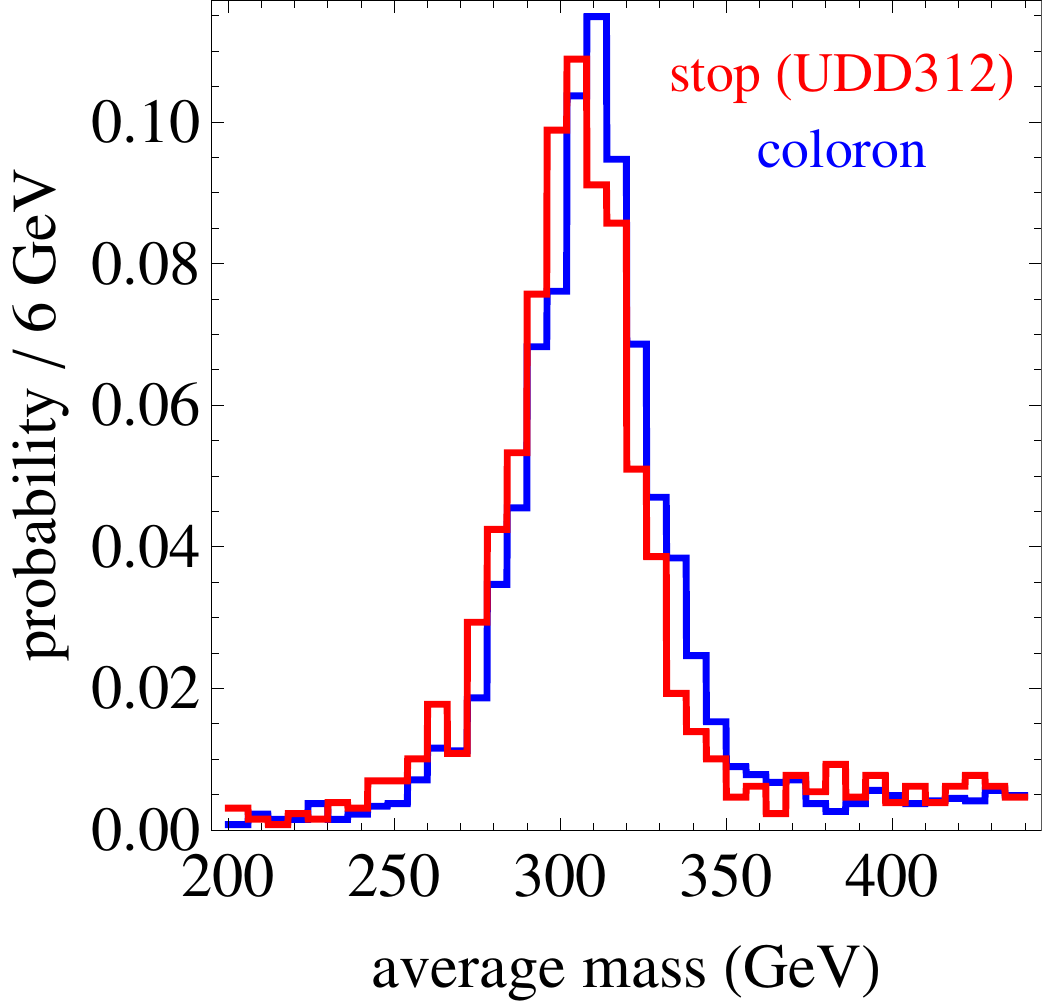}
\caption{Left: Limits on two-body decays of the stops. The thick black curve is the stop pair-production cross section~\cite{Beenakker:2010nq}. For decays to a lepton and a quark (LQD) we present the limit from the CMS search for second generation leptoquarks~\cite{CMS-LQ} for the LQD232 case. A similar limit applies to the LQD131 case based on the search for first generation leptoquarks~\cite{CMS-LQ}. These limits cover the analogous cases with $b$-jets (LQD233 and LQD133) as well. We also present the limit from the CMS search for third generation leptoquarks~\cite{CMS-PAS-EXO-12-002} relevant to the LQD333 case. In the LQD332 case, which does not have a dedicated search, we obtain a limit from the ATLAS searches for $2\tau$+jets+MET~\cite{1203.6580,:2012ht} and the CMS search for opposite-sign (OS) dileptons+MET (with $\tau$'s)~\cite{CMS-OS-DIL}, and at low masses from the $t\bar t$ cross section measurements in the dilepton channel~\cite{ATLAS:2012aa,:2012bt}. For decays to pairs of jets (UDD), the limits from the ATLAS searches~\cite{Aad:2011yh,ATLAS-CONF-2012-110} and the CMS search~\cite{CMS-PAS-EXO-11-016} are shown (the dashed lines are the expected limits). None of the other searches we examined has appreciable sensitivity to these UDD$3jk$ decays. Right: in the context of the CMS paired dijets search~\cite{CMS-PAS-EXO-11-016}, distributions (normalized to 1) of the average mass for the coloron model~\cite{Dobrescu} used in~\cite{CMS-PAS-EXO-11-016} and our UDD312 stop signal, both for $m = 320$~GeV.
\label{fig:2body}}
\end{center}
\end{figure}

In the absence of large hierarchies in the RPV couplings, the direct two-body decays through couplings involving the stop will dominate over multi-body decays that require virtual superpartners. In this subsection, we assume that these decays also dominate over $R$-parity-conserving decays to lighter superpartners (if such decays are available), which may or may not be the case, depending on the size of the relevant couplings. The other possibility will be addressed in the next subsection.

The $\lambda'_{i3k}$ (LQD) couplings mediate stop decays to a charged lepton ($e$, $\mu$ or $\tau$) and a quark (which may be a $b$ quark). If we assume for simplicity that a single $\lambda'$ coupling dominates, the flavors of the lepton and the quark are fixed. These RPV stops are identical to scalar leptoquarks in both production and decay (in the case where the branching ratio of the leptoquark to a neutrino and a quark vanishes). The leptoquark searches~\cite{CMS-LQ,CMS-PAS-EXO-12-002} are directly applicable to the corresponding RPV stop scenarios, excluding them from the natural range of masses. However, since the third-generation leptoquark search~\cite{CMS-PAS-EXO-12-002} requires a $\tau$+$b$ final state, signatures with a $\tau$ and a light-quark jet (from $\lambda'_{331}$ or $\lambda'_{332}$) are not covered.  Nonetheless, searches for hadronic taus+jets+MET, and for the low masses the $t\bar t$ cross section measurements in the dilepton channel, are sensitive,\footnote{For much of the mass range, the stop cross section is only slightly larger than our limit (see LQD332 in figure~\ref{fig:2body}, left), which does not account for the systematic uncertainties of our simplified detector simulation. Therefore one should not be overly confident about our exclusion range in this case.  A more dedicated experimental analysis is desirable.} and the range of excluded masses is the same (within the uncertainty of our simulation) as that of~\cite{CMS-PAS-EXO-12-002}. These results are included in figure~\ref{fig:2body} (left).

The $\lambda''_{3jk}$ (UDD) couplings facilitate stop decays to a pair of quarks (one of which may be a $b$-quark). Searches for pair-produced particles which each decay to two jets have been done in other contexts by ATLAS~\cite{Aad:2011yh,ATLAS-CONF-2012-110} and CMS~\cite{CMS-PAS-EXO-11-016}. In the right side of figure~\ref{fig:2body}, we show our simulated distribution of the average reconstructed mass of the two dijets, the peak being searched for in~\cite{CMS-PAS-EXO-11-016}. We show the distribution for the stop as well as for the coloron model~\cite{Dobrescu} that was actually used as a signal hypothesis in that CMS search. The limits derived in that search can only be translated onto stops (after including the selection efficiencies) because these distributions have approximately the same shape.\footnote{From our communication with CMS, we understand that future versions of their analysis may use a different model for the coloron. If this changes the width of the coloron bump, then it will be impossible to do this kind of re-interpretation. For the ATLAS search, whose updated version~\cite{ATLAS-CONF-2012-110} has appeared just recently, we did not do full simulation, and the presented limit assumes the cut efficiencies and width of the stop bump to be identical to those of the ``sgluon'' of~\cite{ATLAS-CONF-2012-110}. Ideally, CMS and ATLAS would include the RPV stop as one of the benchmark models for which they optimize the searches in this final state.} As we show in figure~\ref{fig:2body} (left), this search sets no limits on the stop (whose cross section is much smaller than that of the coloron, due to both spin and color), neither do the ATLAS searches at lower masses. These decays to two jets, even if one of them is a $b$-jet, also receive no appreciable limits from any of the other searches we examined.

% ------------------------------------------------------------------------------------------------------------------------
\subsection{Decays via other superpartners\label{sec:4body}}
% ------------------------------------------------------------------------------------------------------------------------

\subsubsection{Definition of simplified models}

Let us start by considering decays, shown in figure~\ref{fig:3b4bstop} (right), that proceed through two intermediate particles: an ``ino'' $\tilde X$ (gluino, Higgsino, bino or wino), and a sfermion $\tilde f$ (squark, sneutrino or slepton).

In addition to affecting the kinematic distributions of the decay products, the ino mass sometimes determines the branching ratios into the various final states. For a detailed discussion of this effect, see appendix~\ref{app:helicity}. To cover the different possibilities, we examine three cases for the mass of the ino: much heavier than the stop (in practice, we set $m_{\tilde{X}} = 2$~TeV), slightly heavier than the stop ($m_{\tilde{X}}=1.1 m_\st$) and on shell (with the masses discussed below).\footnote{The behavior in the transition region between light and heavy inos depends somewhat on the stop mixing angle. In our models we assume the stop to be an equal mixture of left and right.}

For the gluino, we will only treat the heavy case, since a light gluino is disfavored experimentally, and only those cases with LQD or UDD couplings, as the gluino does not couple to sleptons. As the Higgsinos couple most strongly to the third generation sfermions, we will only consider Higgsino decays via RPV couplings that contain third generation indices. For on-shell Higgsinos or winos, we will assume $m_{\tilde{X}}=m_\st-100$~GeV, which allows $\st\to b\cho^+$ but forbids $\st\to t\cho^0$. This is a sensible choice of simplified model since the latter final state would be phase-space suppressed even if allowed. In the bino case, where the chargino is absent, we will allow $\st\to W^+b\cho^0$, with, again, $m_{\tilde{X}}=m_\st-100$~GeV. For couplings that allow the stop to decay directly (LQD$i3k$ and UDD$3jk$), we set the bino mass to $m_{\tilde{X}}=m_\st-200$~GeV, allowing for both the top and the bino to be on-shell; this allows $\st\to t\cho^0$ to dominate over the two-body RPV decay for reasonable parameter choices. We will not include contributions from direct electroweak production of wino or Higgsino pairs, even though such processes can lead to further limits on scenarios in which these particles are sufficiently lighter than the stop.

In some cases, there are several contributing sfermion mediators, $\tilde f$, whose masses are generically unrelated. In particular, the masses of the right-handed sleptons, the left-handed sleptons (and sneutrinos), and the squarks, are independent, so their relative contributions may vary. The relative contributions may depend also on $\tan\beta$ (in $\Ho$-mediated scenarios) since it affects the couplings of the different sfermions to the Higgsino differently. However, apart from kinematics, the choice of the dominant sfermion does not affect the possible final states or their branching ratios (so we will assume all sfermions to have the same mass and set $\tan\beta = 40$). The LLE233 and LQD323 $\Ho$-mediated scenarios are exceptions, where different choices of sfermion mediators give rise to different simplified models, as indicated in tables~\ref{tab:LLE} and~\ref{tab:LQD}.

Additionally, scenarios where an on-shell ino can decay through the stop, as well as through some other sfermions, can be sensitive to the sfermion masses. This happens because we allow for significant splittings and mixings within the stop-sbottom sector. Let us consider the LQD couplings first. Note from table~\ref{tab:LQD-i3k} that for any ino, the final state for decays through $\sbo_L$ differs from those of decays through $\st$. The sleptons (and sneutrinos), in the wino and bino cases, contribute the same mixture of final states as would degenerate left-handed stop and sbottom (while in the Higgsino case, the decays through sleptons give the same final state as the decays through the stop). At a typical point in parameter space, the final state composition will be somewhere between the stop-only and the mixed final states. It is less typical, although possible, to have the $\sbo_L$-mediated processes dominate (if the stop is largely right-handed). In the analogous cases with UDD couplings (bottom half of table~\ref{tab:UDD}), the final state composition may depend on whether the Higgsino decays via $\st$ or $\sbo_R$.

For the purpose of event generation, we assume the sfermions $\tilde f$ to be heavy (in practice, we set $m_{\tilde f} = 2$~TeV), except when the sfermion is the stop itself, in which case we use the actual stop mass. The precise value of the sfermion mass does not affect the branching ratios to the various possible final states,\footnote{Scenarios with LQD$i3k$ couplings (table~\ref{tab:LQD-i3k}) with $\tilde f = \snu,\slep_L$ are an exception. Since the $\slep_L$ decays produce a top, decays via an off-shell $\slep_L$ are phase-space suppressed relative to decays via an off-shell $\snu$, while in an on-shell case the rates would be comparable.} and even though the details of the kinematics would have a small effect on the efficiencies of the searches involved, our results would still approximately apply for scenarios with lighter sfermions as well (which, in some cases, is actually necessary in order for the decay to be prompt).

\begin{figure}[t]
\begin{center}
\includegraphics[scale=0.59]{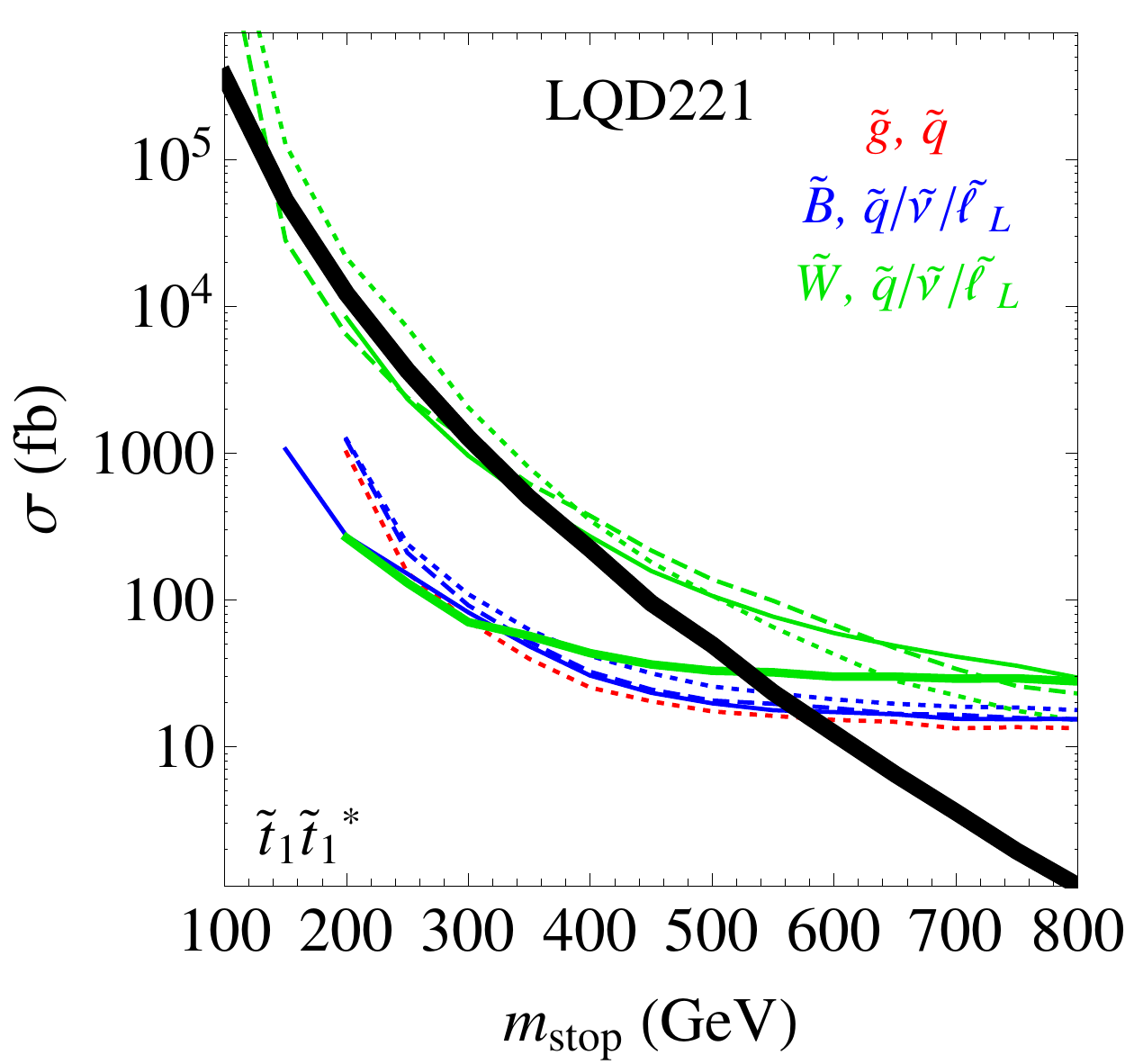}\q
\includegraphics[scale=0.59]{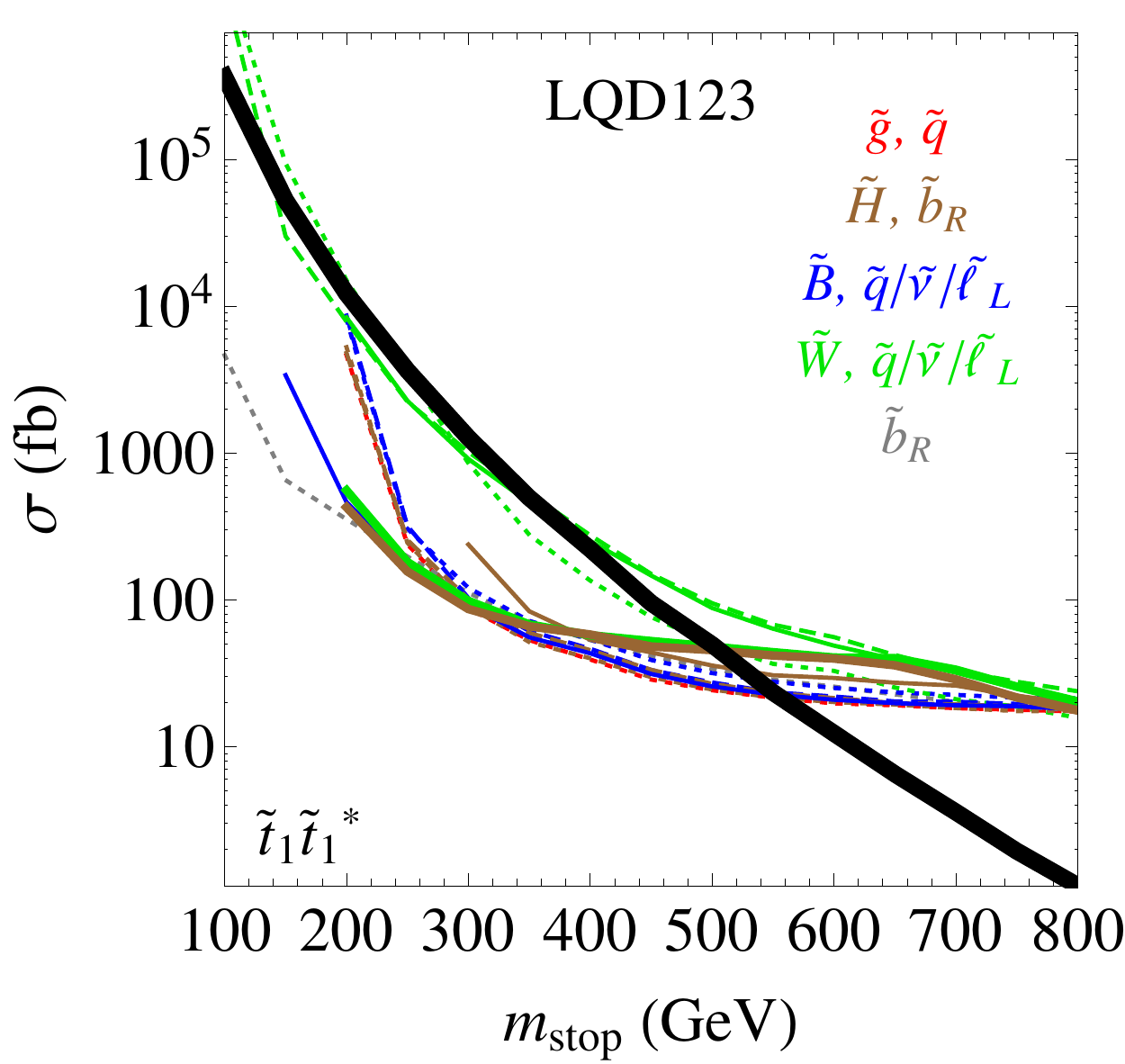}\\\vskip 5pt
\includegraphics[scale=0.59]{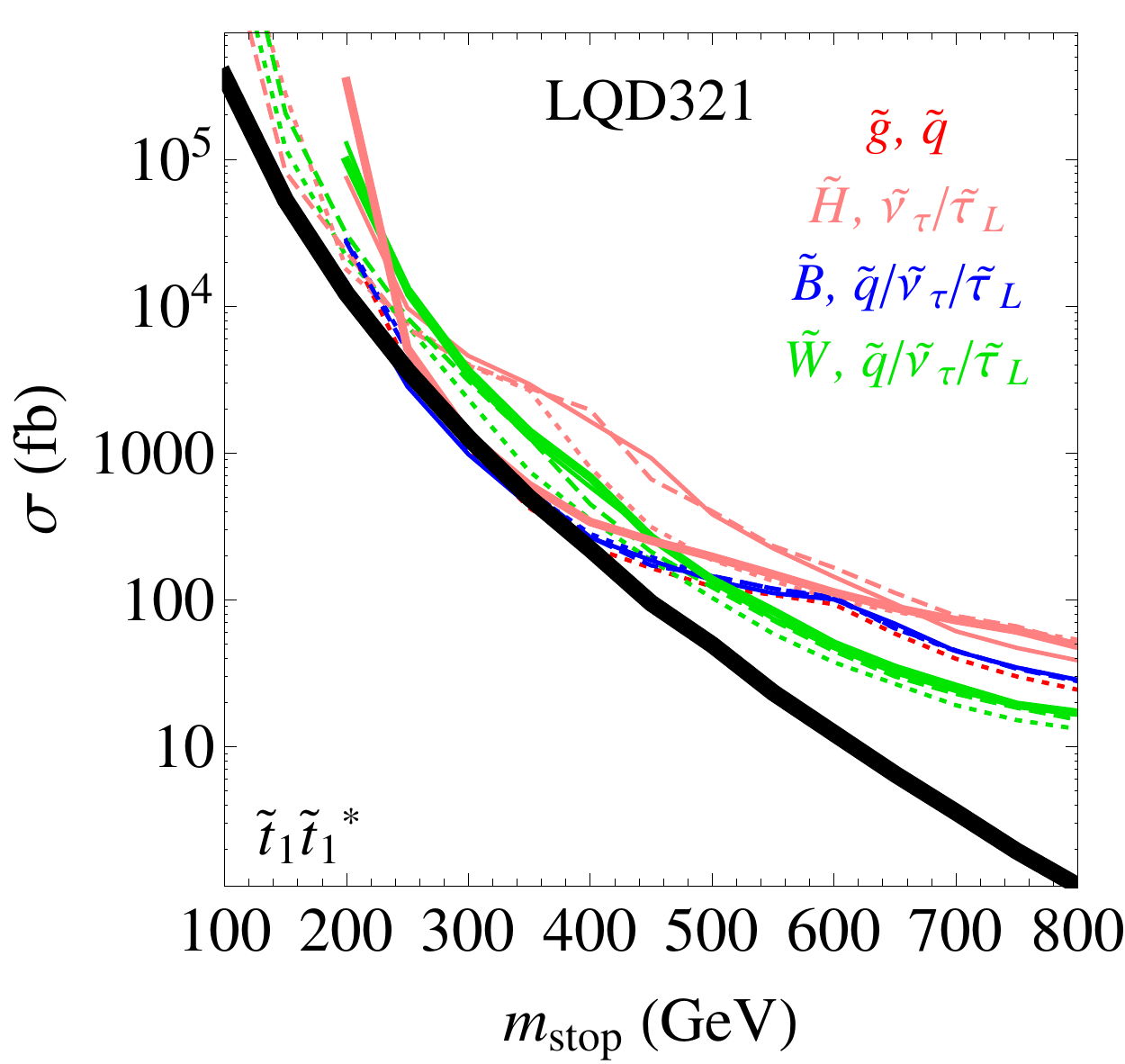}\q
\includegraphics[scale=0.59]{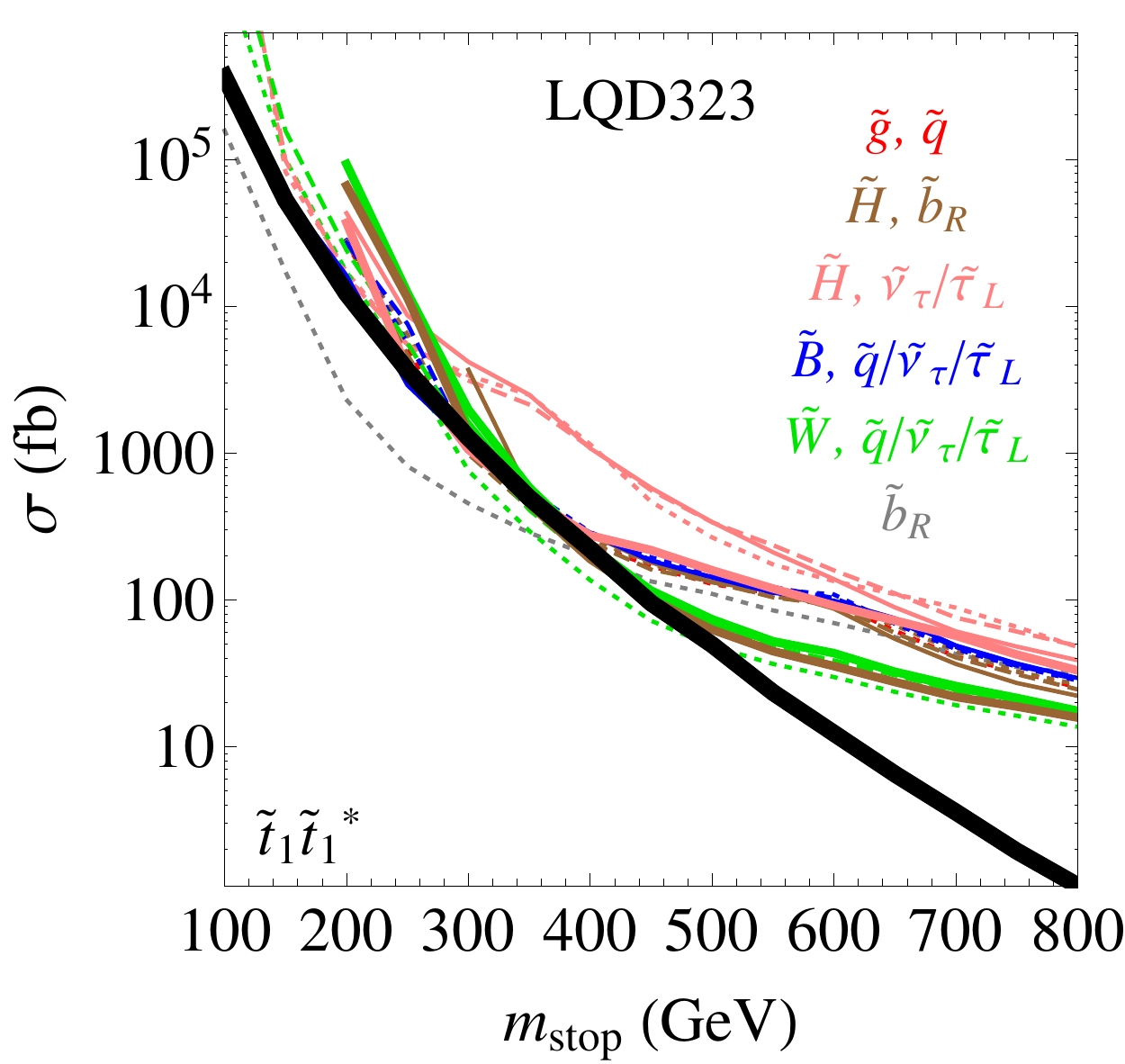}
\caption{Limits on stops decaying via other superpartners (see legend) in the presence of LQD operators (continued in figure~\ref{fig:LQD-i3k}). The first intermediate particle is either very heavy (dotted curves), $10\%$ heavier than the stop (dashed) or 100~GeV lighter than the stop (thin solid: $\cho^0,\cho^\pm\to\mbox{RPV}$, thick solid: $\cho^\pm\to\cho^0\to\mbox{RPV}$). For decays via $\go$, $\Bo$ and $\Ho$ (except for cases mentioned below), the limits are set by searches for SS dileptons~\cite{CMS-SS-DIL,CMS-SS-DIL-b} and $b'$~\cite{Chatrchyan:2012yea,ATLAS-CONF-2012-130}, except in 321 and 323 cases at very low $m_{\rm stop}$ where the most sensitive analyses are the dilepton $t\bar t$ cross section measurements~\cite{ATLAS:2012aa,:2012bt}. In $\Ho$-$\snu_\tau/\stau_L$ cases with $\cho^0,\cho^\pm\to\mbox{RPV}$, the most sensitive analysis is the $t\bar t$ cross section in the $\ell+\tau_h$ channel~\cite{Chatrchyan:2012vs}. In the 323 $\Ho$-$\sbo_R$ case with $\cho^\pm\to\cho^0$ transitions, the limit is set by the search for multiple $b$-jets+MET~\cite{ATLAS-3b}. In most of the $\Wo$ cases, the best limits are set by the ($b$-)jets+MET searches~\cite{ATLAS-3b,:2012rg,ATLAS:2-6jets}. In 221 and 123 cases with $\cho^0,\cho^\pm\to\mbox{RPV}$, $\ell$(+$b$)+jets+MET~\cite{CMS-PAS-SUS-12-010,ATLAS-CONF-2012-140,ATLAS:2012ah}, and $t\bar t$-like searches~\cite{CMS-ttbarjet,Aad:2011wc} are comparably important (except for heavy $\Wo$). In 221 and 123 cases with $\cho^\pm\to\cho^0$ transitions, the limits are set by SS dileptons~\cite{CMS-SS-DIL,CMS-SS-DIL-b,Chatrchyan:2012yea}. For $\sbo_R$-mediated scenarios, the best limits are set by SS dileptons~\cite{CMS-SS-DIL} and (in the 123 case) multileptons~\cite{CMSmultileptons,ATLAS3lMET}, and (in the 323 case, at low $m_{\rm stop}$) $t\bar t$ cross section~\cite{ATLAS:2012aa,:2012bt}.}
\label{fig:LQD}
\end{center}
\end{figure}

\begin{figure}[t]
\begin{center}
\includegraphics[scale=0.6]{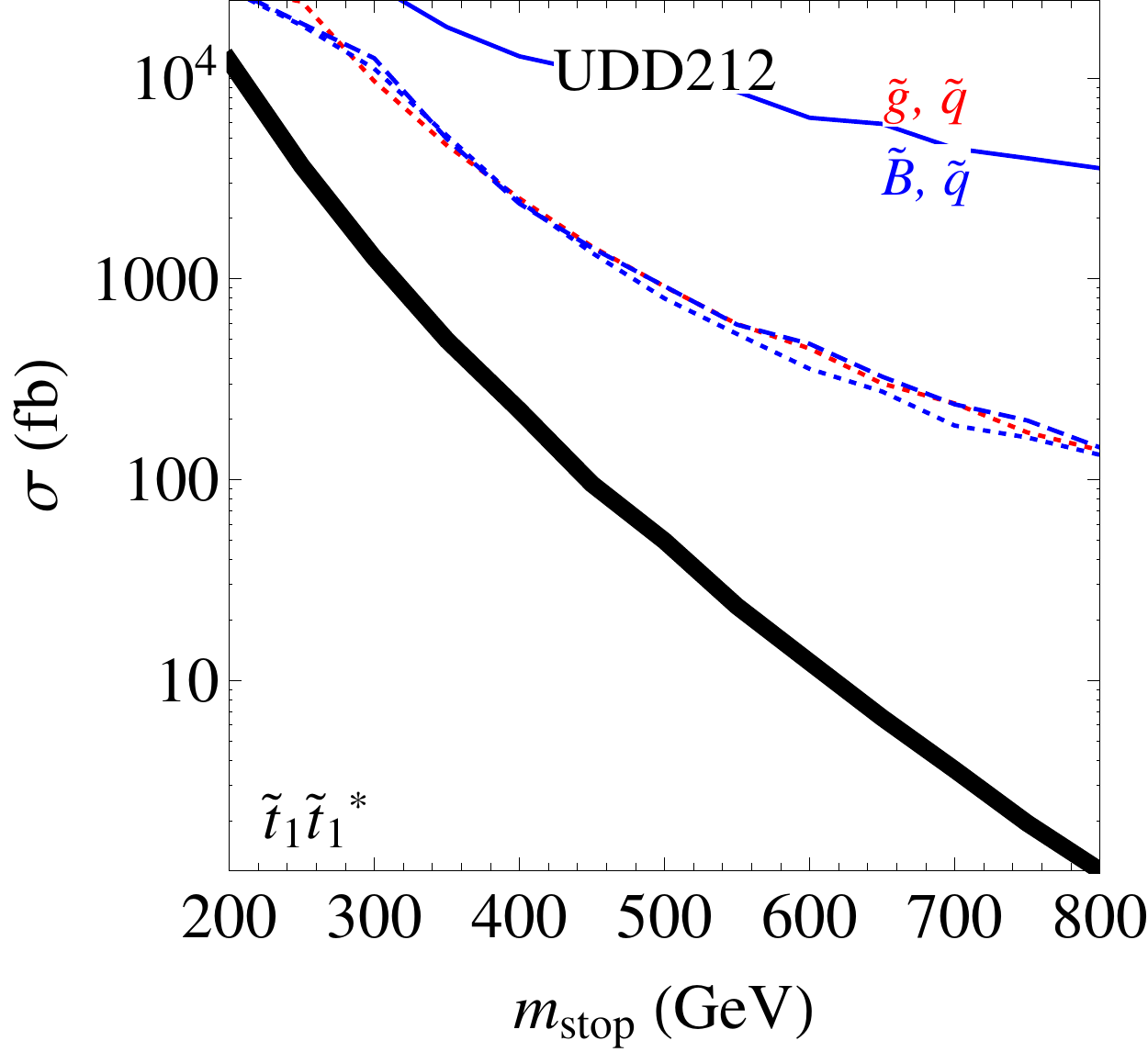}\q
\includegraphics[scale=0.6]{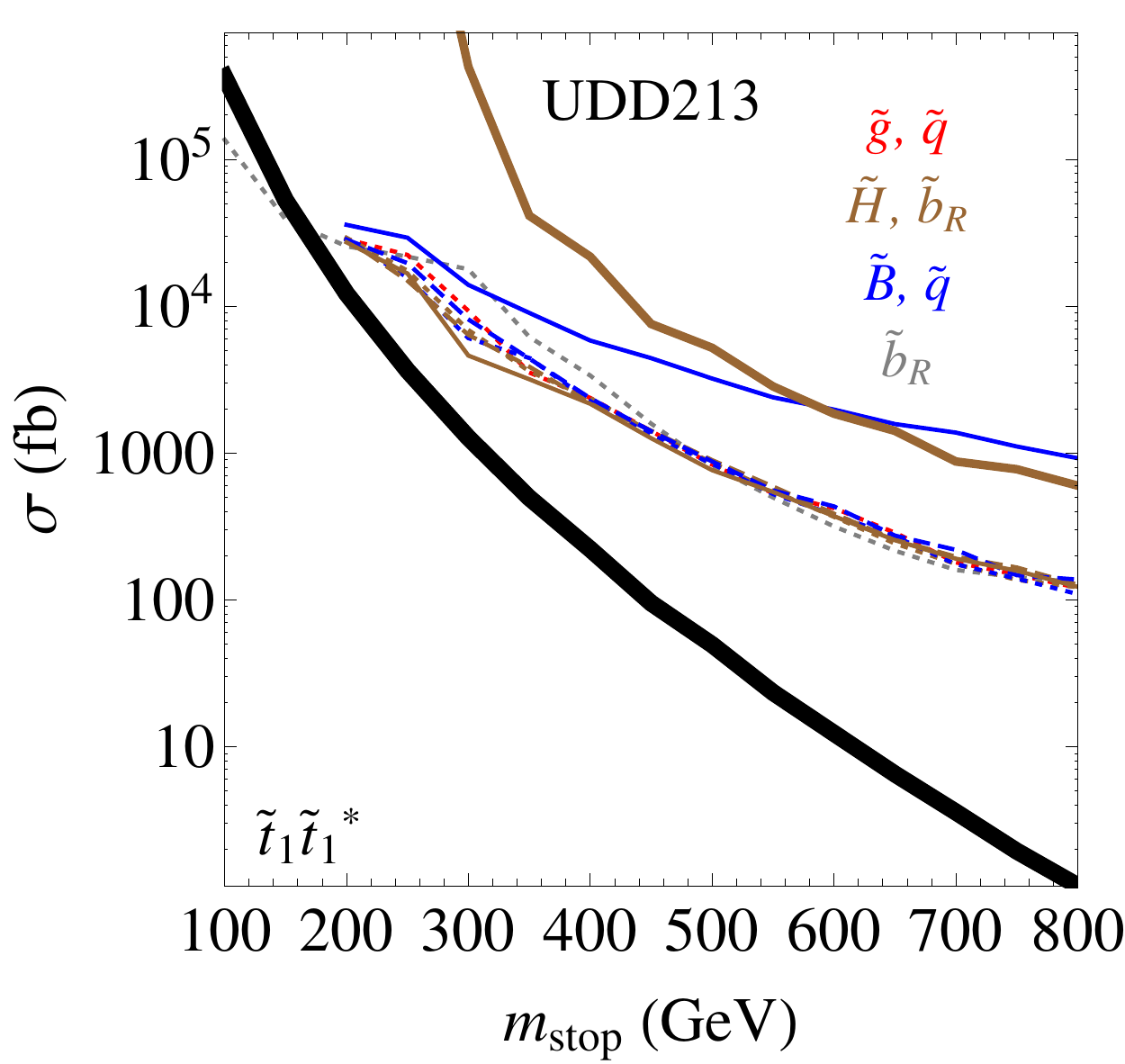}\\\vskip 10pt
\includegraphics[scale=0.6]{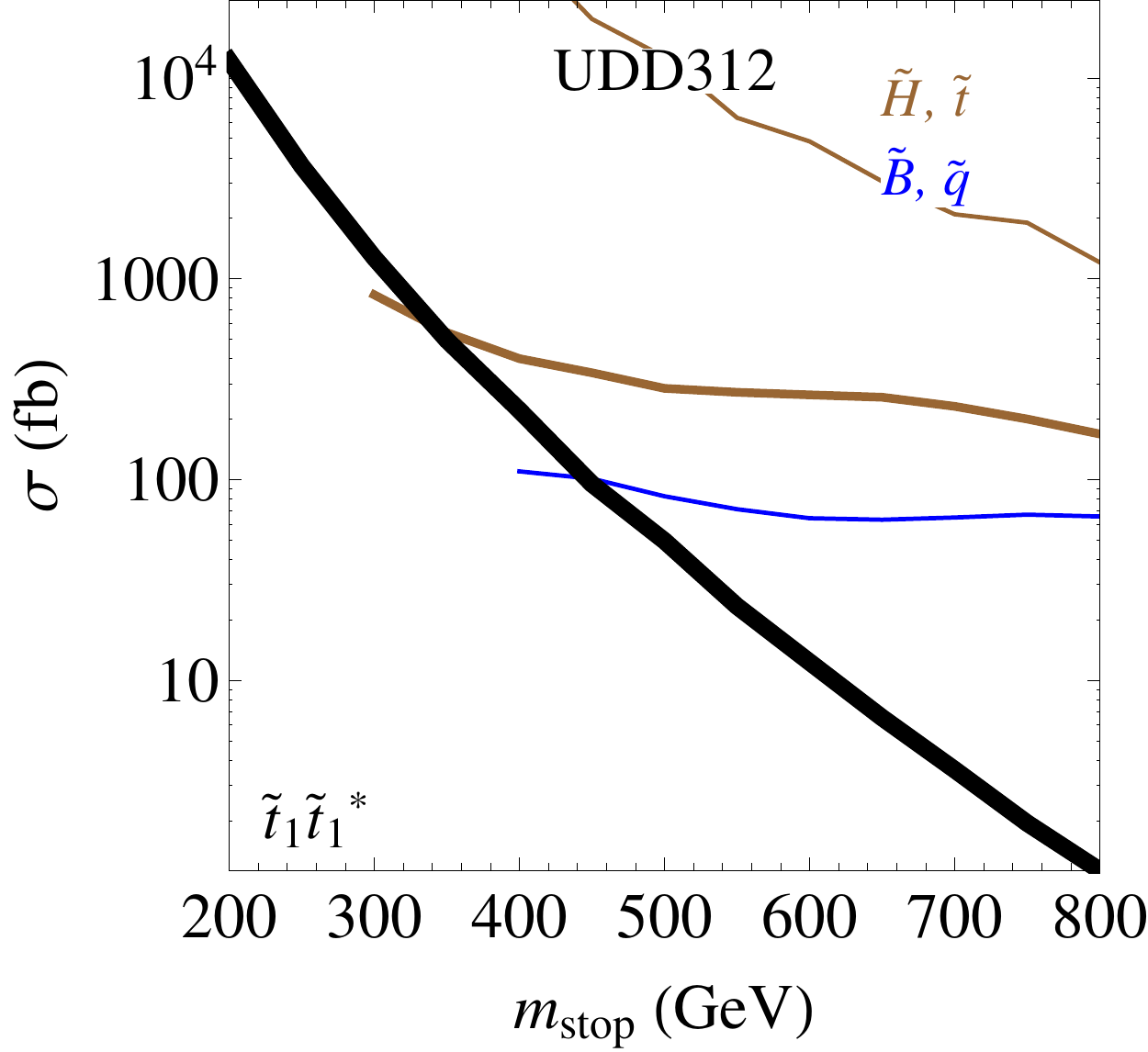}\q
\includegraphics[scale=0.6]{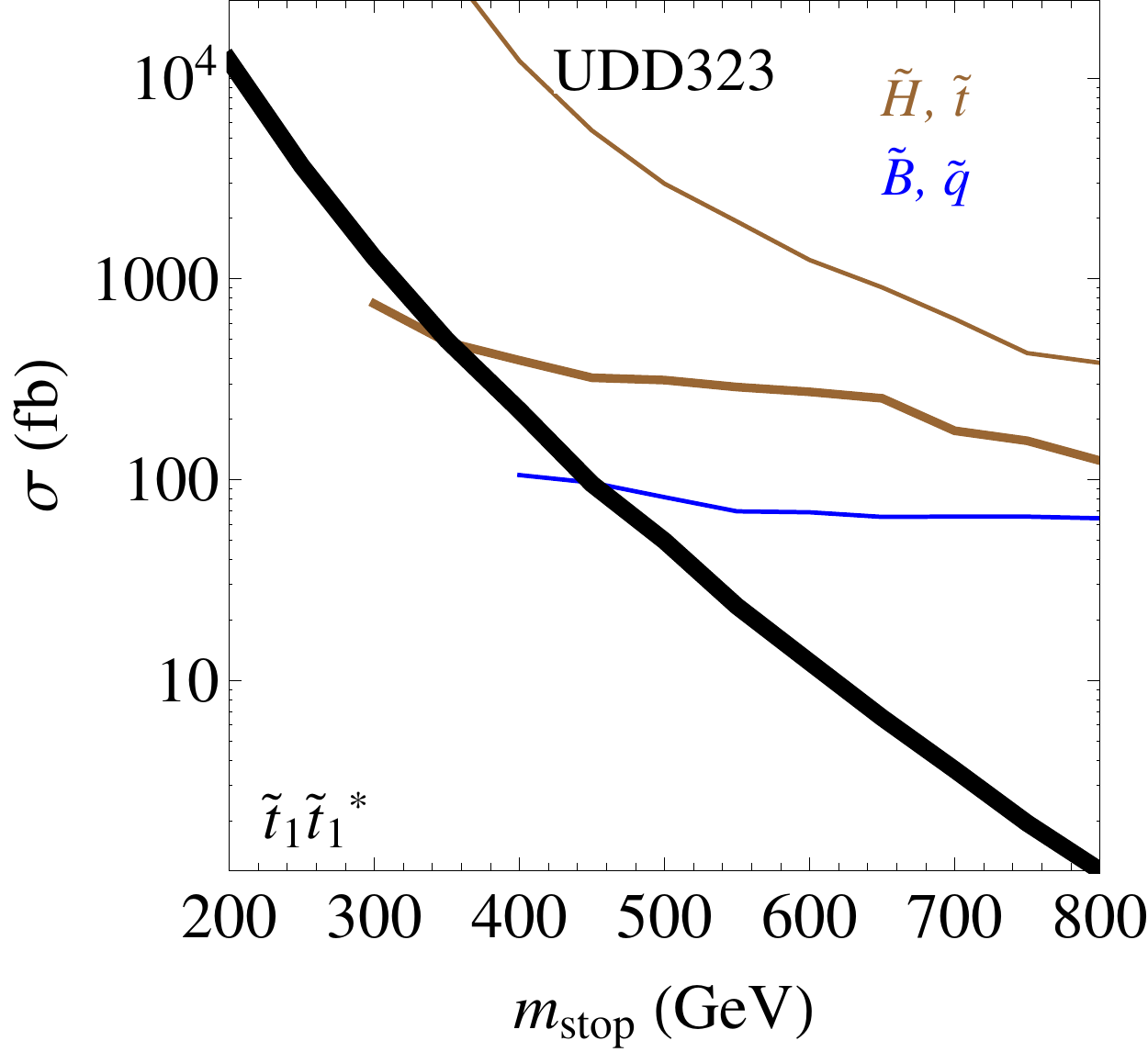}
\caption{Limits on stops decaying via other superpartners (see legend) in the presence of UDD operators. For the 212 and 213 couplings, the first intermediate particle is either very heavy (dotted curves), $10\%$ heavier than the stop (dashed) or 100~GeV lighter than the stop (thin solid: $\cho^0,\cho^\pm\to\mbox{RPV}$, thick solid: $\cho^\pm\to\cho^0\to\mbox{RPV}$). For 312 and 323 couplings (through which the stop can decay directly), the Higgsino (bino) is taken to be 100~GeV (200~GeV) lighter than the stop. For 212 and 213 couplings, the $t\bar t$ cross section measurements~\cite{ATLAS:2012aa,:2012bt} have the best sensitivity at low masses (except for the Higgsino case with $\cho^\pm\to\cho^0$). For the 312 and 323 couplings, the limits on the $\Ho$-mediated cases with $\cho^\pm\to\cho^0$, and $\Bo$-mediated cases, are set by SS dileptons~\cite{CMS-SS-DIL,CMS-SS-DIL-b,Chatrchyan:2012yea}.}
\label{fig:UDD}
\end{center}
\end{figure}

For couplings involving the right-handed bottom superfields, i.e., LQD$ij3$ and UDD$ij3$ (except for LQD$i33$ and UDD$3j3$ which allow the stop to decay directly), there are viable three-body decay paths for the stop through a $W$ boson and an off-shell sbottom (see figure~\ref{fig:3b4bstop}, left). The corresponding simplified models are included in tables~\ref{tab:LQD} and~\ref{tab:UDD} and resulting limits are included in figures~\ref{fig:LQD} and~\ref{fig:UDD}. One may also consider an analogous diagram with a charged Higgs, $H^+$, instead of the $W^+$.  For simplicity, we assume that the charged Higgs is too massive for this to be relevant.  Even if it were sufficiently light to appear on-shell, the corresponding process would generally be subdominant to the $W^+$ process as the more massive $H^+$ is suppressed by the available phase space.

\begin{figure}[t]
\begin{center}
\includegraphics[scale=0.6]{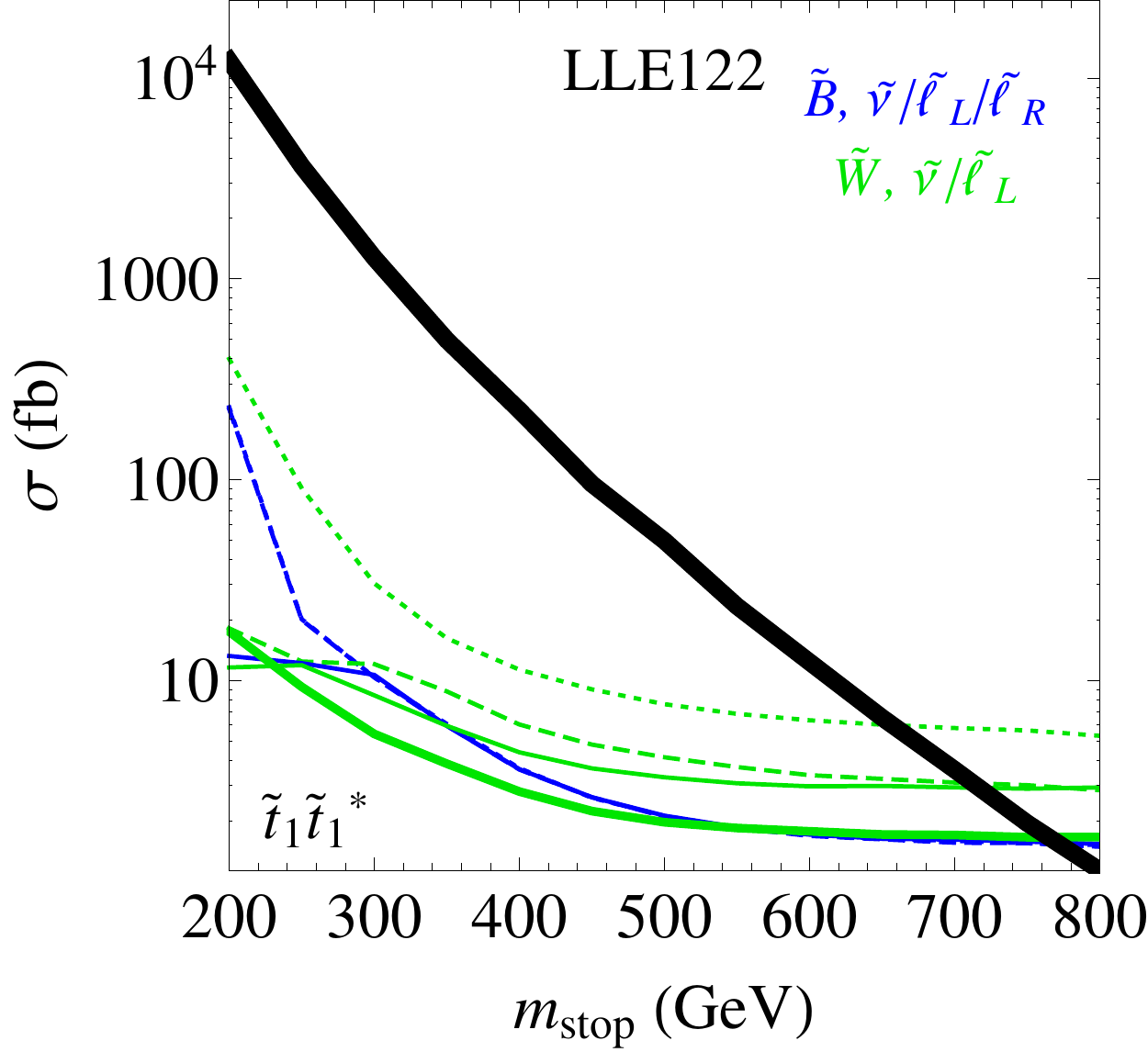}\q
\includegraphics[scale=0.6]{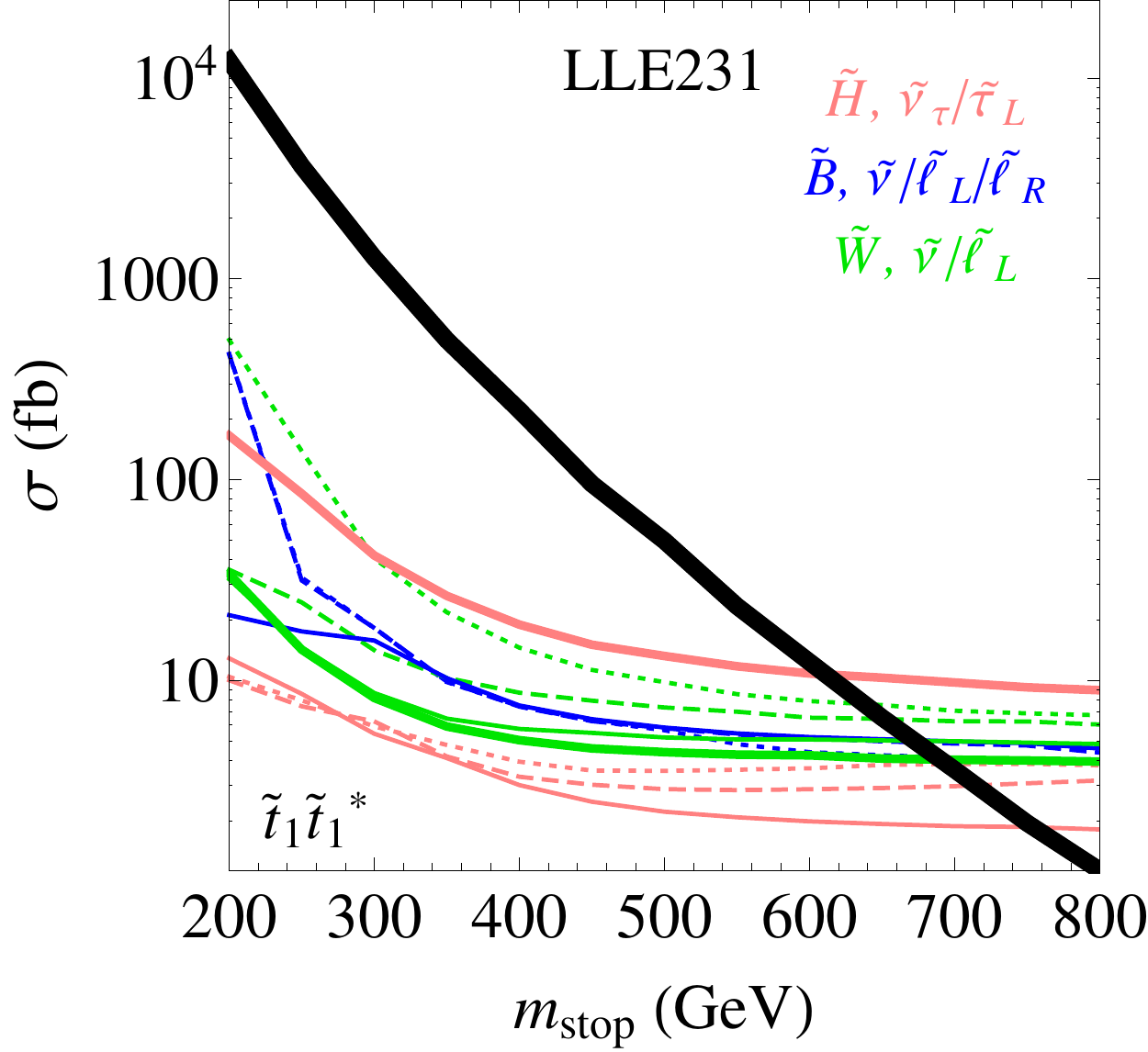}\\\vskip 10pt
\includegraphics[scale=0.6]{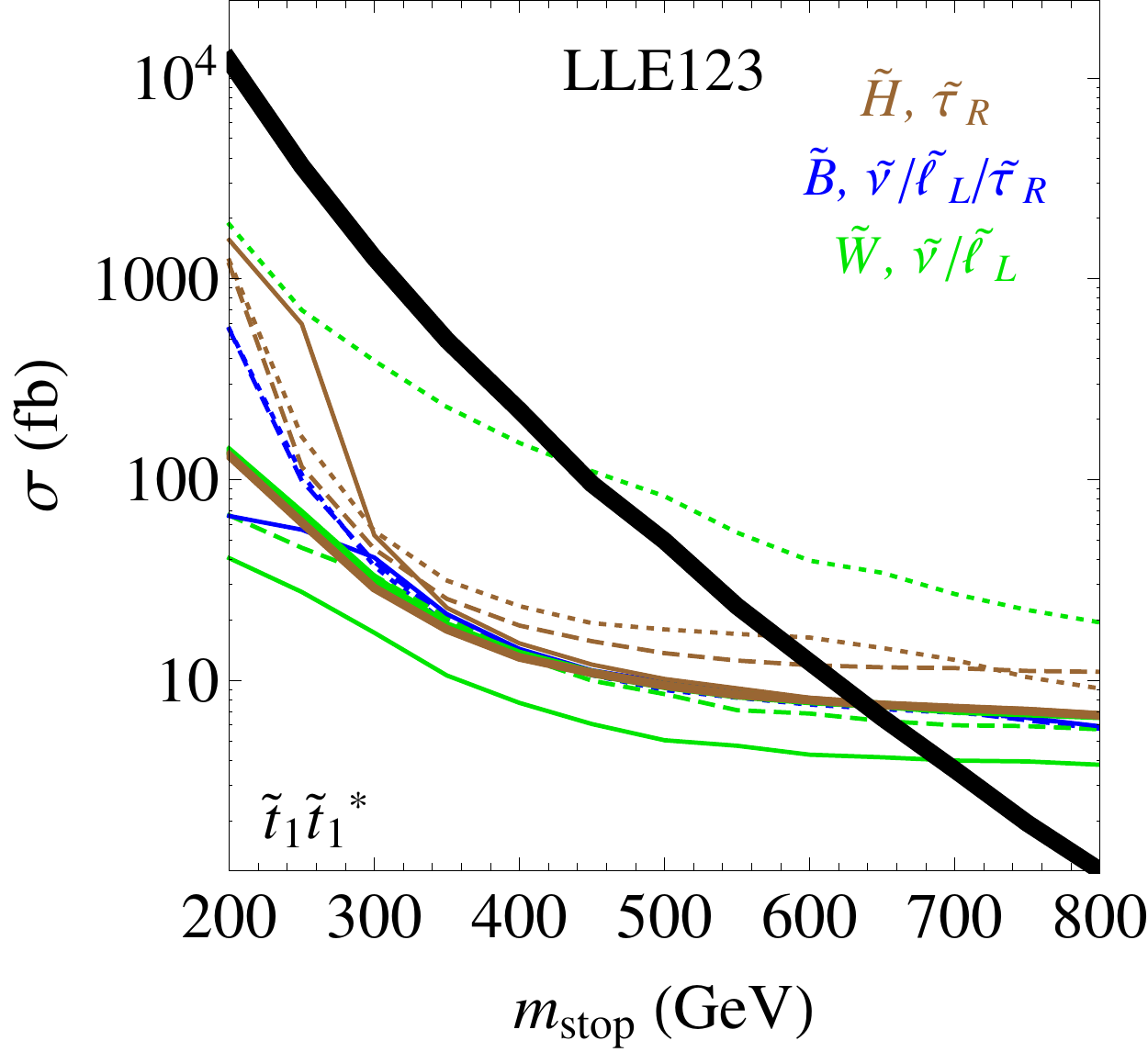}\q
\includegraphics[scale=0.6]{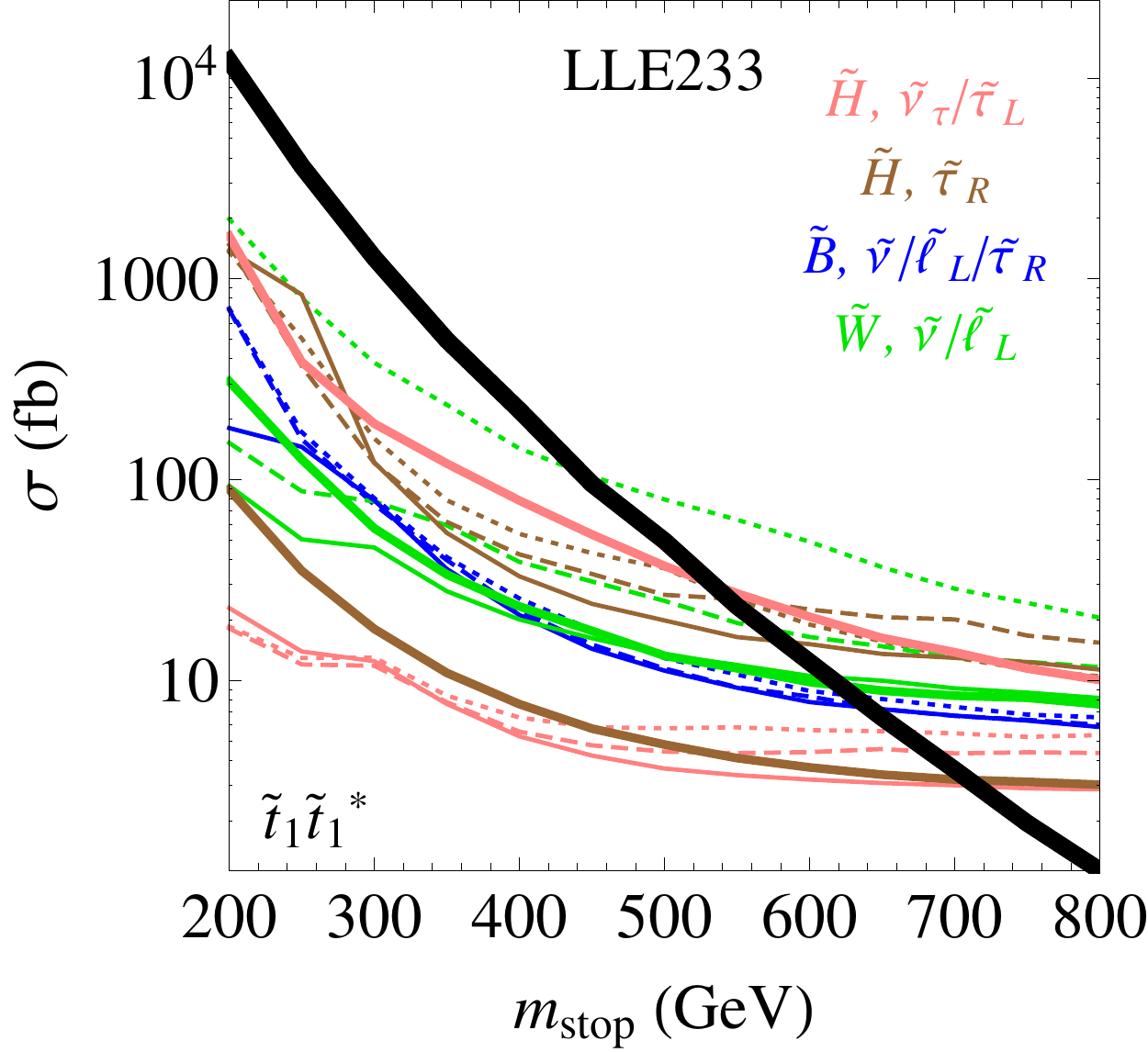}
\caption{Limits on stops decaying via other superpartners (see legend) in the presence of LLE operators. The first intermediate particle is either very heavy (dotted curves), $10\%$ heavier than the stop (dashed) or 100~GeV lighter than the stop (thin solid: $\cho^0,\cho^\pm\to\mbox{RPV}$, thick solid: $\cho^\pm\to\cho^0\to\mbox{RPV}$). In most cases, the best limits are set by multilepton or same-sign (SS) dilepton searches, in particular~\cite{CMSmultileptons,CMS-SS-DIL,Chatrchyan:2012yea,ATLAS-CONF-2012-130,ATLAS-CONF-2012-001}. In heavy $\Wo$ cases with 122 or 231, and $\Ho$-$\stau_R$ cases with $\cho^0,\cho^\pm\to\mbox{RPV}$, the best (or comparable) limits are set by the leptonic $m_{T2}$ search~\cite{:2012uu}, supplemented by the $t\bar t$ cross section measurements~\cite{ATLAS:2012aa,:2012bt,Chatrchyan:2012vs} at low $m_{\rm stop}$. Searches for 1 or~2$\tau$+jets+MET~\cite{1204.3852,1203.6580,:2012ht,CMS-OS-DIL} are comparable to the SS dilepton searches in the 123 heavy $\Wo$ case and provide the best limits on the 233 heavy $\Wo$ case. The $2\tau$+jets+MET searches~\cite{1203.6580,:2012ht} also sets the best limits in the 233 $\Ho$-$\snu_\tau/\stau_L$ case with $\cho^\pm\to\cho^0$ transitions.}
\label{fig:LLE}
\end{center}
\end{figure}

\begin{figure}[t]
\begin{center}
\includegraphics[scale=0.6]{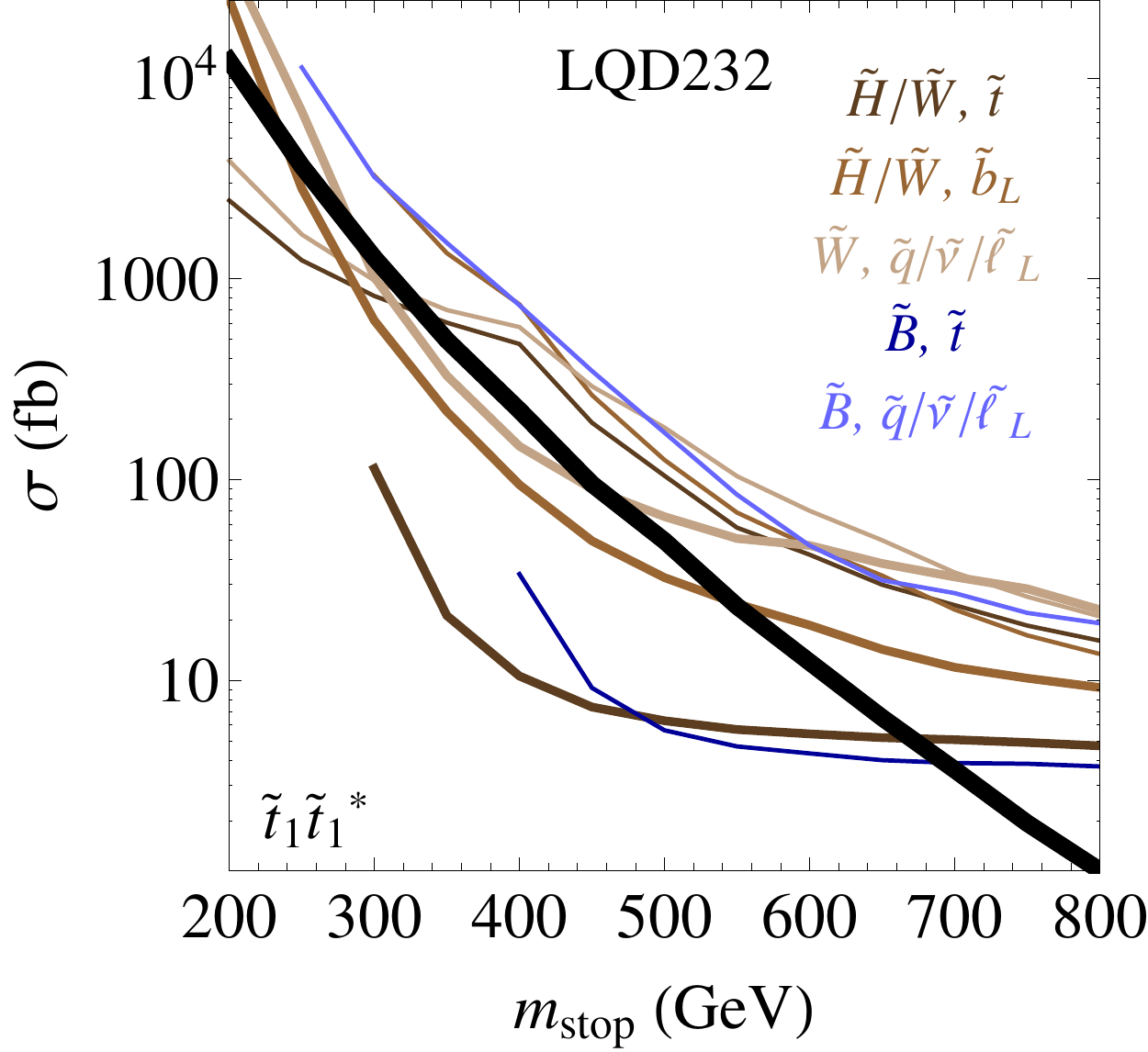}\q
\includegraphics[scale=0.6]{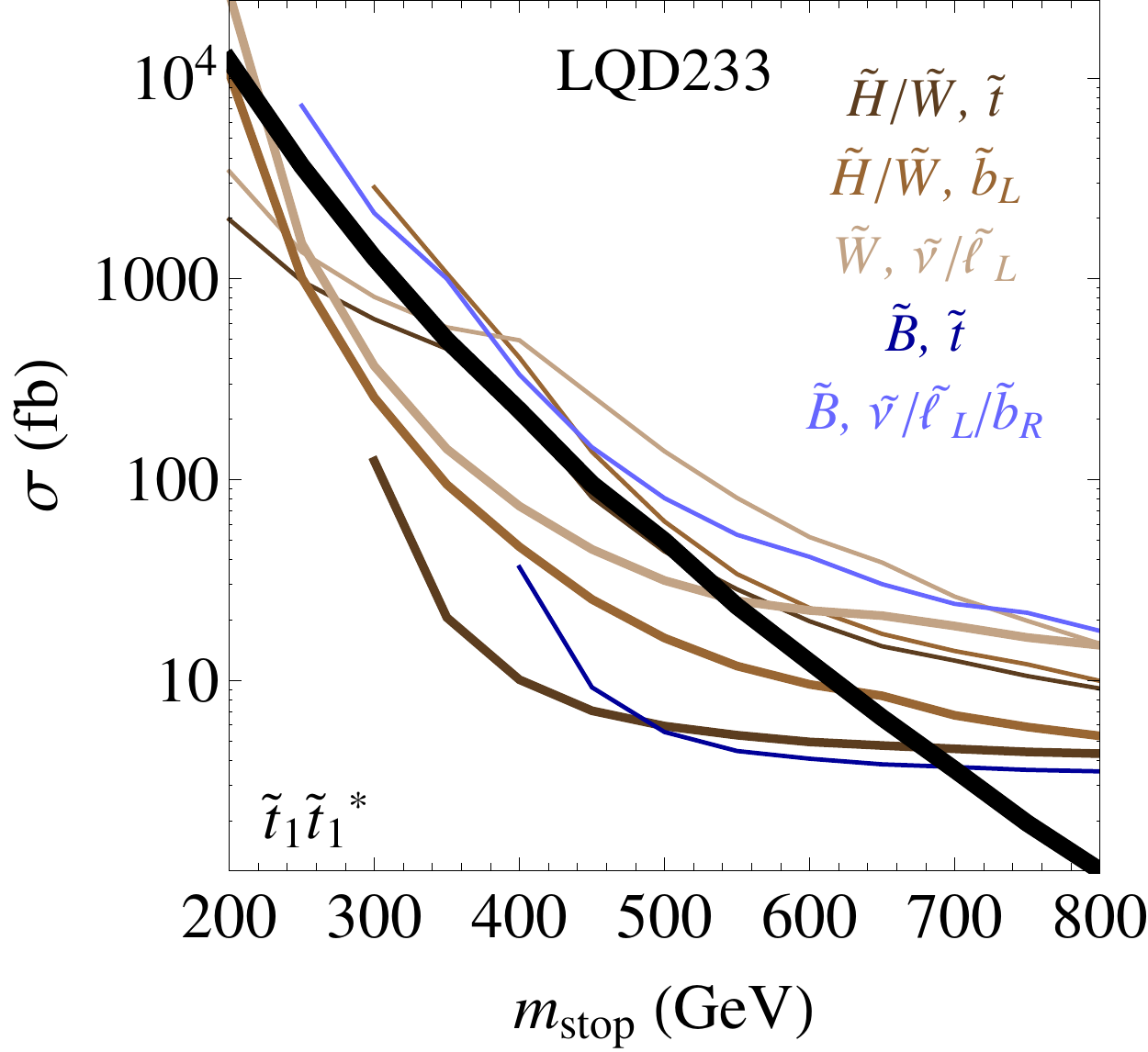}\\\vskip 10pt
\includegraphics[scale=0.6]{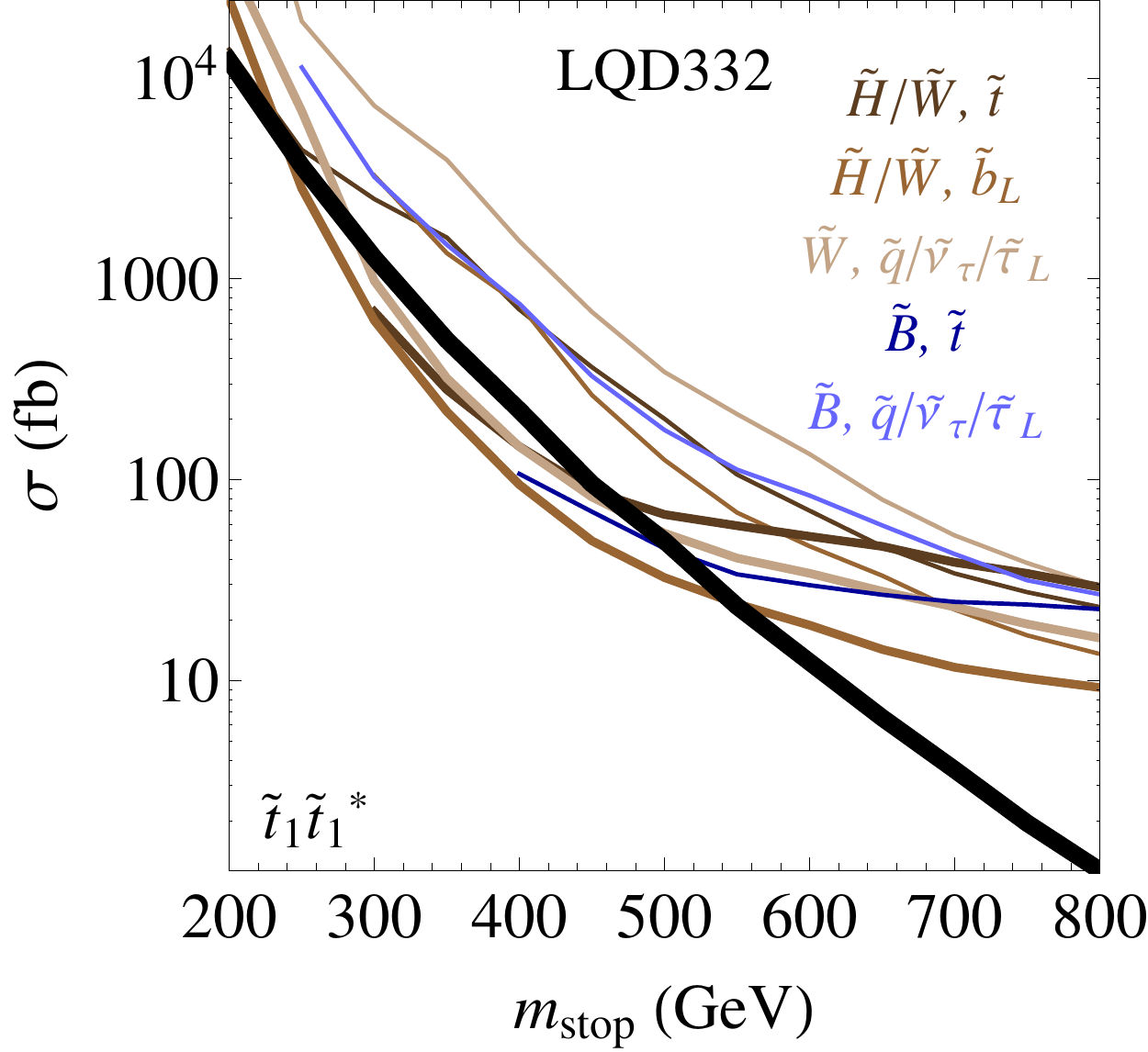}\q
\includegraphics[scale=0.6]{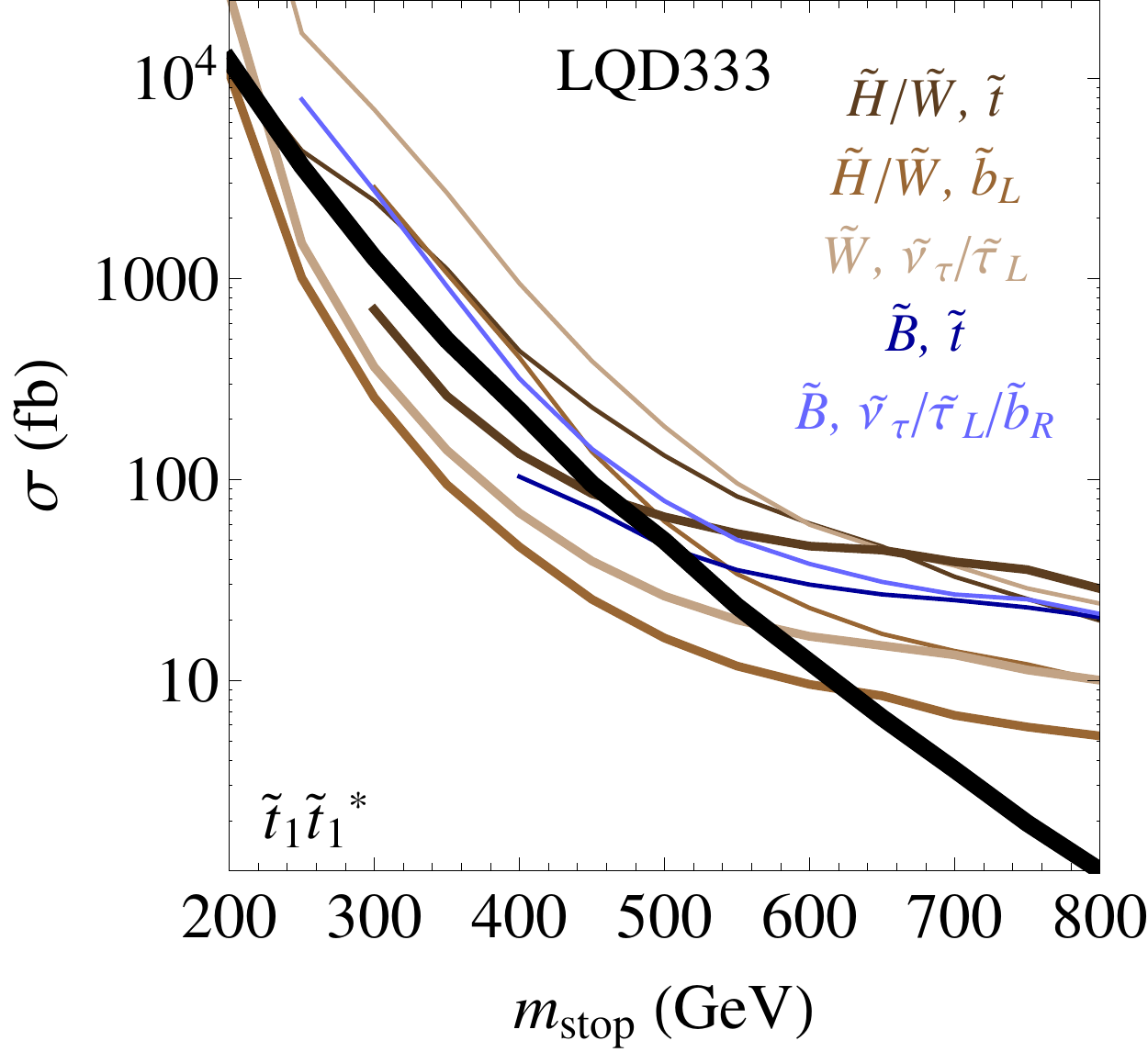}
\caption{Limits on stops decaying via on-shell inos (see legend) in the presence of LQD operators through which the stop can decay directly. The Higgsinos and winos are assumed to be 100~GeV lighter than the stop (thin curves: $\cho^0,\cho^\pm\to\mbox{RPV}$, thick curves: $\cho^\pm\to\cho^0\to\mbox{RPV}$) and the binos 200~GeV lighter than the stop. For $\Ho$ and $\Wo$ mediators with $\cho^\pm\to\cho^0$, and $\Bo$ mediators, the best limits on the $\st$-dominated cases are set by the searches for SS dileptons~\cite{CMS-SS-DIL,CMS-SS-DIL-b} and $b'$~\cite{Chatrchyan:2012yea,ATLAS-CONF-2012-130}, while for the other sfermion choices presented, most powerful are searches requiring multiple $b$-jets+MET (without leptons~\cite{ATLAS-3b}, or in the $\Bo$ case also with a lepton~\cite{CMS-PAS-SUS-11-028}). For $\Ho$ and $\Wo$ mediators with $\cho^\pm\to\mbox{RPV}$, the limits on the $\st$-dominated cases are set by the $t\bar t$ cross section measurements~\cite{:2012bt,Chatrchyan:2012vs} and are close to being extended to higher masses by the leptonic $m_{T2}$ search~\cite{:2012uu} in the 232 and 233 cases, and the search for $b$-jets+$\ell$+MET~\cite{CMS-PAS-SUS-11-028} in the 333 case; in the $\sbo_L$-dominated cases, the searches for $b$-jets(+$\ell$)+MET~\cite{ATLAS-3b,CMS-PAS-SUS-11-028} are closest to setting limits.}
\label{fig:LQD-i3k}
\end{center}
\end{figure}

\subsubsection{Discussion of the results}

The limits we obtain for the various scenarios are presented in figures~\ref{fig:LLE}--\ref{fig:UDD}. The different mediator assumptions are represented by different colors, as indicated in the legends. The style of the curve describes the mass of the mediator: dotted for a mediator much heavier than the stop (which is the only option considered for a gluino and a sbottom), dashed for a mediator $10\%$ heavier than the stop, and solid for an on-shell mediator.
For Higgsinos and winos, the on-shell mediator scenario is further separated to cases where $\cho^+ \to \{$RPV$\}$ (thin solid line) and where, via the process~(\ref{eq:ino-cascade}), $\cho^+ \to \cho^0 \to \{$RPV$\}$ (thick solid line). In the latter case, we conservatively assumed the chargino to be only 5~GeV heavier than the neutralino so that the decay products of the off-shell $W$ are essentially undetectable.\footnote{For splittings $\gtrsim 20$~GeV, transitions (between the chargino and neutralino, or between the two neutralinos of the $\Ho$ case) will sometimes add detectable objects to the events, and in some cases this will make the discovery easier.}
For decays through on-shell $\Ho$ or $\Wo$, we do not present limits for $m_{\rm stop} < 200$~GeV since charginos lighter than 100~GeV are excluded by LEP;\footnote{However, models with stop-chargino splitting smaller than our benchmark value of 100~GeV are possible, so scenarios with even lighter stops are also possible.} in on-shell $\Bo$ cases we include neutralinos as light as $50$~GeV. For 4-body decays through off-shell $\Bo$ and $\go$ mediators, we do not present limits for $m_{\rm stop} < 200$~GeV since the final state always contains a top, and with an off-shell top the five-body decay would likely be displaced (assuming in the $\Bo$ case that the sleptons are not much lighter than 200~GeV).

For LLE couplings (table~\ref{tab:LLE}, figure~\ref{fig:LLE}), in almost all cases, we find that the lower bounds on the stop mass are as high as 600-700~GeV, beyond the natural range for stops. These strong limits indicate that the experimental coverage of these scenarios is very good, as even before taking into account branching ratios, acceptances, identification efficiencies and cuts, a 750~GeV stop would have only $\sim 10$~events in 5~fb$^{-1}$ of 7~TeV data. The limits are so strong because all of the events contain either four or more charged leptons, or at least two charged leptons and several neutrinos (a source of $\met$). The weakest limits are obtained for the 123 and 233 couplings when the mediator is a heavy wino, where the dominant signature is a pair of opposite-sign taus, two $b$-jets and $\met$ (in appendix~\ref{app:helicity}, we explain why this final state dominates for winos much heavier than the stop).

For LQD couplings that do not involve the stop (table~\ref{tab:LQD}, figure~\ref{fig:LQD}), the limits are weaker than in the LLE case because there are fewer leptons. For light-lepton LQD couplings (221 and 123), gluino, bino, Higgsino, and sbottom-mediated decays, as well as scenarios with on-shell $\cho^\pm\to\cho^0$ transitions, have relatively strong limits from SS dilepton searches. Indeed, decays of neutralinos and gluinos can produce leptons of either sign. Decays via the sbottom or the chargino will produce opposite-sign leptons, but additional leptons can arise from leptonically-decaying $W$s (in the sbottom case) or tops (in a chargino case where a top is produced in the chargino decay, as happens in the 123 Higgsino scenario). On the other hand, SS dilepton events are rare in the wino scenarios since decays via $\cho^+ + b$ dominate over decays via $\cho^0 + t$ (due to phase space). Furthermore, hard neutrinos (i.e., $\met$) are available only at the price of losing a lepton. Consequently, the limits on the wino scenarios are relatively weak (with the exception of the $\cho^\pm\to\cho^0$ scenarios mentioned above). The picture changes drastically for couplings where the lepton is a tau (321, 323). Hadronically decaying taus have low identification efficiencies and high fake rates, while leptonically decaying taus are suppressed by their branching ratios. As a result, the signal gets spread over many different final states, making the limits on all the scenarios weak.

For LQD couplings that involve the stop (table~\ref{tab:LQD-i3k}, figure~\ref{fig:LQD-i3k}), the inos have multiple decay paths with branching fractions that depend on the mediating sfermion masses, as discussed in the previous subsection.\footnote{For binos, we do not present the $\sbo_L$-mediated case separately since the corresponding final state is the same as the one dominating in the slepton-mediated case. Decays of Higgsinos through the sleptons or $\sbo_R$ are also covered by the other scenarios that we present (see table~\ref{tab:LQD-i3k}).} The decays of the bino through the stop give rise to events with two leptons and four tops, leading to very strong limits from SS dileptons. However, if the bino also may decay via other superpartners, the decays $\cho^0\to\nu bj$ dominate over $\cho^0\to\ell tj$ due to phase space suppression from the heavy top. This leads to final states with only a single $t\bar t$ pair and no leptons, resulting in very weak limits. In Higgsino and wino scenarios with $\cho^\pm\to\cho^0\to\{$RPV$\}$, events with two leptons and two tops, from the decays of $\cho^0$ through the stop, again allow strong limits to be set by SS dileptons. The limits remain strong when the neutralino can decay through other superpartners due to the searches for multiple $b$-jets+MET, unlike in the bino case where these searches are not sufficiently powerful. In Higgsino and wino scenarios with $\cho^\pm\to\{$RPV$\}$ decays, the dominant final states contain either two opposite-sign leptons (and jets, but no significant MET) or a $t\bar t$ pair (accompanied by jets and MET), so the limits are weak.

For UDD couplings (table~\ref{tab:UDD}, figure~\ref{fig:UDD}),\footnote{In the 323 Higgsino case, we present results only for decays via $\st$ since the final state of decays via $\sbo_R$ is highly phase-space suppressed, except in the case that the chargino transitions to a neutralino before decaying, where the final state is simply the same as with $\st$.} the limits on all the 212 and 213 cases (and the 312 and 323 Higgsino cases with $\cho^\pm\to\cho^0$) are very weak.  Indeed, the signatures $t\bar t$+jets, $W^+W^-$+jets, or just jets, have large SM backgrounds, and there are no searches that make use of the large jet multiplicities or the extra $b$-tags that are available in these low-$\met$ stop events. On the other hand, for 312 or 323 couplings, the 4-top events of bino-mediated decays, and the same-sign di-top events of Higgsino-mediated decays with $\cho^\pm\to\{$RPV$\}$, allow for limits to be set by SS dilepton searches.

The least constrained cases, and possible methods to search for some of them, will be discussed in more detail in section~\ref{sec:Discussion}.

% =============================================================================
\subsection{Scenarios with stop-antistop oscillations}
\label{sec:limits-oscillation}
% =============================================================================

\begin{figure}[t]
\begin{center}
\includegraphics[scale=0.6]{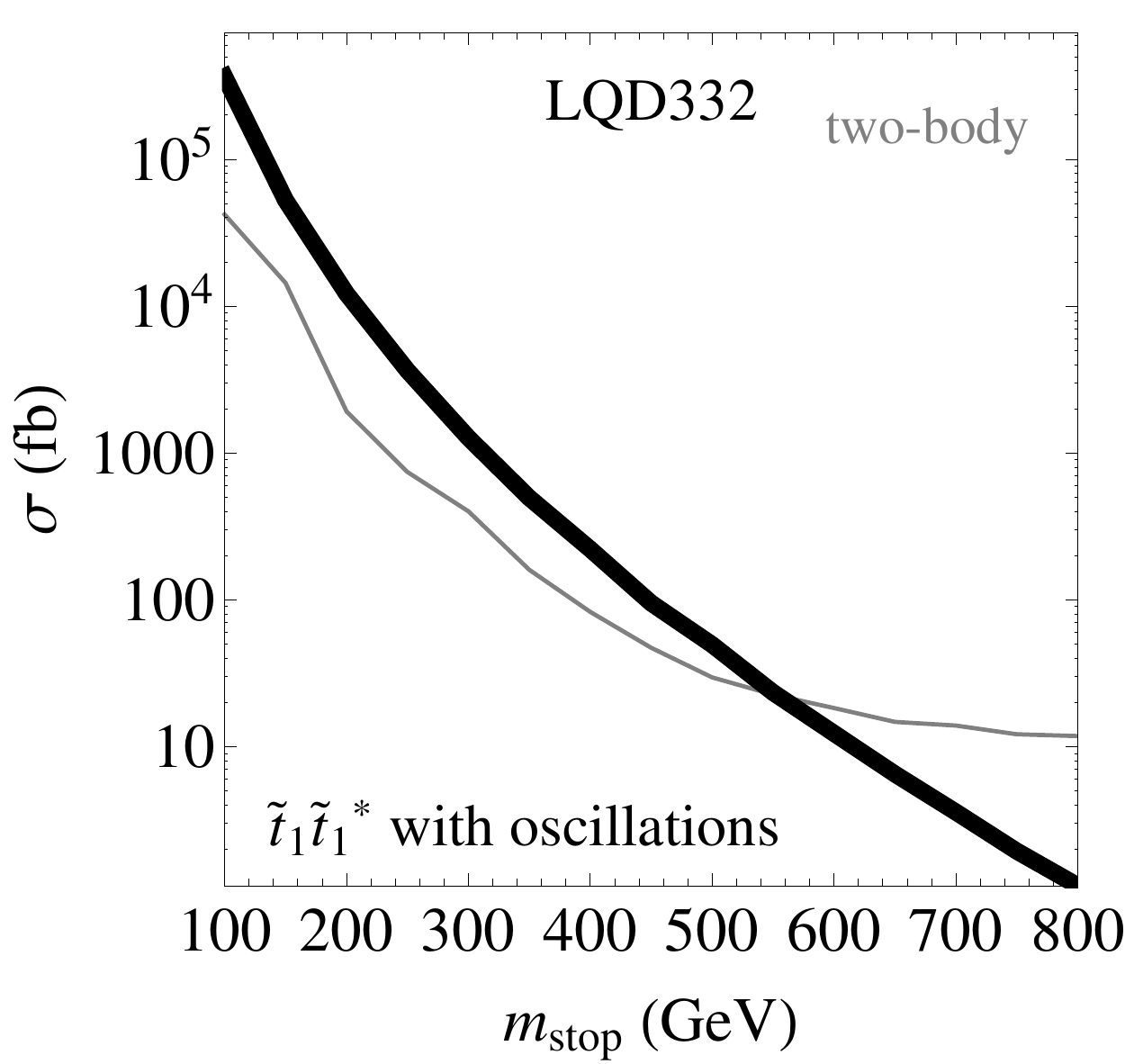}
\caption{Limit on $\st\to\tau^+ j$ decays (LQD332) in the presence of stop-antistop oscillations. The best limit is set by the search for $2\tau$+jets+MET~\cite{:2012ht}, supplemented by SS dilepton searches~\cite{ATLAS:2012mn,CMS-SS-DIL,CMS-SSSF-DIL} at low masses.}
\label{fig:LQD332osc}
\end{center}
\end{figure}

\begin{figure}[t]
\begin{center}
\includegraphics[scale=0.6]{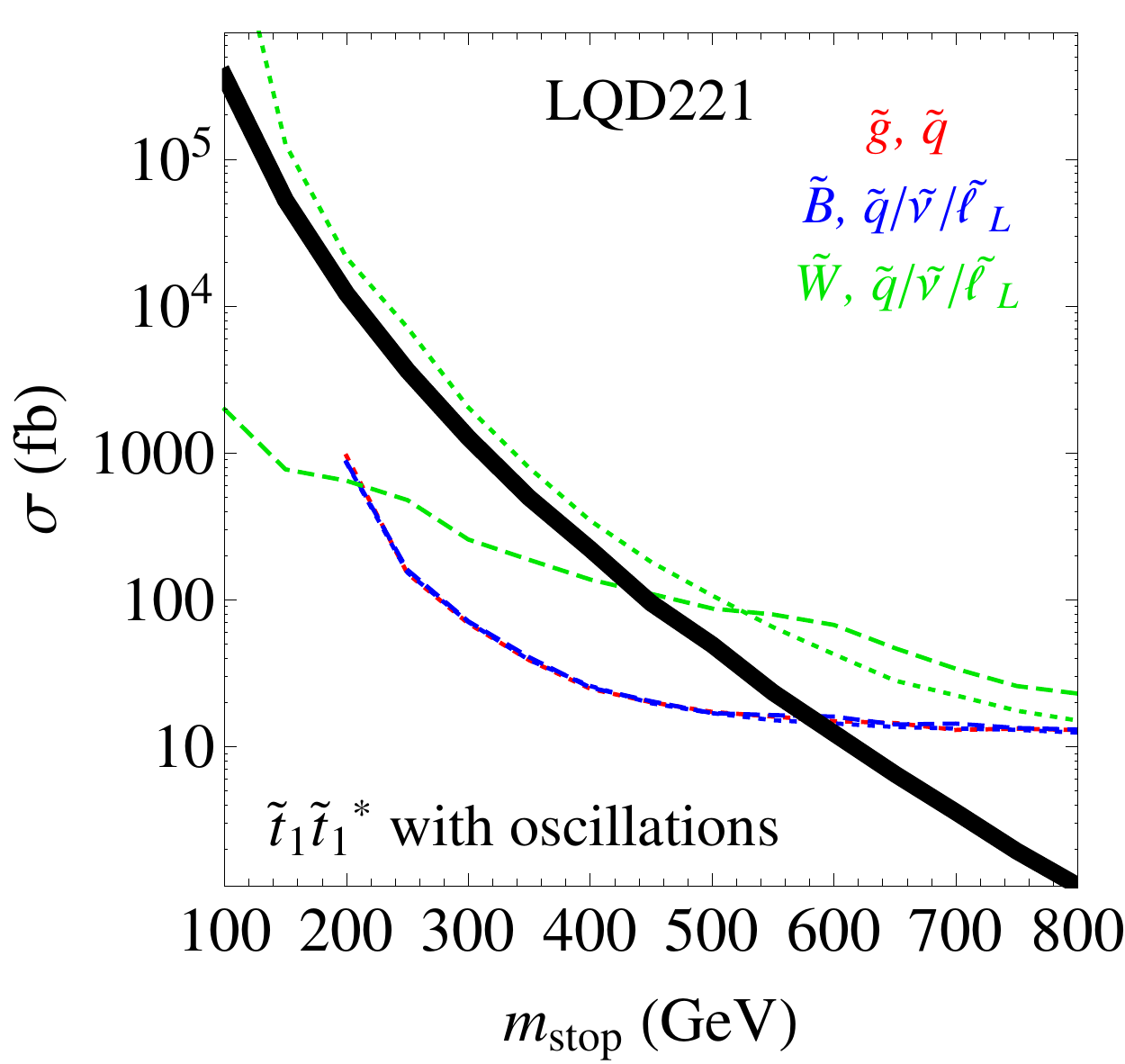}\q
\includegraphics[scale=0.6]{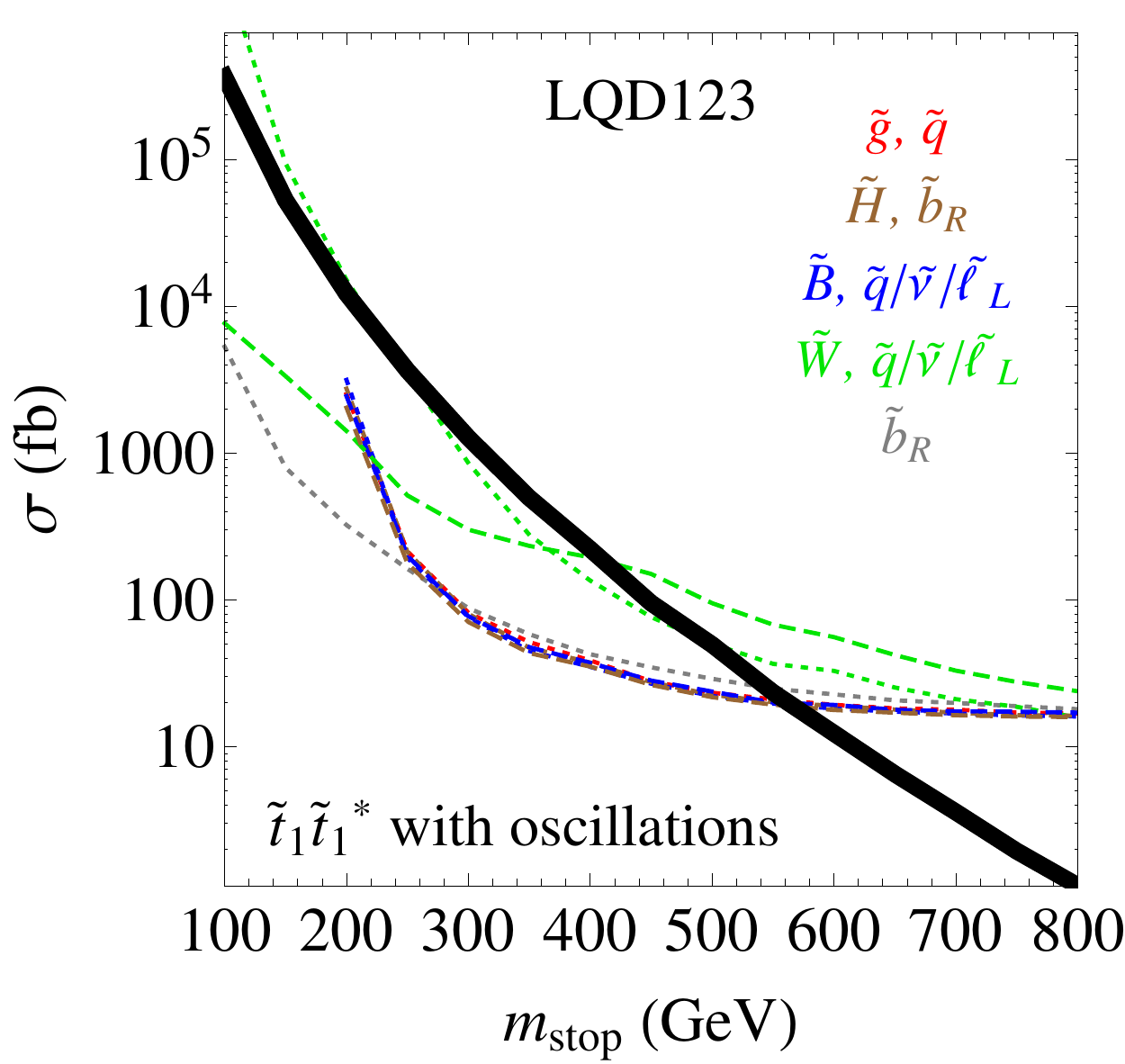}\\\vskip 10pt
\includegraphics[scale=0.6]{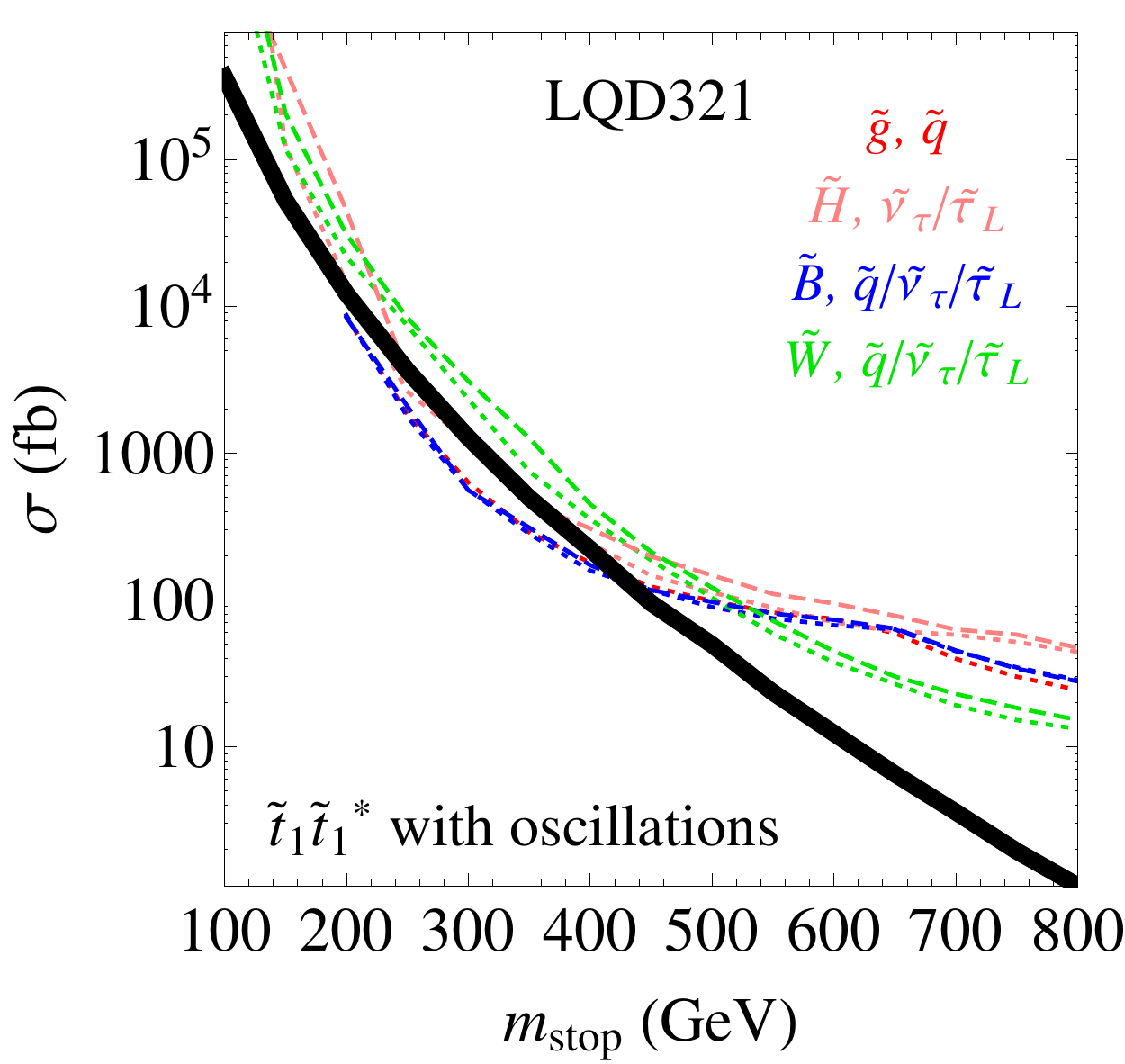}\q
\includegraphics[scale=0.6]{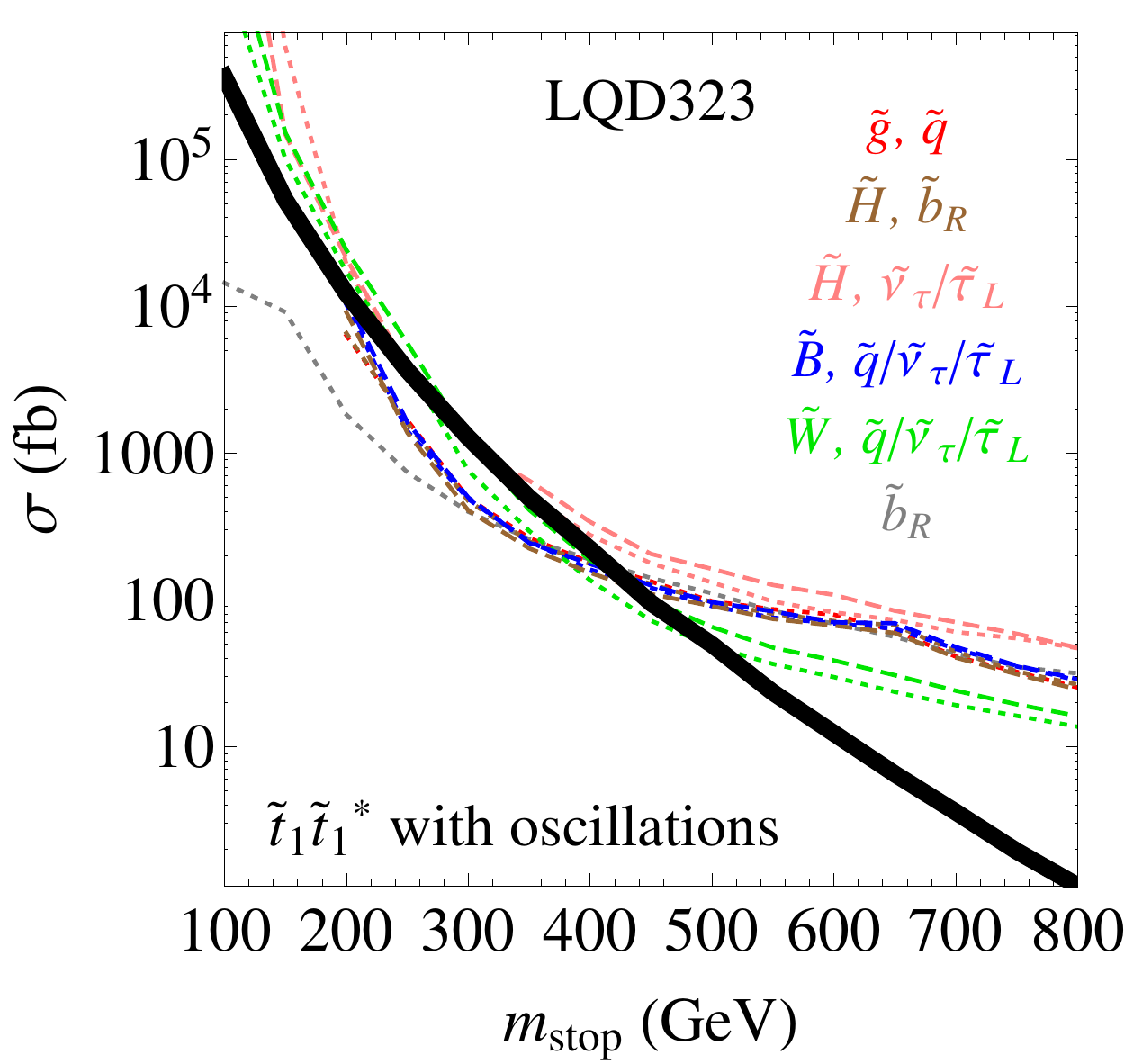}
\caption{Limits on LQD scenarios with stop-antistop oscillations. In most cases, the best limits are set by SS dilepton searches~\cite{ATLAS:2012mn,CMS-SS-DIL,CMS-SSSF-DIL,CMS-SS-DIL-b,Chatrchyan:2012yea,ATLAS-CONF-2012-130}. However, the $\Wo$ scenarios (except for light $\Wo$ in the 221 and 123 cases) are still constrained primarily by ($b$-)jets+MET searches~\cite{ATLAS-3b,:2012rg,ATLAS:2-6jets}, as in the case without oscillations.}
\label{fig:LQDosc}
\end{center}
\end{figure}

\begin{figure}[t]
\begin{center}
\includegraphics[scale=0.6]{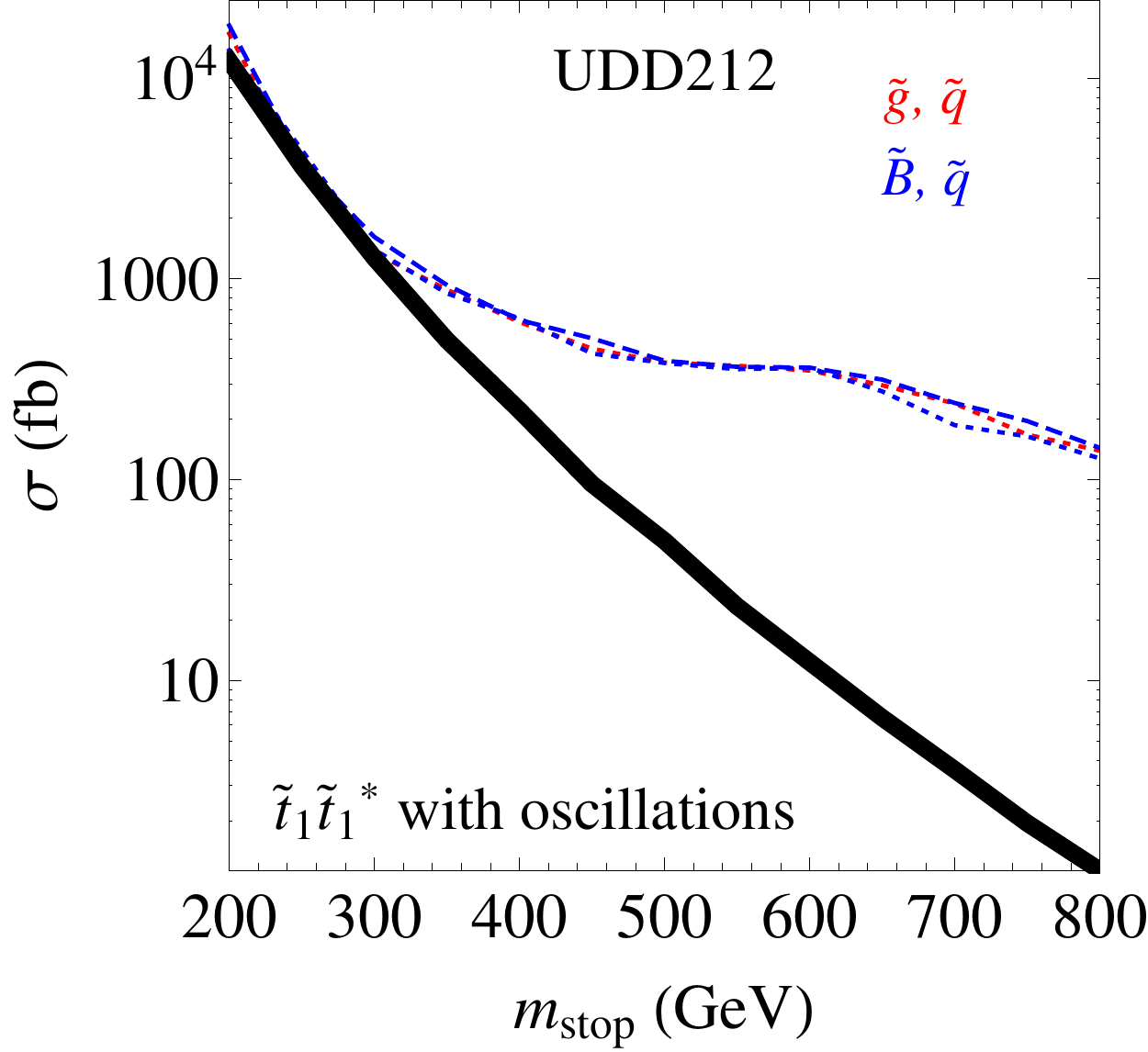}\q
\includegraphics[scale=0.6]{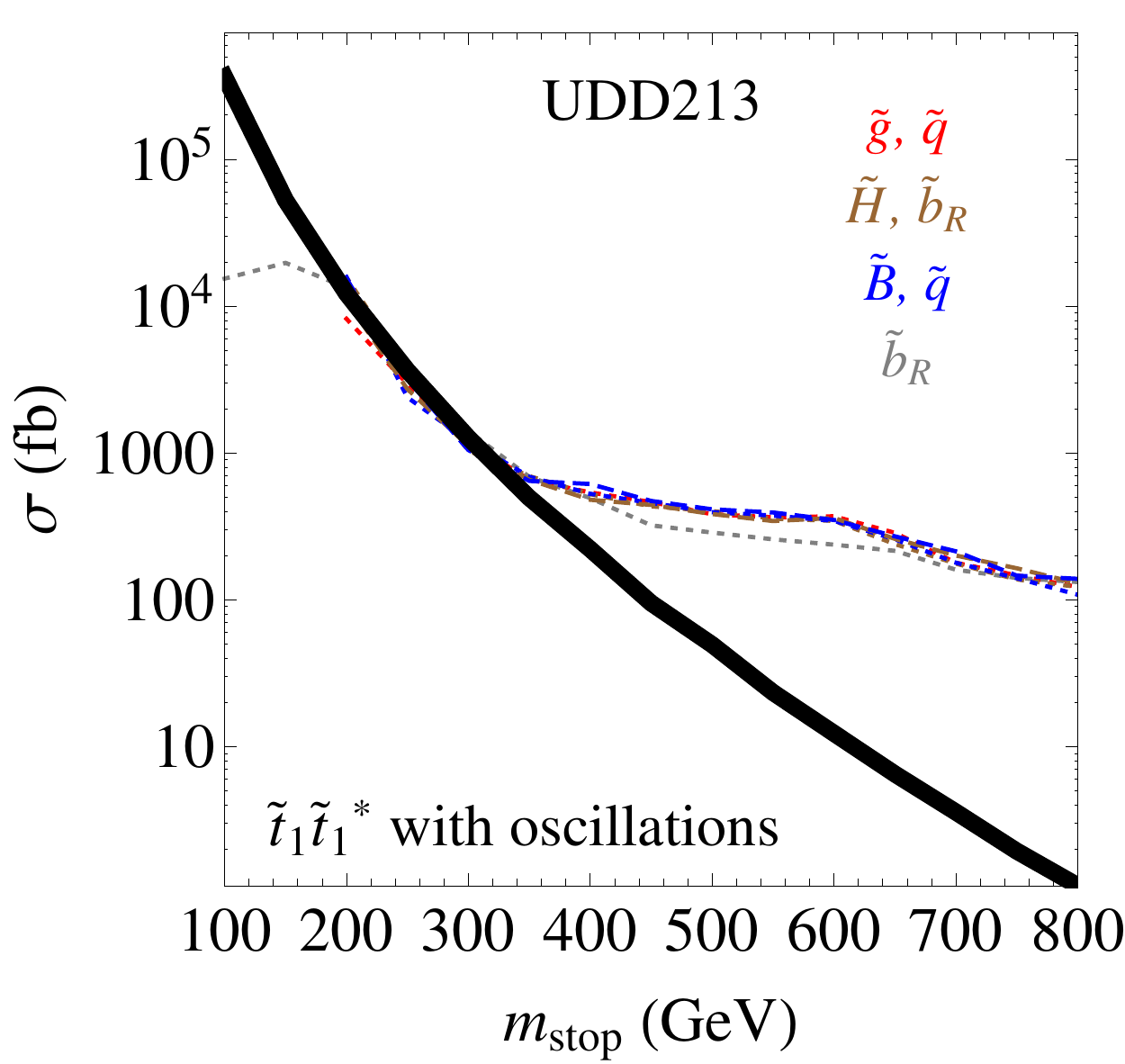}\\
\caption{Limits on UDD scenarios with stop-antistop oscillations. In all cases, the best limits are set by SS dilepton searches~\cite{ATLAS:2012mn,CMS-SS-DIL,CMS-SS-DIL-b,Chatrchyan:2012yea}.}
\label{fig:UDDosc}
\end{center}
\end{figure}

Sufficiently long-lived stops (i.e., $\Gamma_\st \lesssim \Lambda_{\rm QCD}$) will hadronize.  If the stop hadronizes into a neutral ``mesino,'' it may oscillate to an antistop before decaying~\cite{Sarid:1999zx}.  The oscillation period is determined by the rate of the relevant flavor violating processes, which is model-dependent, however it is plausible to have a situation in which the stop oscillates and still decays promptly.  Oscillation can enhance the detectability of the stops, due to a higher abundance of same-sign lepton pairs in cases where the leptons would have opposite signs otherwise.  Our implicit assumption so far has been of a regime where the oscillations are not important.  Now we will examine the situation with oscillations.

The probability for a stop to form a neutral mesino is $\frac 13 \lesssim f_0 \lesssim \frac 12$ (the uncertainty is due to the strange quark)~\cite{Sarid:1999zx}. The probability for a neutral mesino to be found in the anti-mesino eigenstate at the time of the decay reaches ${\cal P}=\frac12$ in the rapid oscillation regime. To illustrate the possible effects of oscillations, we will assume the maximal possible value for $P_{\rm osc} = {\cal P} f_0 = \frac14$. This implies there is a 3/8 probability for an event to oscillate to $\st\st$ or $\st^*\st^*$, allowing for SS dilepton signatures.

The resulting limit on the least constrained two-body LQD decay, $\st\to\tau^+ j$, is shown in figure~\ref{fig:LQD332osc}. Except at low masses, it has not improved relative to the case without oscillations (figure~\ref{fig:2body})~-- the limit from SS dileptons is weaker than the limit from the search for 2$\tau$+jets+MET. The limits on the two-body UDD decays (not shown) do not improve either as their final states do not contain leptons.

The limits on decays through the inos or sbottom are shown in figures~\ref{fig:LQDosc} and~\ref{fig:UDDosc}, where we have not included LLE scenarios since the limits there are usually very strong even without oscillations, nor have we included the cases in which the stop decays to an on-shell ino, assuming that the stop would be too short-lived to hadronize and oscillate. As comparison with figures~\ref{fig:LQD} and~\ref{fig:UDD} demonstrates, SS dilepton searches improve limits significantly in many of the cases. In fact, these searches now set the best limits in all cases, except for LQD heavy-wino mediated decays (where final states with leptons are suppressed by helicity considerations) and LQD321/323 light-wino mediated decays (where $b$-jets+MET final states yield stronger limits than final states with SS taus).

% =============================================================================
\section{Limits on simplified models with $\st_1$, $\st_2$, $\sbo_1$ production}
\label{sec:limits-twostops}
% =============================================================================

Having derived limits under the conservative assumption that they are dominated by $\st_1\st_1^\ast$ production, we now examine scenarios containing the additional production of $\st_2\st_2^\ast$ and $\sbo_1\sbo_1^\ast$\footnote{If the sbottom has a significant right-handed component and the UDD coupling $\lambda''_{123}$ is relatively large, \emph{single} sbottom production may be observable, as has been studied in~\cite{Kilic:2011sr}. We will not explore this possibility here.} (since naturalness also predicts $\st_2$ and $\sbo_1$ to be light). Let us consider a case where the masses of $\st_2$ and $\sbo_1$ are
\be
m_{\st_2} = m_{\sbo_1} = m_{\st_1} + 100\mbox{ GeV}
\label{eq:st2-sb1-masses}
\ee
and they decay as\footnote{For sufficiently large $\st_1$--$\st_2$ splittings, which we will not examine, the decay $\st_2 \to \st_1 h$ also becomes possible, with a (model-dependent) branching ratio comparable to that of $\st_2 \to \st_1 Z$.}
\be
\st_2 \to \st_1 Z\,,\q
\sbo_1 \to \st_1 W^-
\label{eq:st2-sb1-decays}
\ee
This leads to the same final states as found in the $\st_1\st_1^\ast$ case, but with additional $ZZ$ or $W^+W^-$.

\begin{figure}[t]
\begin{center}
\includegraphics[scale=0.6]{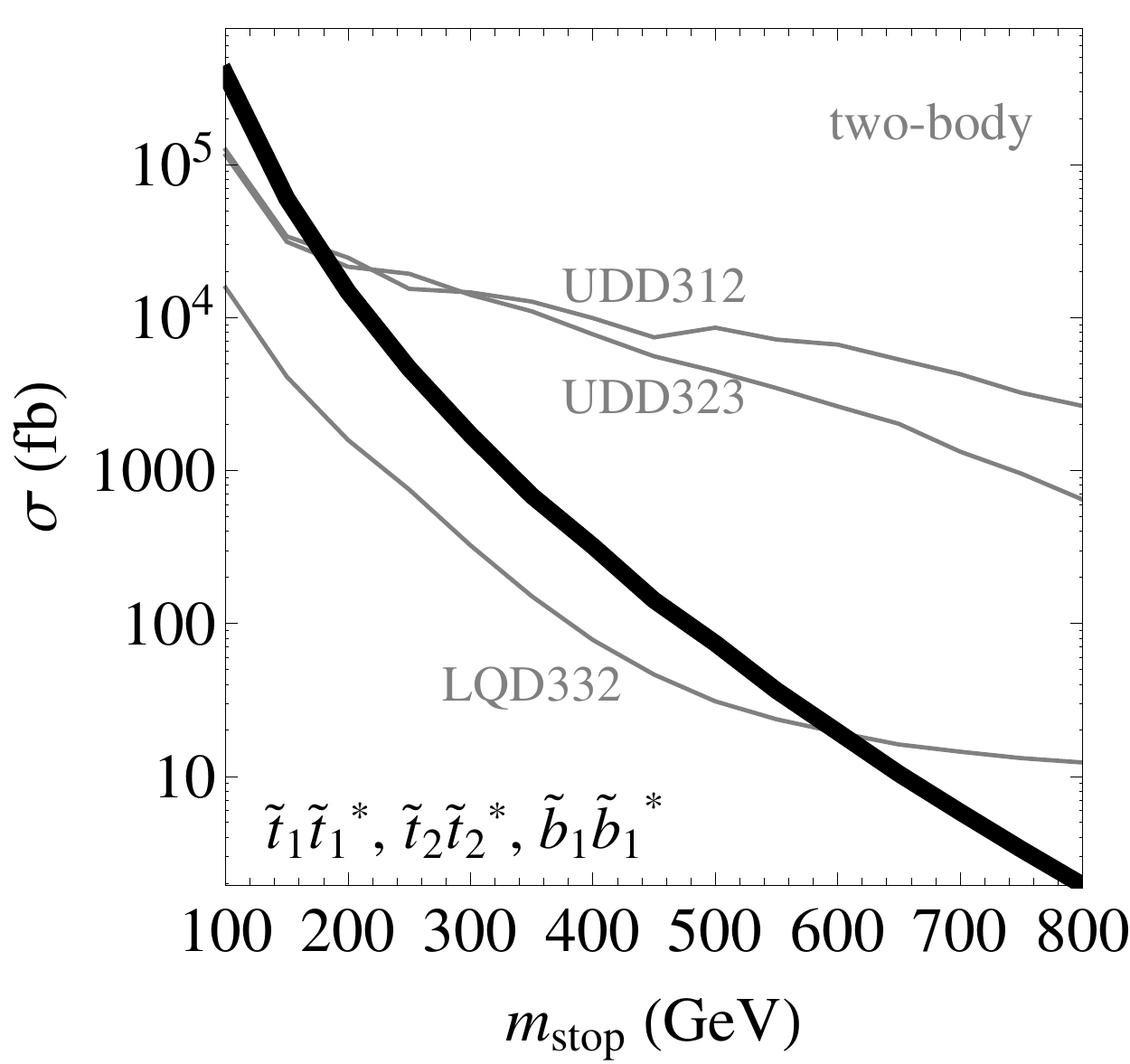}
\caption{Limits on scenarios with $\st_2$ and $\sbo_1$ (both 100~GeV heavier than $\st_1$) with two-body decays of $\st_1$---compare to the $\st_1$-only limits of figure~\ref{fig:2body}. In the UDD cases, the limits are set by the searches for multileptons~\cite{CMSmultileptons} and $Z$+jets+MET~\cite{Chatrchyan:2012qka}. In the LQD case, the limits are set by the searches for $2\tau$+jets+MET~\cite{1203.6580,:2012ht} and OS dileptons+MET (with $\tau$'s)~\cite{CMS-OS-DIL}, as with $\st_1$-only production, except at low masses where they are strengthened by searches for SS dileptons~\cite{CMS-SS-DIL} and multileptons~\cite{CMSmultileptons}.}
\label{fig:2bodyx2}
\end{center}
\end{figure}

\begin{figure}[t]
\begin{center}
\includegraphics[scale=0.6]{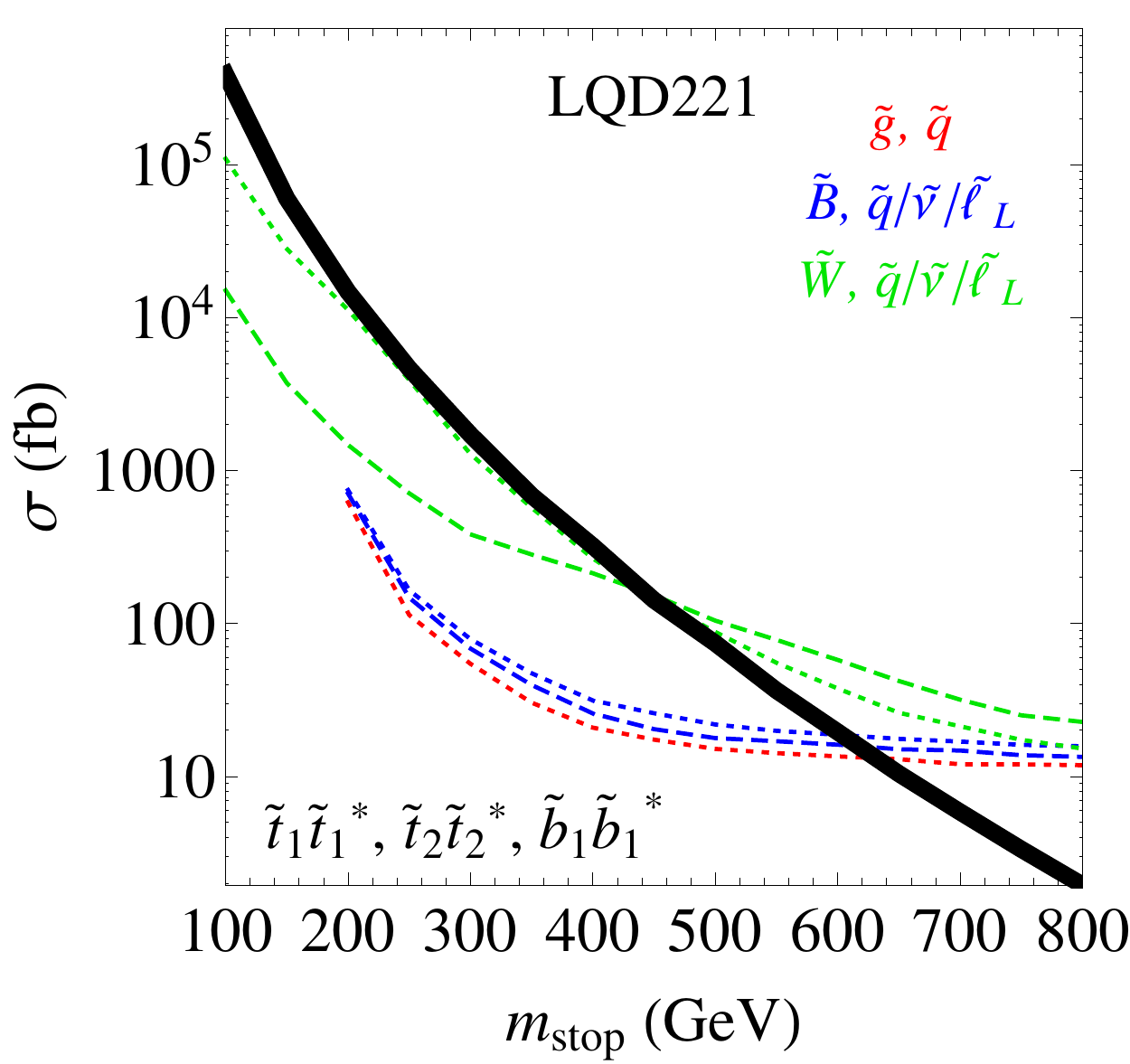}\q
\includegraphics[scale=0.6]{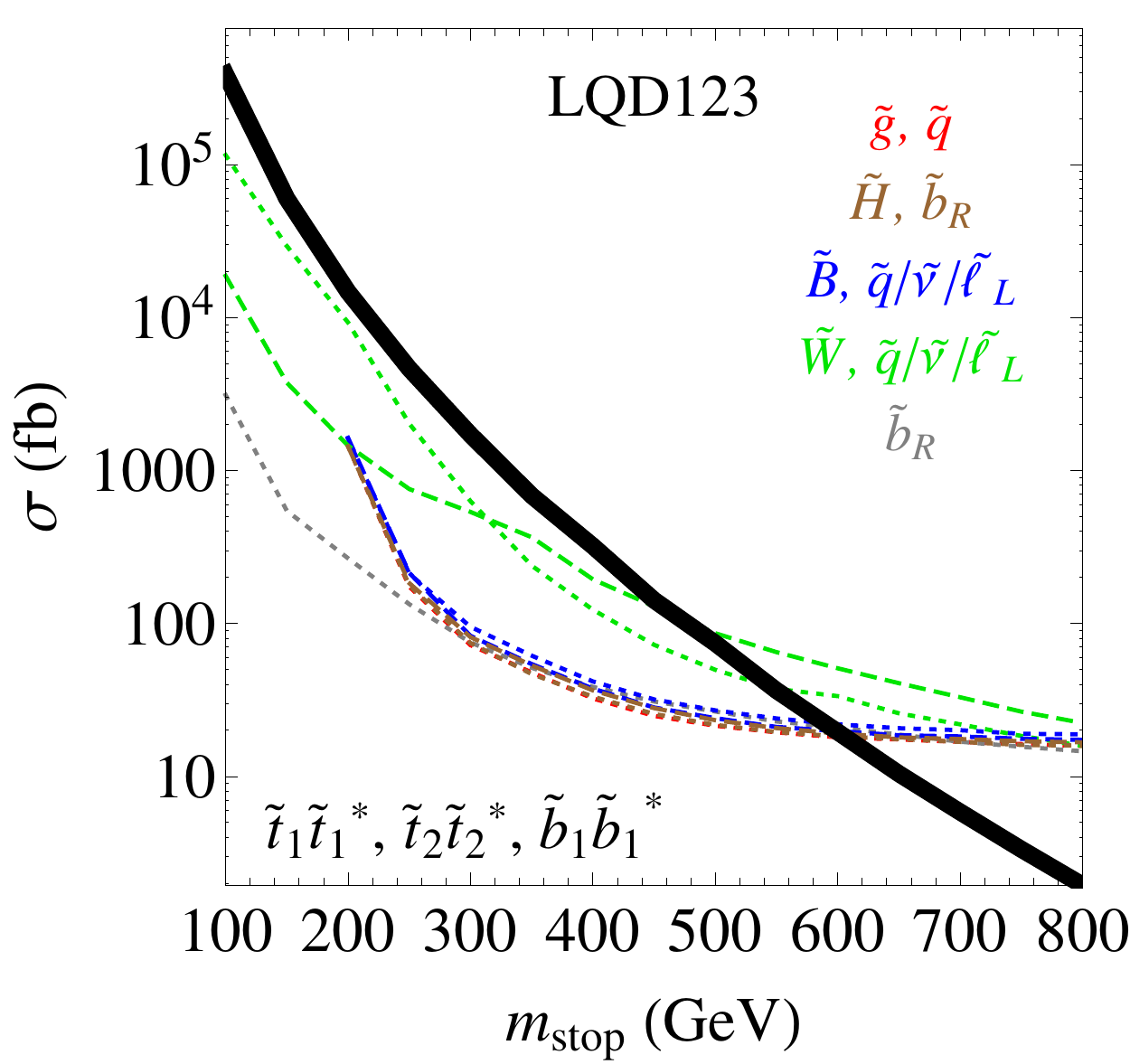}\\\vskip 10pt
\includegraphics[scale=0.6]{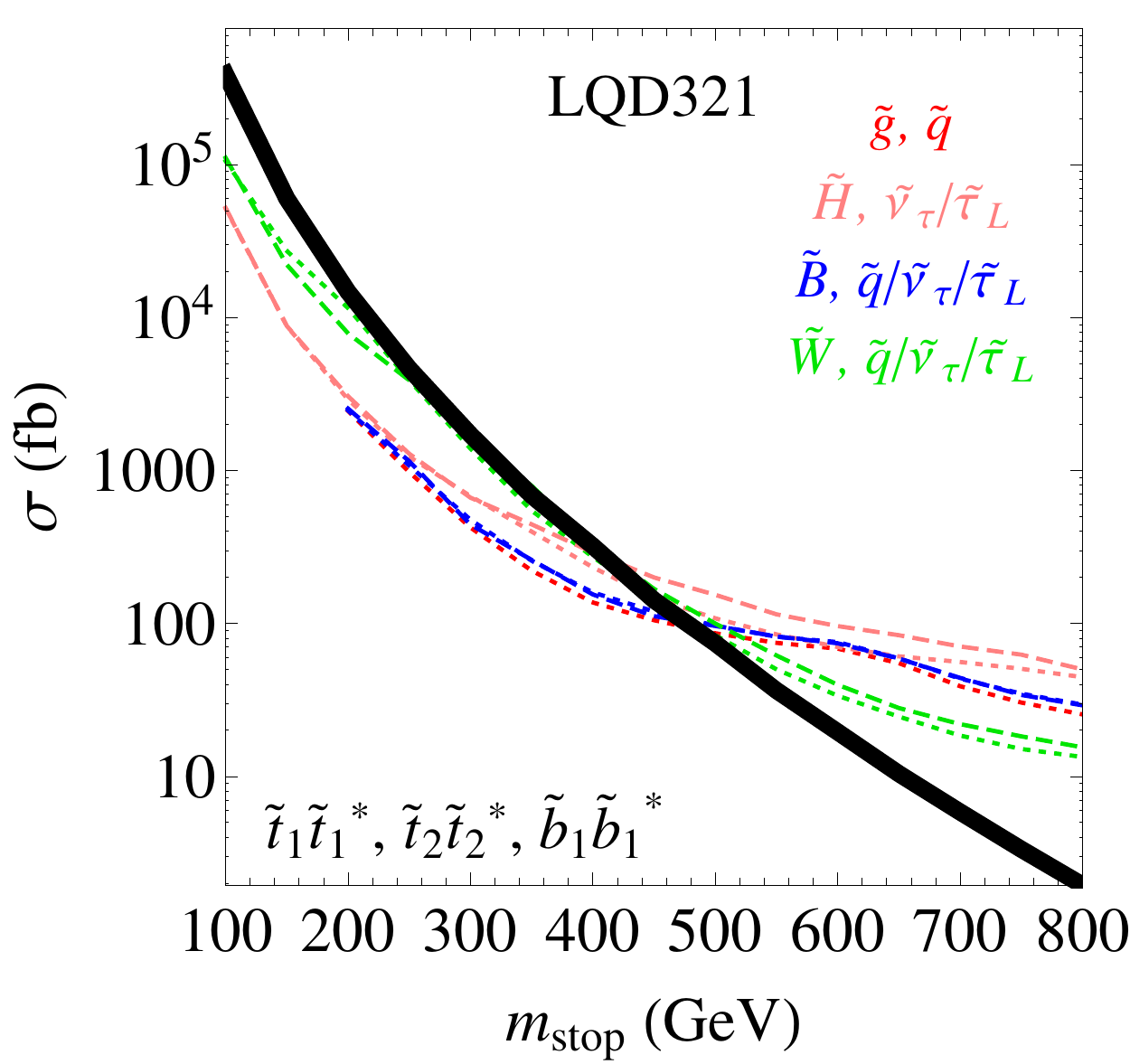}\q
\includegraphics[scale=0.6]{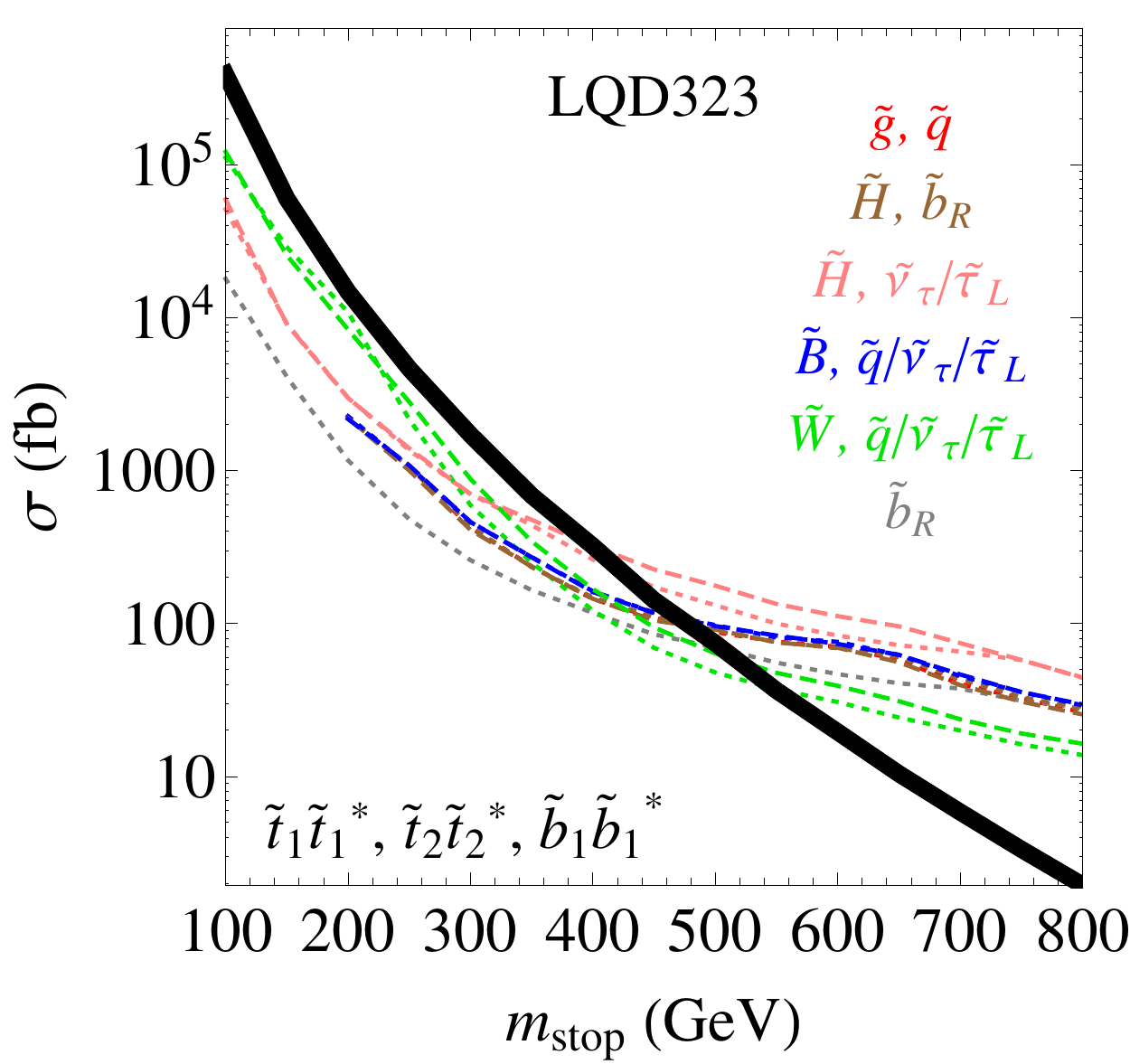}
\caption{Limits on scenarios with $\st_2$ and $\sbo_1$ (both 100~GeV heavier than $\st_1$) in the presence of LQD operators---compare to the $\st_1$-only case of figure~\ref{fig:LQD}. In the $\go$-, $\Ho$-, $\Bo$-, and $\sbo_R$-mediated scenarios, as well as 221 and 123 scenarios with light $\Wo$, the strongest limits are obtained primarily from searches for SS dileptons~\cite{CMS-SS-DIL,CMS-SS-DIL-b} and $b'$~\cite{Chatrchyan:2012yea,ATLAS-CONF-2012-130}. In the other $\Wo$-mediated scenarios, the limits at high $m_{\rm stop}$ are set mostly by ($b$-)jets+MET searches~\cite{ATLAS-3b,ATLAS:2-6jets,ATLAS:6-9jets}, similarly to the $\st_1$-only case, and at low $m_{\rm stop}$ by $Z$+jets+MET searches~\cite{Chatrchyan:2012qka,Aad:2012cz}.}
\label{fig:LQDx2}
\end{center}
\end{figure}

\begin{figure}[t]
\begin{center}
\includegraphics[scale=0.6]{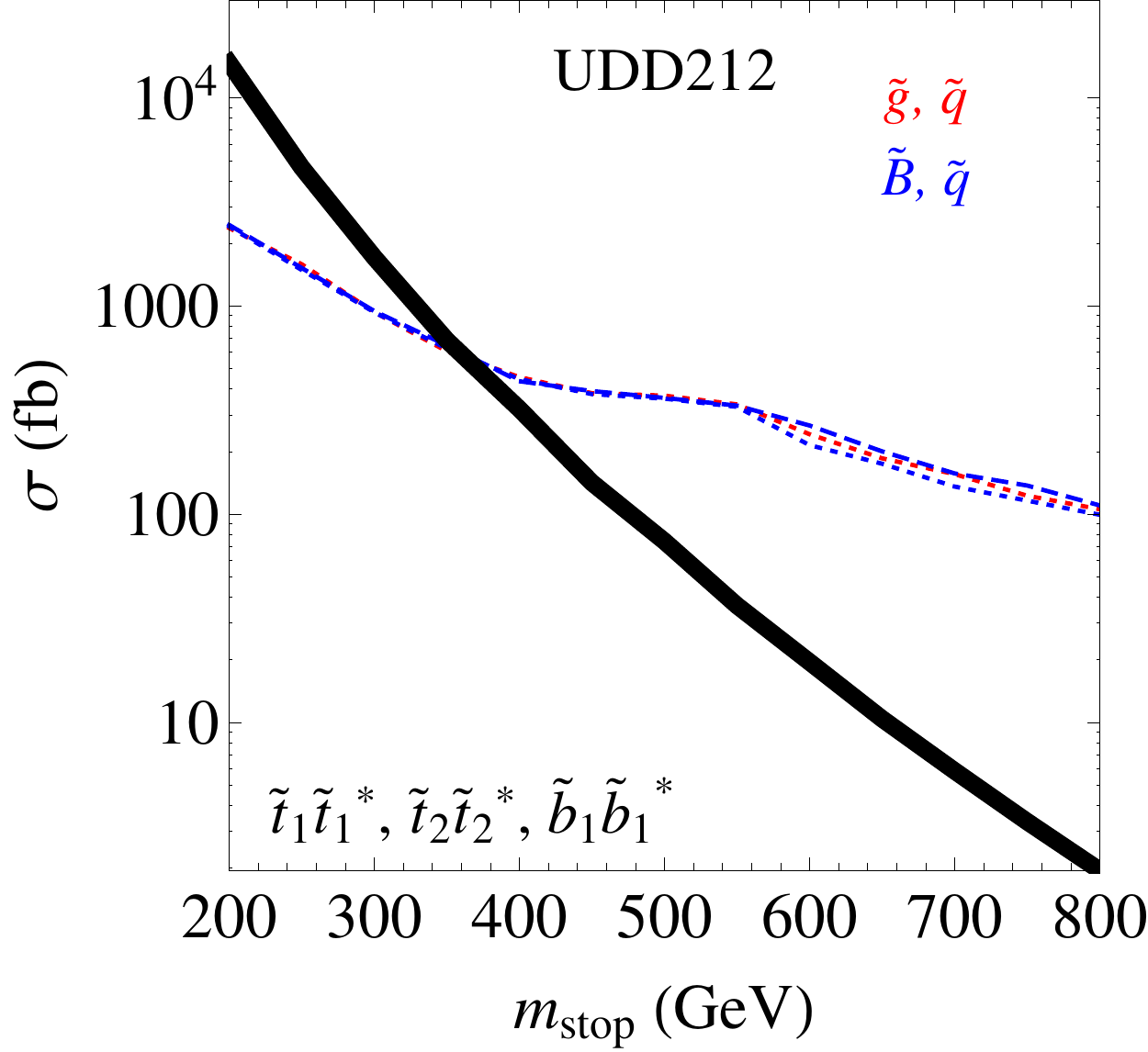}\q
\includegraphics[scale=0.6]{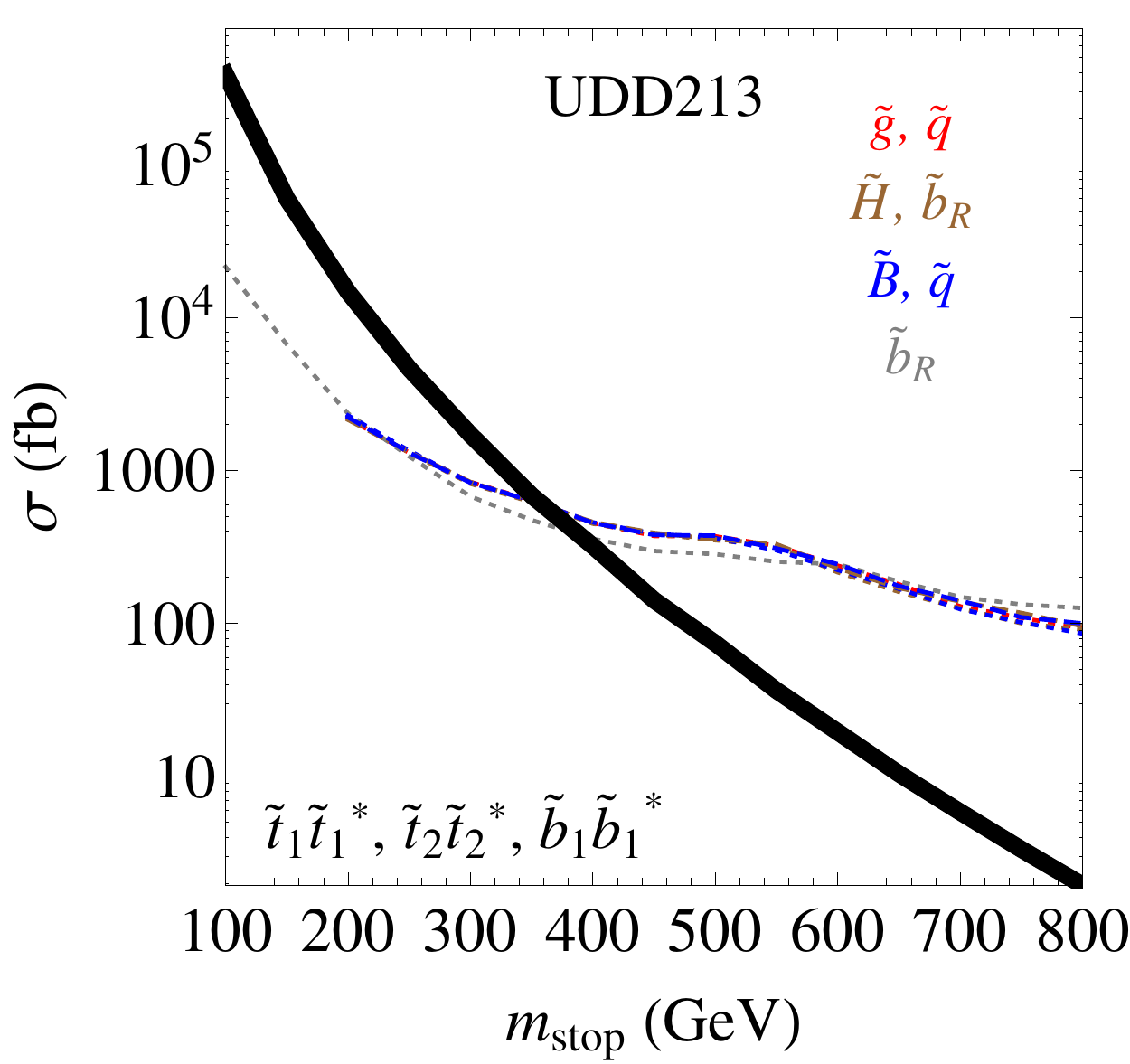}
\caption{Limits on scenarios with $\st_2$ and $\sbo_1$ (both 100~GeV heavier than $\st_1$) in the presence of UDD operators---compare to the $\st_1$-only cases of figure~\ref{fig:UDD}. The limits are set by the search for SS dileptons~\cite{CMS-SS-DIL} and (except for the $\sbo_R$-mediated case) the $b'$ searches~\cite{Chatrchyan:2012yea,ATLAS-CONF-2012-130} and (in the 213 case) the search for SS dileptons with $b$-tags~\cite{CMS-SS-DIL-b}.}
\label{fig:UDDx2}
\end{center}
\end{figure}

As shown in figure~\ref{fig:2bodyx2}, UDD$3jk$ scenarios, where stops decay to pairs of jets, which are completely unconstrained with $\st_1\st_1^\ast$ production alone (figure~\ref{fig:2body}), now obtain limits from searches for multileptons and $Z$+jets+MET.\footnote{The utility of the CMS multileptons search for scenarios of this type has been also studied earlier in~\cite{Brust:2012uf}.} For LQD decays to $\tau j$, the lower bound on the stop mass is still set by the same searches as before, even though the searches for SS dileptons and multileptons strengthen the exclusion at lower masses.

The results for decays via off-shell intermediate particles are shown in figures~\ref{fig:LQDx2} and~\ref{fig:UDDx2}.\footnote{LLE limits are not presented as most of them are very strong even in the case with only a single stop.} The limits on scenarios that were least constrained with $\st_1$ alone (LQD 321 and 323, and all the UDD cases) improve the most (compare to figures~\ref{fig:LQD} and~\ref{fig:UDD}), in some cases even to the extent of almost excluding the stops from the natural mass range.

However, in the presence of inos lighter than $\st_2$ and $\sbo_1$, the decays (\ref{eq:st2-sb1-decays}) can be highly suppressed due to the possibility to decay directly to the inos via
\be
\st_2 \;\to\; b\cho^+\,,\;   t\cho^0
\ee
\be
\sbo_1 \;\to\; b\cho^0\,,\;  t\cho^-
\ee
Such decays of $\st_2$, especially to inos that are lighter than $\st_1$, would typically lead to signatures similar to those arising from the decays of $\st_1$ via the same ino, so they would effectively just increase the size of the signals we studied in section~\ref{sec:limits-singlestop} by a small amount. For the sbottom, the dominant decay would typically be $\sbo_1 \to b\cho^0$,\footnote{However, in the Higgsino case, the decay $\sbo_1 \to t\cho^-$ may dominate (to the extent allowed by phase space) due to the large Yukawa coupling of the top.} leading to a signature similar to that of $\st_1 \to b\cho^+$ followed by a transition of $\cho^+$ to $\cho^0$ as in (\ref{eq:ino-cascade}). We have already included simplified models with the corresponding final states in section~\ref{sec:limits-singlestop}.\footnote{In the bino case, which does not have a $\cho^+$, the analogy does not work. However, bino decays almost always have the same final states as wino and/or Higgsino decays with the same RPV coupling.} While studying these classes of scenarios in full generality would involve scanning a multidimensional parameter space, it is clear that the signatures and the corresponding limits would often be similar to those of the $\st_1$-only scenarios from section~\ref{sec:limits-singlestop}. Therefore, there is a stronger motivation for designing searches for the more conservative $\st_1$-only scenarios. This is relevant also to scenarios with off-shell mediators (the limits on which were presented in figures~\ref{fig:LQDx2}--\ref{fig:UDDx2}) since it is quite plausible for the $\st_2$ and $\sbo_1$ to be much more massive than in (\ref{eq:st2-sb1-masses}).  With a significantly reduced production cross section, their final states can be irrelevant relative to those of $\st_1$.

One should also consider scenarios in which two-body RPV decays of the sbottom dominate over its decays to lighter superpartners. Sbottom decays via LQD$i33$ operators result in a $\nu b$ final state, which is covered by searches for $\sbo_1 \to b\cho_1^0$ with a stable and massless $\cho_1^0$~\cite{ATLAS-CONF-2012-106} or third generation leptoquarks~\cite{1210.5627}, where the current lower bound on the sbottom mass is 490~GeV~\cite{ATLAS-CONF-2012-106}. If the sbottom has a sizable right-handed component, decays to $\ell t$ or $\tau t$ will also be present and are likely to be detectable in SS dilepton or multi-lepton searches. In the LQD$i3k$ ($k\neq 3$) cases, the sbottom decays to $\nu j$. Even though it is a leptoquark-like final state, it is not covered by the existing leptoquark searches since those require at least one of the leptoquarks to give a charged lepton~\cite{CMS-LQ,Aad:2011ch,ATLAS:2012aq}, or the jets to be $b$-jets~\cite{1210.5627}. However, we find a limit of $\sim 400$~GeV (not shown) from the jets+MET search with $\alpha_T$~\cite{CMS-PAS-SUS-11-022}. If the sbottom is partially right-handed, its decays through UDD$ij3$ ($i\neq 3$) would give dijet pairs, a final state we have already analyzed in section~\ref{sec:2body}. On the other hand, UDD$3j3$ would give rise to tops (pairs of $tj$ resonances), joining the $t\bar t$+jets signatures of some of the stop decays from table~\ref{tab:UDD}. In the LQD$ij3$ ($i,j\neq 3$) case, the sbottom has $\ell j$ and $\nu j$ final states, a signature covered by the standard leptoquark searches~\cite{CMS-LQ,Aad:2011ch,ATLAS:2012aq}. The combination of $\tau j$ and $\nu j$ final states in the LQD$3j3$ ($j\neq 3$) case in not covered since the third generation leptoquark searches assume the jet to be a $b$-jet~\cite{CMS-PAS-EXO-12-002,1210.5627}. However, general searches for taus+jets+MET~\cite{:2012ht,CMS-PAS-SUS-12-004} are likely to be effective.

% =============================================================================
\section{Summary and discussion}
\label{sec:Discussion}
% =============================================================================

Motivated by naturalness of the MSSM Higgs potential, which favors light stops, we have spanned the $R$-parity violating stop decay topologies. We considered most plausible prompt\footnote{It is important to remember that the decays of the stop may be displaced, since the RPV couplings may be very small.   In such a case, our limits do not apply, though our classification of the various possible decay signatures may be useful for ensuring that future searches for displaced decays take these possibilities into account.} decays through the various possible mediators in the MSSM. We utilized the ansatz of single RPV coupling dominance to pose the RPV parameter space in a way amenable to a comprehensive, signature-based exploration. Considering both the direct pair production of a single stop and of two stops and a sbottom, we recast existing experimental searches in the 7~TeV LHC data to derive bounds on these scenarios.

We found that stops within the natural mass range are strongly constrained from decaying via LLE operators, moderately constrained from decaying via $e$ or $\mu$ LQD operators, and quite unconstrained from decaying via LQD operators with taus or UDD operators. Including the production of the second stop and a sbottom or stop-antistop oscillations improves limits in some scenarios, but neither of these factors has to be present. Let us therefore focus on the more conservative $\st_1\st_1^\ast$ scenarios without oscillations, and explore the cases in which a naturally light stop within the RPV MSSM is still allowed. In what follows, we will discuss search strategies that could address these unconstrained or underconstrained final states.

The various leptoquark searches have placed very strong limits on the two-body stop decays through LQD$i3k$ operators.  An exception is the $\tau j$ final state (where the jet is not a $b$-jet), which does not have a dedicated search despite the simplicity of the signature. Such a search would likely be able to set strong limits on stops decaying in this manner, since even the generic searches for $\tau$s in association with jets and MET are able to set moderately strong limits on this signature (see LQD332 in figure~\ref{fig:2body}).

Some of the most difficult final states to address experimentally are stop decays into pairs of jets.  As we have seen (figure~\ref{fig:2body}), the existing searches for paired dijet resonances~\cite{Aad:2011yh,ATLAS-CONF-2012-110,CMS-PAS-EXO-11-016} currently place no limits on the stop mass. One simple modification which may allow for sensitivity to the stop is to require one or two $b$-tagged jets (i.e., to form $jb$ resonances). Indeed, two out of the three UDD couplings that could facilitate such a decay include a $b$-quark.  In some scenarios, such as MFV SUSY~\cite{Csaki:2011ge}, couplings involving the third generation are even predicted to dominate.  The presence of $b$-jets may also be useful for triggering, allowing for sensitivity at low masses, even with high instantaneous luminosities. Even without $b$-jets, it may be possible to save bandwidth by recording partial event information as has been done recently in the CMS search for low-mass (single) dijet resonances~\cite{CMS-PAS-EXO-11-094}. It may also be useful to utilize multiple triggers to capture more signal events into the sample, to use pre-scaled triggers, or to extend the analyses down to masses where the trigger efficiency is not flat. More generally, it would likely be beneficial to optimize the searches for the stop, rather than the sgluon or the coloron, which differ in the size of the cross section, the invariant mass distribution of the pairs (i.e., the boost of the stops, on which these searches cut), and radiation (which differs between quarks and gluons and affects combinatoric ambiguities and the signal shape).

While the signature of dijet pairs is one of the most well-known examples of difficult stop decays in the RPV MSSM, there are many well-motivated signatures which have received significantly less attention. For example, it is quite plausible for an electroweak gaugino or the Higgsino to be lighter than the stop~-- a natural MSSM even requires the Higgsino to be fairly light. The $R$-parity conserving decays of the stops to these particles can then easily dominate over the direct RPV decays of the stop. The gaugino or Higgsino would then undergo an RPV decay through a diagram with an off-shell stop or another sfermion. As we have seen, this and other types of scenarios lead to a diverse spectrum of possible final states, the current limits on many of which are very weak or non-existent.

\begin{table}[t]
\begin{center}\small{$\begin{array}{|c|c|c|}\hline
\mbox{Final state} & \mbox{$b$-jets} & \mbox{Scenario(s)} \\\hline\hline
(\tau^+j)(\tau^-j) & 0 & \mbox{LQD332} \\\hline
(jj)(jj) & 0, 2 & \mbox{UDD312/323} \\\hline
8j & 4, 6 & \mbox{UDD312/323 with $\Ho$ decaying via $\st$; UDD213 with $\Ho^\pm\to\Ho^0$} \\\hline
\ell^+\ell^- + 6j & 2, 4, 6 & \begin{tabular}{c} LQD232/233 with $\Ho$/$\Wo$ (unless decays via $\sbo_L$ or $\sbo_R$) \\ LQD221/123 with $\Wo$ \end{tabular} \\\hline
\tau^+\tau^- + 6j &\, 2, 4, 6 \,& \begin{tabular}{c} LQD332/333 with $\Ho$/$\Wo$ (unless decays via $\sbo_L$ or $\sbo_R$) \\ LQD321/323 with $\Ho$-$\snu_\tau/\stau_L$ or $\Wo$ (with or without $\cho^\pm\to\cho^0$) \end{tabular} \\\hline
\tau^\pm\tau^\pm + 6j & 2, 4 & \mbox{LQD321/323 with $\Ho$-$\snu_\tau/\stau_L$ or $\Wo$, with $\cho^\pm\to\cho^0$} \\\hline
t\bar t + 6j & 2, 4 & \mbox{UDD212/213 with $\go$/$\Bo$; UDD213 with $\Ho$} \\\hline
t\bar t + 4j + \mbox{MET} & 2, 4, 6 & \begin{tabular}{c}
LQD321/323 with $\go$/$\Bo$ \\
LQD323/233/333 with $\Ho$ decaying via $\sbo_R$ \\
LQD232/233/332/333 with $\Ho$/$\Wo$ decaying via $\sbo_L$ \\
LQD232/233/332/333 with $\Bo$ (unless decays via $\st$) \end{tabular} \\\hline
\mbox{($tt$ or $t\bar t$) + 6$j$} & 4, 6 & \mbox{UDD312/323 with $\Ho^\pm\to\Ho^0$} \\\hline
t\bar t + 2\tau + 4j & \multirow{2}{*}{2, 4} & \multirow{2}{*}{LQD321/323 with $\go$/$\Bo$; LQD323 with $\Ho$-$\sbo_R$} \\
t\bar t + \tau + 4j + \mbox{MET} & &  \\\hline
\tau^+\tau^-W^+W^- + 2j & \multirow{3}{*}{0} & \multirow{3}{*}{LQD323 with $\sbo_R$} \\
\tau + W^+W^- + 2j + \mbox{MET} & & \\
W^+W^- + 2j + \mbox{MET} & & \\\hline
\mbox{4 tops + 4$j$} & 4, 6 & \mbox{UDD312/323 with $\Bo$} \\\hline
6j + \mbox{MET} & 2, 4 & \begin{tabular}{c}
LQD221/123/321/323 with $\Wo$ \\
LQD321/323 with $\Wo^\pm\to\Wo^0$ \\
LQD232/332 with $\Wo^\pm\to\Wo^0$ (unless decays via $\st$) \\
LQD323 with $\Ho^\pm\to\Ho^0\to\sbo_R$ \end{tabular} \\\hline
\ell + 6j + \mbox{MET} & 2, 4 & \mbox{LQD221/123 with $\Wo$} \\\hline
\tau + 6j + \mbox{MET} & 2, 4 & \begin{tabular}{c} LQD321/323 with $\Wo$ (with or without $\Wo^\pm\to\Wo^0$) \\ LQD323 with $\Ho^\pm\to\Ho^0\to\sbo_R$ \end{tabular} \\\hline
\mbox{$\tau^+\tau^-$ + 2$b$ + MET} & 2 & \mbox{LLE123/233 with heavy $\Wo$} \\\hline
W^+W^- + 4j & 0 & \mbox{UDD213 with $\sbo_R$} \\\hline
\end{array}$}\end{center}
\caption{Dominant final states in scenarios for which the coverage is insufficient (for $m_{\rm stop} \lesssim 500$~GeV). See tables~\ref{tab:LLE}--\ref{tab:UDD} for more detailed descriptions of the scenarios mentioned. The chargino is assumed to decay directly via a sfermion and its RPV coupling (rather than transition to a neutralino first), except where explicitly noted otherwise. As before, couplings related by interchanging electrons and muons, or first and second generation quarks, are listed just once. The second column indicates the possible number of $b$-jets in each scenario (including those coming from top decays, where relevant).}
\label{tab:unconstrained-final-states}
\end{table}

In table~\ref{tab:unconstrained-final-states}, the scenarios for which the limit on the stop mass does not exceed 500~GeV are classified according to the dominant final states for the whole event. Why do these signatures remain elusive?  Unsurprisingly, they typically contain: multiple jets, taus, little or no $\met$, and no more than two leptons with any sizable branching ratio. However, while some of the cases may be genuinely difficult, many of these final states have unique characteristics that are not being exploited by existing searches:\footnote{A generic search for high-multiplicity final states is the search for microscopic black holes~\cite{Chatrchyan:2012taa}. However, its lowest possible $S_T$ cut is $S_T > 1200$~GeV, making it inefficient for stop masses below $\sim 600$~GeV. For 600~GeV stops, the exclusion limits with $S_T > 1200$~GeV are too weak by an order of magnitude relative to the production cross section.}
\begin{itemize}
\item
Most of the final states contain at least two $b$-jets. This happens because the decays of the stop through a chargino (neutralino) always produce a bottom (top). In some of these cases there are even four or more $b$-jets overall. These scenarios provide important motivation for including search regions with a large number of $b$-tagged jets. So far, only a small number of final states have been studied with three $b$-tags (jets+MET~\cite{:2012rg,ATLAS-3b,CMS-PAS-SUS-11-022}, lepton+jets+MET~\cite{CMS-PAS-SUS-11-028}, and SS dileptons+MET~\cite{CMS-SS-DIL-b}) and there have been no searches requiring four or more $b$-tags. Some final states have not been searched for with $b$-tagging at all. Aside from the (quite specific) search for third generation leptoquarks~\cite{CMS-PAS-EXO-12-002} and one of the $t\bar t$ cross section measurements~\cite{Chatrchyan:2012vs}, there have been no new physics searches requiring both hadronic taus and $b$-jets, which would be relevant for a large fraction of the scenarios with weak or no bounds in table~\ref{tab:unconstrained-final-states}. It would also be interesting to explore to what extent the requirement of an unusually large number of $b$-tags can replace the $\met$ requirement for both triggering and background rejection purposes. We should note that reduction in the signal efficiency due to $b$-tagging (or other requirements) would typically not be an issue as light stops have huge production cross sections.
\item
Many of the final states contain a $t\bar t$ pair (where the tops may or may not be on-shell) or a $W^+W^-$ pair (with extra jets). Therefore, searches based on $t\bar t$ cross section measurements (without harsh cuts on $\slash E_T$, jet $p_T$s, etc.) could potentially constrain some of these scenarios.\footnote{Such $t\bar t$-like searches can also be useful for light stops ($m_\st \lesssim m_t$) decaying to $W$, $b$, and an invisible particle in $R$-parity conserving scenarios~\cite{Kats:2011it,Kats:2011qh}.} The stop signal will not stand out on top of the uncertainty on the $t\bar t$ cross section unless the stops are lighter than $\sim 150$~GeV (which is still possible in some cases).  However, requiring multiple additional jets or even extra $b$-tags (which are available in many of the scenarios), or looking at the invariant mass distribution, is likely to be helpful.
\item
Some of the stops decay into final states that mimic top decays. For example, $\st\to b\Ho^{+(\ast)} \to b\tau^+\snu_\tau^{(\ast)} \to b\tau^+ j j$ (where the sneutrino decays via LQD321) has a final state similar to $t\to b W^+ \to b\tau^+\nu_\tau$ (which may contain extra jets from radiation). This, again, motivates searches based on the $t\bar t$ cross section measurements, in this case in the dilepton channel with $\tau$s. These can be extremely efficient since the $t\bar t$ background is suppressed by the dileptonic branching ratio while the stop signal is not. Indeed, in this and several other cases we found the $t\bar t$ cross section measurements to be more sensitive than any existing new physics searches that we examined (despite the fact that we used cross section measurements with 1-2~fb$^{-1}$ while most of the new physics searches were based on 5~fb$^{-1}$ of data). Even when the stop final states do not contain neutrinos, the contribution to $\met$ from $\tau$ decays and/or the multiple jets in the event (jet mismeasurement, neutrinos from $b$ decays) is often sufficient for passing the very mild $\met$ cuts of the $t\bar t$ cross section measurements. It is clear that with further optimization of the selection criteria, the limits on such $t\bar t$-like scenarios, many of which are not yet sufficiently constrained, can be improved significantly.
\item
Leptons produced in the RPV decays of stops or inos are generally hard.  Designing searches with hard cuts on lepton $p_T$s may be useful for improving limits. An example of this gap in coverage can be seen in the LQD221 scenario with on-shell winos, where the dominant process is $\st\to b\Wo^+$, $\Wo^+\to\nu jj$ or $\ell^+ jj$. The strongest bounds there come from single $\ell$+jets+$\met$ searches, while a quarter of the events contain two very high $p_T$ leptons (and many hard jets). Utilizing hard leptons in such a case may be more efficient than $\met$. This could be effectively implemented by a cut on the variable $\lambda_T = \sum_\ell p_T$, or simply requiring the leptons to pass harder $p_T$ cuts. This is essentially done in searches for leptoquarks with $(\ell^+j)(\ell^-j)$ final states~\cite{CMS-LQ} by cutting on the variables $M_{\ell\ell}$, $M_{\ell j}^{\rm min}$ and $S_T^{\ell\ell}$, and may work in a similar way for the $(\ell^+jjj)(\ell^-jjj)$ final states of stops. Additionally, multilepton searches in high $\lambda_T$ regions could provide even tighter limits on some of the simplified models already constrained, increasing the reach for scenarios in which decays through lepton-rich RPV couplings (LLE) compete with less spectacular, but more common decays.
\item
Final states without hard neutrinos lack significant $\met$, making conventional SUSY searches inefficient. However, many of these final states contain one or two leptons and a large number of jets. Such scenarios may be accessible by replacing the large $\met$ requirement with the requirement of a large jet multiplicity, as has been studied in~\cite{Lisanti:2011tm}. The recent CMS search for heavy quarks~\cite{Chatrchyan:2012af} (in the single-lepton channel) demonstrates the viability of such an approach, and it would be useful for its results to be presented in a manner amenable to re-interpretation to other scenarios.
\item
Unlike in typical $R$-parity conserving scenarios, where all the superpartner decay chains include invisible particles, RPV scenarios often have fully visible decays. In such cases, the decay products form resonances, which can be utilized for improving sensitivity. In our context, this can be relevant to the stops themselves and/or other particles through which the decays proceed. In particular, LQD scenarios may contain $(jj)$ resonances from intermediate $\snu$ or $\slep_L$ and/or $(\ell jj)$ resonances from $\cho^0$ or $\cho^\pm$, and UDD scenarios may contain $(jjj)$ resonances due to $\cho^0$ or $\cho^\pm$ (the jets may be $b$-jets and the leptons $\tau$s, depending on the coupling). Most of the final states from table~\ref{tab:unconstrained-final-states} can contain such resonances (when the corresponding intermediate particles are on-shell).
\item
The structure of RPV couplings can easily introduce a preference for one lepton flavor over another.  This means that keeping separate search regions for the different lepton flavors can improve sensitivity.  Furthermore, the lepton flavor universality of most of the SM backgrounds can be used for data-driven background estimates (e.g., doing a measurement in the electron channel for estimating the expected number of events in the tau channel).
\end{itemize}

It is our hope that the experimental community will take these suggestions into consideration in order to maximize the potential for constraining or discovering the light stop of the $R$-parity violating MSSM, or any new physics beyond the Standard Model for that matter, at the LHC.

\section*{Acknowledgments}
\noindent
We thank Andrey Katz, Markus Luty and Scott Thomas for useful conversations.  We are especially grateful to David Shih for useful discussions, collaboration in the early stages of this work and comments on the draft. We also thank R.~Gray, E.~Halkiadakis, A.~Hoecker, K.~Kaadze, G.~Karapostoli, A.~Lath, I.~Melzer-Pellmann, N.~Pietsch, T.~Potter, P.~Pralavorio, F.~Salvatore, and D.~Stuart for answering our questions about some of the ATLAS and CMS searches used in this work.  We thank J.~Alwall and O.~Mattelaer for assistance with a \textsc{MadGraph} complication.  We thank B.~Fuks for help with \textsc{FeynRules}.   This research was supported by DOE grant DE-FG02-96ER40959.

% ------------------------------------------------------------------------------------------------------------------------
% ------------------------------------------------------------------------------------------------------------------------
% ------------------------------------------------------------------------------------------------------------------------
% ------------------------------------------------------------------------------------------------------------------------
% ------------------------------------------------------------------------------------------------------------------------
\appendix

%---------------------------------------------------------------------------------------------------------
\section{Dependence on the off-shell ino mass\label{app:helicity}}
%---------------------------------------------------------------------------------------------------------

In this appendix, we discuss the effect responsible for the dependence of the branching ratios on the ino mass for decays via an off-shell ino. In particular, we will explain why some final states (those presented in parentheses in tables~\ref{tab:LLE} and~\ref{tab:LQD}) in $\Wo$-mediated decays disappear when the wino is much heavier than the stop. The effect is present for other ino mediators as well, but is less dramatic. The essence of the effect, in terms of the right diagram in figure~\ref{fig:3b4bstop}, is that the ino propagator can either preserve or flip the helicity, and since the helicity-flipping propagator is proportional to the ino mass, it dominates the process for heavy inos.

As an example, consider the $\Wo$-mediated decays with LLE couplings (figure~\ref{fig:LLE}), where the effect on the limits is most significant. Processes with tops are phase-space suppressed for any wino mass, so the dominant mediator is the charged (rather than the neutral) wino, i.e., $\tilde{X}=\cho^+\approx \Wo^+$ in the notation of figure~\ref{fig:3b4bstop} (right). The wino can decay through either a left-handed slepton or a sneutrino ($\tilde f = \slep_L$ or $\snu$, with $m_{\slep_L} \approx m_\snu$), so the final state can be either $b\nu\slep_L^{+(\ast)}$ or $b\ell^+\snu^{(\ast)}$ (with the $\slep_L^+$ or $\snu$ decaying further via RPV). As winos only couple to $SU(2)_L$ doublets, the $b$ is left-handed. Without a helicity flip, the helicity of $f_1$ would need to be right-handed, which allows it to be $\ell^+$, but not $\nu$. With a helicity flip it can be $\nu$, but not $\ell^+$. As a result, since the helicity-flipping diagram is proportional to $m_\Wo$, final states with $\nu$ dominate for $m_\Wo \gg m_\st$. This results in $\Gamma(\st\to b \nu\nu \ell^+) \gg \Gamma(\st\to b \ell^+\ell^+ \ell^-)$ for heavy winos, weakening the limits relative to cases with light winos where the rates of the two processes are comparable. Similarly, for the LQD operators and heavy winos, $\Gamma(\st\to b \nu jj ) \gg \Gamma(\st\to b \ell^+ jj)$.

For the sake of completeness, let us also discuss the effect for processes mediated by the neutral wino $\Wo^0$, even though they are suppressed relative to the $\Wo^+$-mediated ones. The helicity-flipping diagram produces $f_1 = \ell^-$ or $\nu$, while in the helicity-conserving diagram $f_1 = \ell^+$ or $\bar\nu$ (assuming again $\tilde f = \slep_L$ or $\snu$).\footnote{The consequences are perhaps surprising. At first glance, one might assume that, for example, the decays $\st\to t \cho^{0\ast} \to t (e^+ \bar{\nu}_\mu \tau^- \mbox{ \emph{vs.} } e^- \nu_\mu \tau^+)$ (for $\lambda_{123}$) have equal branching fractions because the mediator is neutral. This would be correct for an on-shell $\cho^0$ but is not generally the case for off-shell inos. For $m_{\cho^0} \gg m_\st$, the helicity flip will make $t e^+ \bar{\nu}_\mu \tau^-$ dominate over the final state in which the decay products of the neutralino are replaced with their antiparticles. A similar effect has been discussed in~\cite{Ambrosanio:1997bq}.} As a result, the leptons from the stop and the antistop (if both decay through this process) will always have opposite signs for $m_\Wo \gg m_{\st}$, but will sometimes have same signs for lighter winos (note though, depending on the case, additional leptons may be coming from the sfermion decays).

For the Higgsinos, bino, and gluino, which couple to both right- and left-handed fermions, one can construct helicity-flipping diagrams with either outgoing helicity (as long as the stop is somewhat mixed), so the effect is typically less drastic than it is for the winos.

%---------------------------------------------------------------------------------------------------------
\section{Details of simulation and limit computation\label{simdetails}}
%---------------------------------------------------------------------------------------------------------

We use the RPVMSSM model~\cite{Fuks:2012im} of \textsc{FeynRules}~\cite{Christensen:2008py} to define the RPV couplings for \textsc{MadGraph~5}~\cite{Alwall:2011uj} via the UFO interface~\cite{Degrande:2011ua}. To make the generation of our (up to) $2 \to 10$ processes feasible, we generate events for stop pair production ($2 \to 2$) and the various stop decays ($1 \to 2,3,4,5$) separately and combine the resulting LHE files while taking the boosts and the color connections of the stops into account.\footnote{Scenarios with $\st_2$ and $\sbo_1$ production are generated in a similar way, with the $\st_2$s or $\sbo_1$s decayed down to $\st_1$s in the first stage, and then combined with $\st_1$ decays. For scenarios with chargino-neutralino transitions, the decay down to the neutralino is included in the first stage, and then combined with $\Ho^0$ or $\Wo^0$ decays.} The combined events are showered and hadronized in \textsc{Pythia~8}~\cite{Sjostrand:2006za,Sjostrand:2007gs} and further processed with a private detector simulator (using the anti-$k_T$ jet algorithm from \textsc{FastJet}~\cite{Cacciari:2005hq}), which uses truth MC information and includes geometric acceptances of the various particles, jet energy resolution (based on~\cite{CMS-PAS-SUS-11-007}), identification of $b$-jets and hadronic $\tau$ candidates, and computation of isolation variables for leptons. We then apply trigger efficiencies, lepton identification efficiencies, lepton isolation requirements and $b$-tagging efficiencies relevant to each search, to the extent that details about them are provided in the experimental publications or obtained via other means. For the scenarios of section~\ref{sec:limits-oscillation}, we also apply a probability for leptons from each stop decay to reverse their charges due to a stop-antistop oscillation. The resulting events are passed through the analysis cuts. We then compare the NLO+NLL production cross section~\cite{Beenakker:2010nq} with the $95\%$~CL excluded cross section (the limit on $\sigma\times\epsilon$ divided by the simulated efficiency) for each search region. In cases where the limits on $\sigma\times\epsilon$ are not provided in the experimental papers, we compute them with the frequentist method~\cite{Conway:2000ju} using the provided backgrounds (and their uncertainties). The single search region giving the best limit is used in each case.

We have validated our detector simulation and analysis code on signal models that were used by the experimental analyses in cases where the experiments provided their simulated events yields for easily reproducible examples of such models. The previous version of our detector simulation code has been also validated in~\cite{Kats:2011qh}. Typically, our event yields agree with those quoted in the experimental papers to within $\sim 30\%$ (although, in a few cases the discrepancy is about a factor of 2).\footnote{Part of the discrepancies are likely related to the fact that many of the searches have not provided information about the $p_T$ dependence of their lepton identification or $b$-tagging efficiencies. Lepton identification efficiencies are important for searches that allow for, and scenarios that contain, soft electrons or hadronic taus, since their efficiencies get small and $p_T$-dependent. Similarly, knowing the $p_T$ dependence of the $b$-tagging efficiency is crucial for simulating searches that require multiple $b$-tags. Also, we do not simulate fake leptons or $b$-jets (as in most cases the fake rates are not available in the experimental publications). Another plausible source of validation discrepancies is not using the same event generators as in the experimental studies.} The reader may shift our exclusion curves by these amounts to estimate by how much such uncertainties may be affecting the mass limit in each case. When specifying which searches set the best limits, we have listed not only the searches that turned out to be the most powerful according to our simulation, but also those that had comparable power within this uncertainty.

We have not explicitly taken the systematic uncertainty on the signal efficiency into account. In most cases, this can indeed be neglected since the overall uncertainty is typically dominated by the background. However, the uncertainty on the signal efficiency becomes very important when the analysis cuts are such that the signal is coming from the tails of the distributions (of $\met$, $H_T$, etc.). The tails are problematic because higher-order QCD corrections and/or imperfect modeling of the detector may change them dramatically. The multiplicity of scenarios and searches that we cover here does not allow us to analyze the range of validity in each case in detail (we also do not check whether there are any cases in which the contribution of the signal to the control regions of the searches has a significant effect on the background estimate). However, we have used the size of the signal efficiency $\epsilon$ as a rough proxy of sensitivity to the tails (ignoring the fact that some of the efficiency reduction comes from branching ratios) and included only limits that are based on efficiencies above $\epsilon_{\rm min} = 10^{-3}$. More specifically, we have implemented this threshold by modifying the excluded cross section as $\sigma \to \sigma\exp(\epsilon_{\rm min}/\epsilon)$. For scenarios that combine the $\st_1$, $\st_2$, and $\sbo_1$ pair production processes, in order to take into account the qualitatively different properties of these three samples and their unequal cross sections, the efficiency threshold is applied to each sample separately. Furthermore, since our main motivation for considering scenarios with the second stop and the sbottom is to show that the single stop scenario is not overly conservative, we have liberally relaxed $\epsilon_{\rm min}$ by the square of the branching ratio of $Z$ or $W$ to leptons, for the $\st_2$ and $\sbo_1$ samples, respectively, in order to avoid artificially penalizing these samples for the smallness of the branching ratios into their most distinctive final states.

% ------------------------------------------------------------------------------------------------------------------------
\small{\bibliography{RPV}}

\end{document}